\documentclass{article}

\usepackage{arxiv_style}

\usepackage{comment}
\usepackage{subcaption}
\usepackage{bookmark}
\usepackage{float}
\usepackage{placeins}
\usepackage{authblk}
\usepackage{booktabs}
\usepackage{amsfonts}       
\usepackage{amsmath}        
\usepackage{amssymb}        
\usepackage{graphicx}       
\usepackage{wrapfig}
\usepackage[table]{xcolor}
\usepackage{mdframed}
\usepackage[font=small,labelfont=bf]{caption}
\usepackage[square,numbers,sort&compress]{natbib}
\usepackage{subfiles}
\usepackage{microtype}
\usepackage{tablefootnote}
\usepackage{tikz}

\usepackage{amsmath,amssymb,amsthm}
\usepackage{mathtools}
\usepackage{subcaption}

\usepackage{graphicx}
\usepackage{booktabs}
\usepackage{url}

\allowdisplaybreaks

\newcommand{\mciga}{Ciga}
\newcommand{\mhipt}{HIPT}
\newcommand{\mretccl}{RetCCL}
\newcommand{\mkangdino}{Kang-DINO}
\newcommand{\mctranspath}{CTransPath}
\newcommand{\mphikon}{Phikon}
\newcommand{\muni}{UNI}
\newcommand{\mvirchow}{Virchow}
\newcommand{\mrudolfv}{RudolfV}
\newcommand{\mkaiko}{Kaiko ViT-B/8}
\newcommand{\mprovgigapath}{Prov-GigaPath}
\newcommand{\mhoptimus}{H-optimus-0}
\newcommand{\mvirchowTWO}{Virchow2}
\newcommand{\mphikonTWO}{Phikon-v2}
\newcommand{\matlas}{Atlas}
\newcommand{\muniTWO}{UNI2-h}
\newcommand{\mconch}{CONCH}
\newcommand{\mconchONEFIVE}{CONCHv1.5}
\newcommand{\mmusk}{MUSK}
\newcommand{\mhmini}{H0-mini}

\begin{document}

\title{\textbf{Towards Robust Foundation Models for Digital Pathology}}
\date{}

\author[1,2,*]{Jonah~Kömen}

\author[3,*,$\dagger$]{Edwin~D.~de~Jong}

\author[1,2,*]{Julius~Hense}

\author[1,2]{Hannah~Marienwald}

\author[1,2,3]{Jonas~Dippel}

\author[1,2]{Philip~Naumann}

\author[4]{Eric~Marcus}

\author[3]{Lukas~Ruff}

\author[3,7]{Maximilian~Alber}

\author[4]{Jonas~Teuwen}

\author[1,5,6,7,$\dagger$]{Frederick~Klauschen}

\author[1,2,8,9,$\dagger$]{Klaus-Robert~Müller}

\affil[1]{Berlin Institute for the Foundations of Learning and Data (BIFOLD), Berlin, Germany}

\affil[2]{Machine Learning Group, Technische Universität Berlin, Berlin, Germany}

\affil[3]{Aignostics GmbH, Berlin, Germany}

\affil[4]{The Netherlands Cancer Institute Amsterdam (NKI), Antoni van Leeuwenhoek Hospital (AvL), Amsterdam, Netherlands}

\affil[5]{Institute of Pathology, Ludwig Maximilian University, Munich, Germany}

\affil[6]{German Cancer Research Center, Heidelberg, and German Cancer Consortium, Munich, Germany}

\affil[7]{Institute of Pathology, Charité Universitätsmedizin, Berlin, Germany}

\affil[8]{Department of Artificial Intelligence, Korea University, Seoul, Korea}

\affil[9]{Max-Planck Institute for Informatics, Saarbrücken, Germany}

\affil[*]{Equal contribution (co-first authorship).}

\affil[$\dagger$]{Corresponding authors. Emails: edwin.dejong@aignostics.com, f.klauschen@lmu.de, klaus-robert.mueller@tu-berlin.de.}

\maketitle

\begin{abstract}

Biomedical Foundation Models (FMs) are rapidly transforming AI-enabled healthcare research and entering clinical validation. However, their susceptibility to learning non-biological technical features --- including variations in surgical/endoscopic techniques, laboratory procedures, and scanner hardware --- poses risks for clinical deployment.
We present the first systematic investigation of pathology FM robustness to non-biological features. Our work (i) introduces measures to quantify FM robustness, 
(ii) demonstrates the consequences of limited robustness, and (iii) proposes a framework for FM robustification to mitigate these issues. Specifically, we developed \textit{PathoROB}, a robustness benchmark with three novel metrics, including the \textit{robustness index}, and four datasets covering 28 biological classes from 34 medical centers.
Our experiments reveal robustness deficits across all 20 evaluated FMs, and substantial robustness differences between them. We found that non-robust FM representations can cause major diagnostic downstream errors and clinical blunders that prevent safe clinical adoption. Using more robust FMs and post-hoc robustification considerably reduced (but did not yet eliminate) the risk of such errors.
This work establishes that robustness evaluation is essential for validating pathology FMs before clinical adoption and demonstrates that future FM development must integrate robustness as a core design principle. PathoROB provides a blueprint for assessing robustness across biomedical domains, guiding FM improvement efforts towards more robust, representative, and clinically deployable AI systems that prioritize biological information over technical artifacts.

\end{abstract}


\section{Introduction}\label{sec:introduction}

Biomedical Foundation Models (FMs) are large-scale AI models pre-trained on increasingly large unlabeled biomedical datasets \cite{bommasani2021fms,azad2023mediafm_survey,khan2025medfm_survey, bilal2025pathofm_review}. They drastically improved performance and generalization capabilities over standalone supervised models and non-biomedical pre-training across domains \cite{ciga2021self, campanella2025benchmark, tiu2022expert, zhang2025multimodal, lee19biobert, gu2021domain, theodoris2023geneformer, fu2025foundation}.
In digital pathology, FM pre-training has been scaled up to millions of Whole Slide Images (WSIs) and billions of model parameters  \cite{zimmermann2024virchow2, alber2025novel_atlas}.
Some of the resulting models demonstrate remarkable capabilities at a wide range of diagnostic tasks, such as pan-cancer classification or rare cancer detection \cite{chen2024uni, vorontsov2024virchow_natmed, dippel2024ad, campanella2025benchmark}. They further advance the prediction of clinically relevant biomarkers from histology that typically require additional molecular or immunohistochemical testing --- such as MSI, HER2, and EGFR \cite{schirris2022deepsmile, Dippel2024RudolfV, jaume2024hest, campanella2025benchmark, campanella2025egfr} --- and enable real-world clinical utility of ML-based biomarkers \cite{campanella2025egfr}.

As the development of pathology FMs is progressing rapidly, measuring their capabilities and differences becomes increasingly challenging \cite{mahmood2025benchmarking}. To this end, many recent efforts have focused on contributing new pathology benchmarks to assess the performance potential of foundation models in various clinically relevant settings \cite{kaiko2024eva, jaume2024hest, campanella2025benchmark, zhang2025pathobench, vaidya2025pathobench, ma2025pathbench, lee2025benchmarking, mulliqi2025foundation, bareja2025evaluating}.
However, one major issue that deserves systematic analysis is the apparent lack of robustness of FMs to technical variability across medical centers (hospitals, laboratories, biobanks, etc.). Such variability (see, e.g., Sup.~Figure~\ref{fig:dataset_samples}) is caused by numerous factors, including biopsy acquisition technique, tissue preparation and sectioning, staining protocols, and whole slide scanning, among other factors. These differences neither reflect medical nor biological tissue characteristics. Nevertheless, machine learning models can be negatively influenced by these types of variation \cite{Howard2021,histo-xai-review}.
Note that such systematic technical data biases, also known as \textit{batch effects} \cite{leek2010batcheffects, goh2017batcheffects, goh2022batcheffects}, are not limited to digital pathology, but pose a fundamental issue across biomedical disciplines, e.g., in radiology \cite{degrave2021ai,unsupervised-clever-hans} or molecular biology \cite{finak2016standardizing, goh2017batcheffects, vcuklina2021proteomics, zindler2020simulating}.

Foundation models might be expected to provide more robust information thanks to their large and diverse pre-training datasets. However, the self-supervised learning methods applied to pre-train pathology FMs are designed to capture any differences in the data \cite{unsupervised-clever-hans}, which includes technical variation. In fact, recent work suggests that pathology FMs encode technical/medical center information in their representations \cite{koemen2024batcheffects, dejong_pathology_fm_robustness, filiot2025distillingfoundationmodelsrobust, gustafsson2024evaluating, ji2025physical, drexlin2025medi}. For example, Filiot et al.\ \cite{filiot2025distillingfoundationmodelsrobust} considered different stainings and scanners applied to the same slides and observed substantial variations in the resulting FM representations. Other factors prevalent in real-world diagnostic slides, such as differences in tissue fixation, section thickness, and quality, were not considered in that study \cite{filiot2025distillingfoundationmodelsrobust}, though.

With the present work, we intend to contribute to the above-described challenge by thoroughly investigating FM robustness, its medical consequences, and strategies for improving FM robustness.
As a part of this endeavor, we constructed \textbf{PathoROB}, an extensive, first-of-its-kind benchmark for systematically measuring pathology foundation model robustness against non-biological variation across medical centers. It consists of four multi-class multi-medical center datasets from three public sources that facilitate comparisons between biological and non-biological signals present in FM representations of histopathology images. We present three novel metrics for assessing FM robustness and its implications: the performance drop in downstream tasks, a clustering score reflecting the global organization of the embedding space, and the {\em robustness index}: a metric measuring the degree to which foundation embeddings represent biological features rather than confounding ones.
Furthermore, we describe a framework to make foundation models more robust without retraining them and compare different ways to achieve this.

We applied our benchmarking approach to 20 current pathology FMs. We identified major performance differences between the models related to pre-training scale and objective, but also found considerable robustness deficits in all FMs. In addition, we discovered that supervised downstream models are prone to exploiting medical center signatures instead of biological signals, causing diminished generalization performance and potentially dangerous failures. Similarly, medical center signatures deteriorated applications like image clustering and diagnostic case search, which are all based on the learned FM representations. We find that in all cases, the performance drops were correlated with a low robustness index, providing evidence for the utility and predictiveness of the proposed metrics.
Using post-training robustification methods like image-space stain normalization \cite{reinhard} and representation-space batch correction \cite{johnson2007combat, behdenna2023pycombat, murchan2024combat} considerably improved robustness and reduced the risk of downstream errors, but could not eliminate them fully.

In summary, our work demonstrates the importance of including robustness criteria in FM development. It further lays the foundation for more robust pathology foundation models and serves as a blueprint for a systematic evaluation and improvement of FM robustness applicable across biomedical domains.

\section{Results}\label{sec:results}

\begin{figure}[h!]
    \centering
    \includegraphics[width=0.95\linewidth]{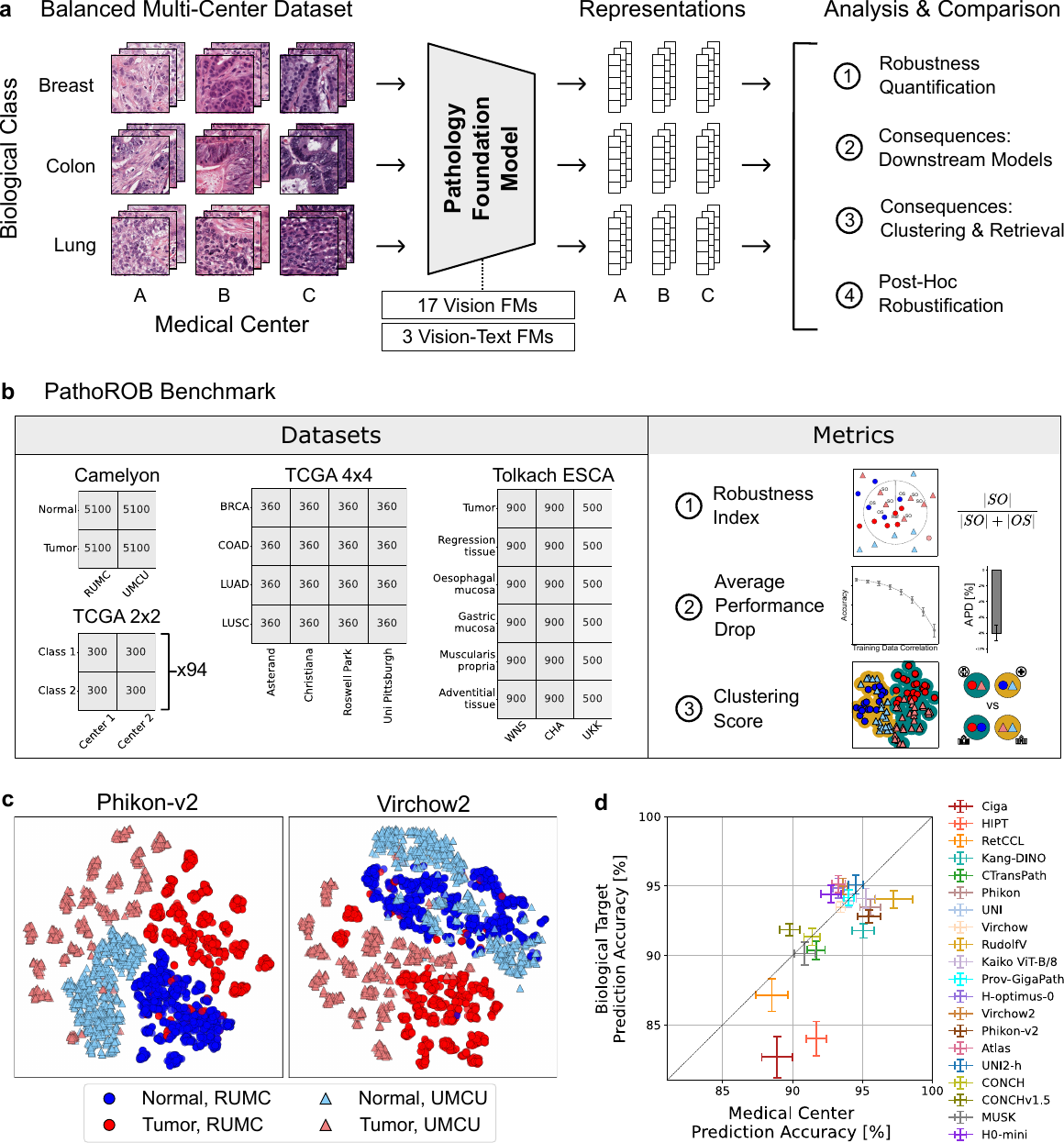}
    \caption{\textbf{PathoROB benchmark and FM representation space exploration}. \textbf{a}~We subsampled balanced multi-center datasets to compare biological class and technical/medical center information, extracted features from 20 pathology foundation models, and analyzed the resulting representations from different perspectives. \textbf{b}~The PathoROB benchmark consists of four datasets from three public sources, together with three metrics to quantify FM robustness and its consequences. Each dataset matrix element depicts the number of patches per combination of biological class and medical center. For TCGA~2x2, we extracted 94 unique class-class-center-center quartets. \textbf{c}~t-SNE plot of the representation spaces of \mphikonTWO{} and \mvirchowTWO{} from a subset of the Camelyon tumor detection dataset (other FMs in Sup.~Note~\ref{sec:apx_feature_space_visus}). The representation space of \mphikonTWO{} is organized by medical center, showing that at the highest level, the model distinguishes images based on the medical center. \mvirchowTWO{}'s representation space is primarily split by biological information (normal/tumor), with a secondary organization by medical center.
    \textbf{d}~Accuracy of predicting medical center vs.\ biological class from the feature vectors via linear probing. We report mean prediction accuracies with 95\% confidence intervals on held-out test sets from three datasets (Camelyon, TCGA~4x4, Tolkach ESCA) with 20 repetitions, respectively. Across all FMs, the medical center origin of most patches could be recovered from the FM representations.}
    \label{fig:sketch-overview}
\end{figure}

Foundation model representations in histopathology encode both biological features (e.g., cell shape and size, tissue architecture, presence of lesions) and technical, non-biological effects (e.g., medical center signatures: staining variations, scanner technology, tissue section thickness). We define robustness as the ability to capture biological features while ignoring confounding technical variations. We argue that foundation models should ideally \textit{only} encode biological information, as technical features compromise generalization and thus reliable clinical usage, as we will show in Section~\ref{sec:consequences}.

In the following, (i) we propose measures for FM robustness and show that a large class of existing foundation models are non-robust. (ii) We then demonstrate that limited robustness can have fatal consequences for downstream tasks such as diagnostics. Finally, (iii) we present a framework for techniques that alleviate these shortcomings and lead to more robust models.
As a basis for demonstrating both the shortcomings and their mitigations, we assembled a data resource composed of four histopathology datasets from three clinical sources \cite{bejnordi2017camelyon, bandi2019camelyon17, komura22tcga_uniform, tolkach2023esca}, each designed to have both multiple medical centers and multiple biological classes (specifically: normal vs.\ malignant, tumor types, tissue compartments) (Figure~\ref{fig:sketch-overview}a). Together, these datasets and our proposed robustness metrics form \textbf{PathoROB}, the first robustness benchmark from real-world multi-center data for pathology foundation models (Figure~\ref{fig:sketch-overview}b). It covers a total of 99,392 patches from 28 biological classes and 34 medical centers (for dataset statistics and example images, see Sup.~Note~\ref{sec:apx_dataset_details}). With PathoROB, we will in the following evaluate 20 popular foundation models covering various architectures, pre-training objectives, pre-training dataset sizes, and model sizes (see Table~\ref{tab:models}), resulting in novel insights into their potential and limitations. 

\subsection{Measuring robustness: the robustness index}\label{sec:robustness_measure}

We distinguish between \textit{biological} features and \textit{confounding} features. Biological features reflect a patient's true condition, e.g.\ whether a tissue sample shows a particular subtype of lung adenocarcinoma; the aim and promise of foundation models is to capture such features reflecting actual underlying biology and morphology. 
We refer to all remaining features as \textit{confounding} features, as they can bias downstream predictions. Examples include features reflecting sample acquisition techniques, such as staining or scanner differences. 

We define robustness as the ability to capture biological features while ignoring confounding features. The \textit{robustness index} quantifies the degree to which this ideal situation is reached. This novel metric results from an analysis of the representation space that can be performed for biomedical foundation models across all domains (see Figure~\ref{fig:robustness_index}a and Methods Section~\ref{ri_derivation}). 

Given a dataset, we collect the $k$ nearest neighbors of all samples. From the set of nearest neighbors, we select the subsets of neighboring samples with either the {\bf S}ame biological class and {\bf O}ther confounding class ($SO$), or the {\bf O}ther biological class and {\bf S}ame confounding class ($OS$). Given these subsets of the $k$ neighbors of the evaluation samples, the robustness index $\mathcal{R}$ is defined as the relation between the sizes of these sets:
\begin{align*} 
\mathcal{R} = \frac{|SO|}{|SO| + |OS|}
\end{align*}
It ranges from 0 (not robust) to 1 (fully robust). Specifically, $\mathcal{R}=0$ / $\mathcal{R}=1$ indicates that technical / biological features completely define the local neighborhoods in the FM representation space. For the motivation behind the metric and a more detailed description, see Methods Section~\ref{sec:methods_robustness} and Sup.~Note~\ref{sec:apx_robustness_index_computation}.

\subsection{Limited robustness of pathology foundation models} \label{sec:robustness_benchmarking}

\begin{figure}[h!]
    \centering
    \includegraphics[width=.86\linewidth]{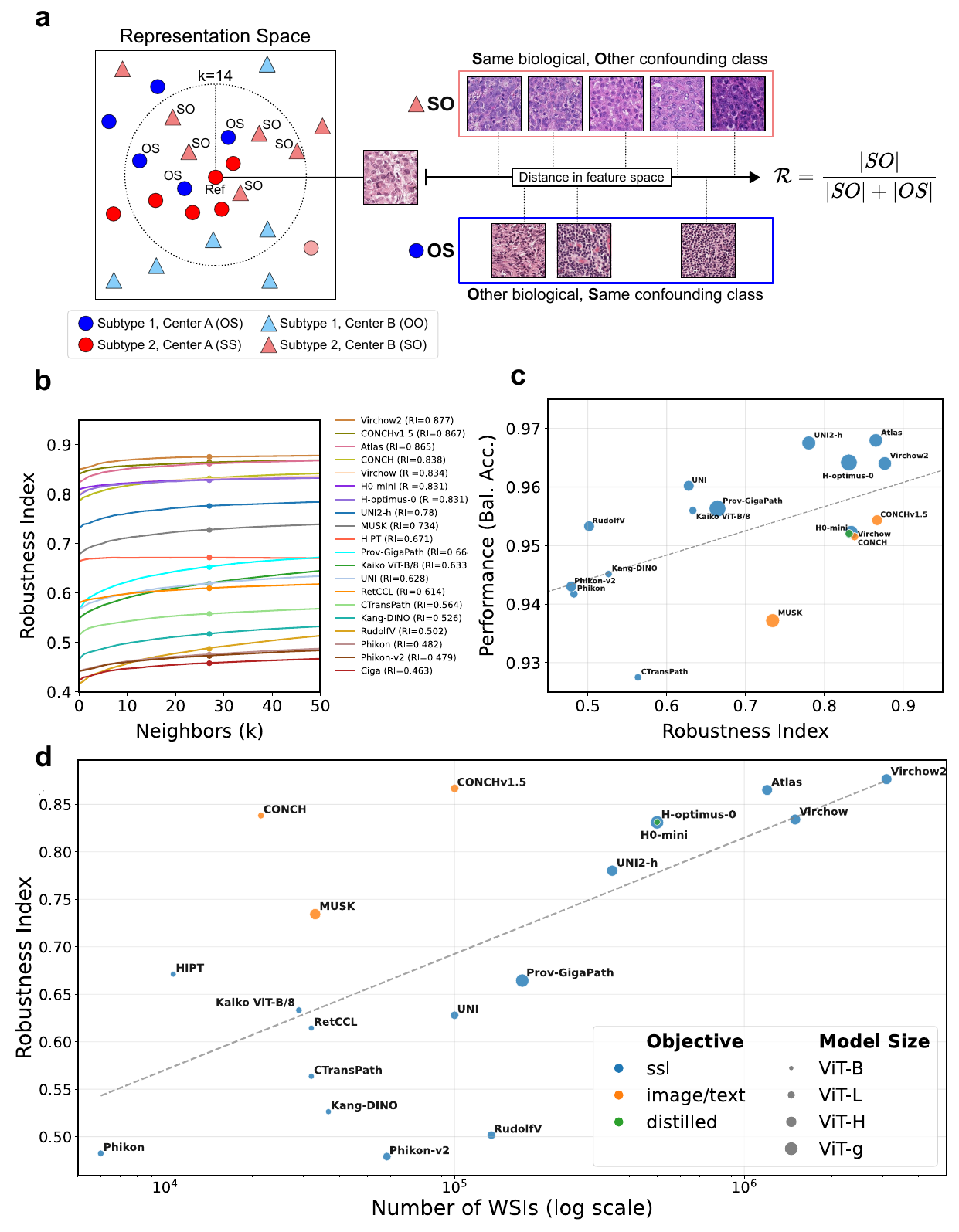}
    \caption{\textbf{Quantification of foundation model robustness}. \textbf{a}~The \textit{robustness index} is computed by comparing the number of nearest neighbors in FM representation space that have the {\bf S}ame biological but {\bf O}ther confounding (``technical'') class (SO) with the number of neighbors having the {\bf O}ther biological and {\bf S}ame confounding (OS) class. Intuitively, having more SO patches in a neighborhood (\textit{=higher robustness index}) indicates that the FM learned to prioritize biological over technical information, while having more OS patches (\textit{=lower robustness index}) shows a stronger influence of technical features on the FM representations. \textbf{b}~Robustness index of all 20 foundation models as a function of the neighborhood size \emph{k}, averaged over the three datasets. The metric shows the models vary widely in robustness. \textbf{c}~Biological $knn$ prediction performance plotted vs.~the robustness index. Only two of the twenty models provide a Pareto-optimal tradeoff between performance and robustness: \mvirchowTWO{} and \matlas{}. \textbf{d}~We can observe a strong correlation ($\rho = 0.692$, $p=0.0047$) between the pre-training dataset size (in number of WSIs) and the robustness index.}
    \label{fig:robustness_index}
\end{figure}

As discussed above, for a foundation model to be robust, its representation space should be organized by \textit{biological} features independent of \textit{confounding} technical features such as scanner type, H\&E staining variations, or section thickness. We note that this is typically not the case (see Figure~\ref{fig:sketch-overview}c, more t-SNE plots in Sup.~Note~\ref{sec:apx_feature_space_visus}): qualitatively, we find in the t-SNE plot that the representation space of the FM \mphikonTWO{}, for example, is organized by medical center, showing that at the highest level, the model distinguishes images primarily based on the medical center origin. In contrast, \mvirchowTWO{}'s representation space is observed as split by biological information (normal/tumor), with a secondary organization by medical center. Interestingly, the medical center of origin could still be reliably predicted from the feature vectors (Figure~\ref{fig:sketch-overview}d) with a mean medical center prediction accuracy between 88\%--98\% averaged over three datasets --- a characteristic that is medically useless and potentially harmful.

On a quantitative level, Figure \ref{fig:robustness_index}b summarizes the main robustness index results. The models vary widely in robustness, and essentially, no models were found to be fully robust, with robustness index scores ranging from 0.463--0.877. Intuitively, this indicates that roughly 53.7\%--12.3\% of the local neighborhoods in embedding space were defined by the medical center instead of biological features. Image/text models (\mconch{}, \mconchONEFIVE{}) and recent large-scale self-supervised models (e.g.\ \mvirchowTWO{}, \matlas{}, \mhoptimus{}) show a higher degree of robustness while smaller-scale models, which have been primarily trained on TCGA, express lower robustness scores (\mciga{}, \mphikon{}, \mphikonTWO{}, \mrudolfv{}, \mkangdino{}, \mctranspath{}).

Figure \ref{fig:robustness_index}c shows the relation between robustness and the prediction accuracy of a $knn$ classifier for predicting the biological class. We argue that robustness should be included as an additional \textit{objective} in medical foundation model evaluation. 
Only two of the twenty models provide a Pareto-optimal tradeoff between prediction accuracy and performance: \mvirchowTWO{} and \matlas{}. All other models score lower than these in either accuracy or robustness, or both. Prediction accuracy and robustness provide different, complementary information, with only limited correlation ($\rho = 0.544$, $p=0.0137$); this highlights the importance of measuring and optimizing both accuracy and robustness, and different models provide different tradeoffs. 

The three image/text models (\mconch{}, \mconchONEFIVE{}, \mmusk{}), while lacking behind top vision-only prediction accuracy, were considerably more robust than many vision-only (SSL) models with comparable or higher prediction accuracy, indicating that additional language supervision (with primarily biological content in the captions) might have helped to suppress the confounding factors, thus increasing robustness.

When only considering SSL models, we can observe a strong correlation ($\rho = 0.692$, $p=0.004$) between the logarithmic number of slides used for FM pre-training and the robustness index (Figure~\ref{fig:robustness_index}d). This highlights that training on larger, diverse datasets also leads to improved robustness.
Yet, despite large-scale pre-training, no foundation model was close to achieving perfect robustness.

Overall, these results highlight that larger SSL models trained on more diverse datasets or the use of image/text objectives yield increased performance; however, robustness remains suboptimal, which may lead to errors in downstream tasks, as we show in the following section.

\subsection{Medical consequences of limited robustness of foundation models} \label{sec:consequences}

\subsubsection{Impact on supervised downstream models} \label{sec:downstream_experiments}

\begin{figure}[h!]
    \centering
    \includegraphics[width=0.92\linewidth]{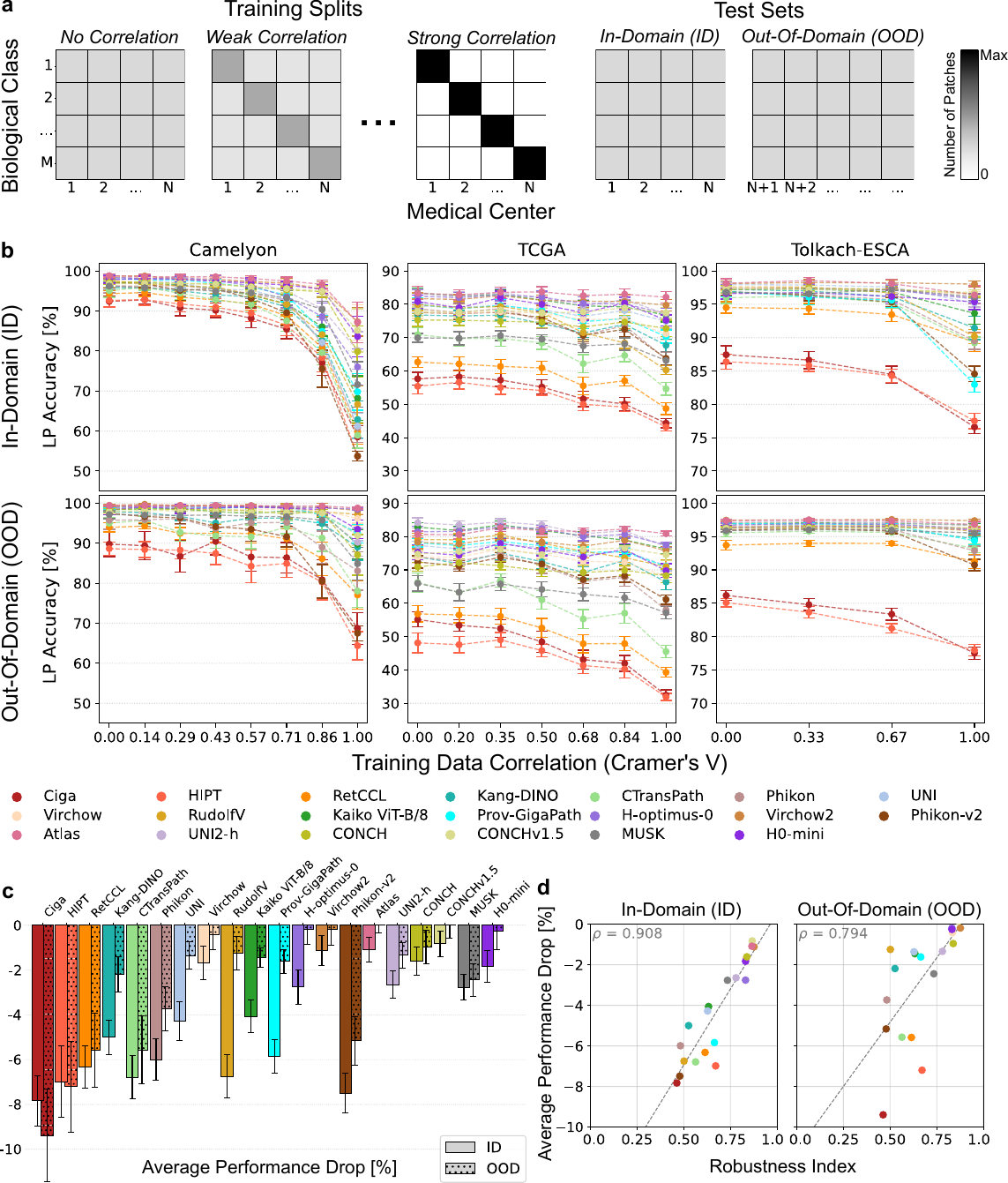}
    \caption{\textbf{Downstream model performance under spuriously correlated training data}. \textbf{a}~Sketch of the experimental setup. We trained supervised linear probing models with heterogeneous data distributions across medical centers, where medical centers and biological targets are increasingly correlated. The models were evaluated on held-out test data from the same and different medical centers. \textbf{b}~In-/out-of-domain generalization performance on data from seen/unseen medical centers. Linear probing accuracies are reported with 95\% confidence intervals over 20 resampling repetitions for each model–dataset combination. As the correlation between the medical center and the biological prediction target (measured by Cram\'er’s V) increased, the generalization performance decreased across foundation models and datasets. \textbf{c}~In-/out-of-domain \textit{average performance drops} per foundation model for FM comparison, aggregated over all datasets and repetitions with 95\% confidence intervals. \textbf{d}~Correlation between the robustness indices and the in-/out-of-domain average performance drops of the FMs. We observed strong correlations (p-values: $0.00004$, $0.00020$) between FM robustness and the stability of downstream model generalization performance.}
    \label{fig:sketch-downstream}
\end{figure}

One of the most important use cases of (upstream) foundation models is the development of task-specific downstream models. For this, we train a shallow neural network on top of the FM representations of a (small) labeled dataset to predict a certain biological target. However, if the FM is not robust, its representations carry medical center signatures. The downstream model may learn to rely on these technical features rather than on true biological signals to make its predictions (aka.\ ``Clever Hans'' / shortcut learning \cite{clever-hans, geirhos2020shortcut, hermannfoundations, unsupervised-clever-hans, Howard2021, mahmood-demographic, brown2023detecting}). Furthermore, there are situations in which the downstream training process cannot distinguish which features reflect true biological signals vs.\ technical artifacts.
These issues will lead to suboptimal generalization and prediction errors on unseen data. In contrast, fully robust foundation models eliminate such risks by exclusively encoding biological signals. An in-depth explanation of Clever Hans learning, its connection with spurious correlation, and a constructed example are provided in Sup.~Note~\ref{sec:apx_downstream_entanglement}.

To corroborate this point, we trained downstream models on data with increasing correlation between medical centers and biological targets (varying degrees of Cram\'er’s V), and measured their accuracy on unseen data from the same medical centers (in-domain generalization, ID) and from previously unseen medical centers (out-of-domain generalization, OOD) (Figure~\ref{fig:sketch-downstream}a; see Methods Section~\ref{sec:methods_downstream} and Sup.~Note~\ref{sec:apx_training_details} for details). Intuitively, such correlations would incentivize the model to exploit medical center information in the FM representations, as they help to solve the training task, even though they are clearly not helpful for generalization. Such ``spurious correlations'' are frequently found in histopathology datasets, particularly for rare diseases \cite{Deangeli2022, Guerrero2024, drexlin2025medi} (see Sup.~Note~\ref{sec:apx_downstream_entanglement} for details).
With increasing spurious correlations, we observed that the generalization performance deteriorated across all foundation models and prediction tasks (Figure~\ref{fig:sketch-downstream}b, top). For Camelyon, where the visual differences between medical centers are strong (see Sup.~Figure~\ref{fig:dataset_samples}), the tumor detection accuracies dropped from $>92\%$ for balanced training data (Cram\'er’s V $=0$) to $53\%$--$87\%$ for fully correlated data (Cram\'er’s V $=1$), indicating that downstream models learned to exploit medical center signatures next to biological features regardless of the foundation model used. In TCGA and Tolkach ESCA --- datasets with more subtle center differences --- the generalization performance was slightly more stable, but the drops observed for most FMs were arguably still unacceptable for clinical applications (TCGA: $-1\%$ to $-25\%$, mean $-12\%$; Tolkach ESCA: $-0\%$ to $-14\%$, mean $-5\%$). Note that decreases in accuracy could even be observed for moderate levels of spurious correlation in the training data, depending on the foundation model.
The out-of-distribution generalization performance also declined as training data correlation increased, although the drop was generally lower than that for in-domain data (Figure~\ref{fig:sketch-downstream}b, bottom).
A closer look into the FM representations revealed a potential explanation for the observed Clever Hans learning effect: irrelevant medical center information is strongly encoded along the directions of greatest variance in the FM representations, making it readily accessible for downstream models to exploit (see Sup.~Note~\ref{sec:apx_feature-space-analysis} for a detailed feature space analysis and elaboration).

We aggregated the diminishing generalization accuracy into an \textit{average performance drop} metric for foundation model comparison (see Methods Section~\ref{sec:methods_downstream} and Sup.~Note~\ref{sec:apx_training_details} for details). The most robust FMs (\mvirchowTWO{}, \mconchONEFIVE{}, \matlas{}) produced the best downstream models, with \mconchONEFIVE{} delivering the most stable performance on in-distribution data ($-0.83\%$ relative decrease on average) and \matlas{} in out-of-distribution settings ($\approx0\%$ performance drop) (Figure~\ref{fig:sketch-downstream}c). Downstream models trained on top of representations from \mhmini{}, \mvirchow{}, \mconch{}, \mhoptimus{}, and \muniTWO{} were generally less stable (ID: $-1.61\%$ to $-2.76\%$, OOD: $-0.19\%$ to $-1.34\%$).  Most other FMs suffered from much more severe relative decreases (up to $-7.83\%$ on ID and $-9.40\%$ on OOD). Overall, the average performance drop was strongly correlated with the robustness index (Figure~\ref{fig:sketch-downstream}d), exhibiting a Spearman correlation of $\rho = 0.908$ ($p=0.00004$) for the in-domain and $\rho = 0.794$ ($p=0.00020$) for the out-of-domain generalization. This indicates that more robust foundation models lead to better downstream models under heterogenous clinical case distributions (aka.~imbalanced multi-site training data), which highlights the importance of foundation model robustness.

\begin{figure}[t!]
    \centering
    \includegraphics[width=0.95\linewidth]{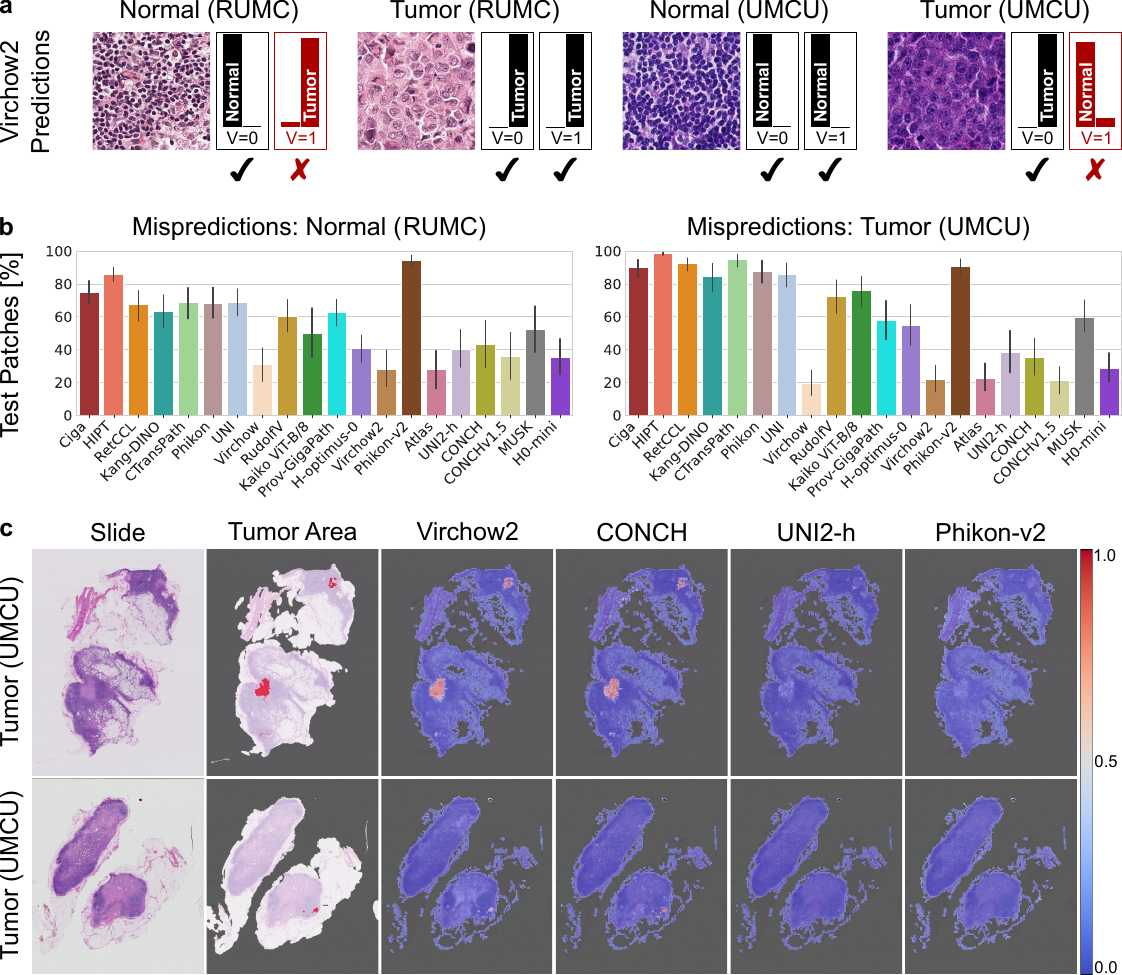}
    \caption{\textbf{Downstream model predictions for the Camelyon tumor detection task}. \textbf{a}~Example predictions for selected Camelyon in-domain test set patches from downstream models trained on \mvirchowTWO{} representations using the balanced training data (Cram\'er’s $\text{V}=0$) and the fully heterogeneous training data (Cram\'er’s $\text{V}=1$). The latter model exploited medical center characteristics as a Clever Hans feature. \textbf{b}~Number of mispredicted RUMC normal and UMCU tumor patches from the in-domain test set per foundation model. The downstream models were trained on the fully heterogeneous training data (Cram\'er’s $\text{V}=1$). We report the mean and 95\% confidence interval over 20 repetitions. We observed systematic medical center-specific mispredictions across foundation models. \textbf{c}~Slide-level downstream model predictions for two held-out Camelyon slides. The downstream models were trained on the fully heterogeneous training data (Cram\'er’s $\text{V}=1$). The ``Tumor Area'' marked in red shows the pathologists' annotation. The four right columns show the softmax patch prediction tumor scores per FM. The less robust models failed to detect the critical tumor areas.}
    \label{fig:mispredictions}
\end{figure}

To further elucidate the medical implications of limited FM robustness and Clever Hans learning, we inspected the predictions of the downstream models trained on fully correlated data (Cram\'er’s $\text{V}=1$) for the Camelyon tumor detection task. Figure~\ref{fig:mispredictions}a showcases the exploitation of medical center signatures as Clever Hans feature. For balanced training data (Cram\'er’s $\text{V}=0$), i.e., homogeneous clinical case distributions across contributing medical centers, the downstream model derived from the \mvirchowTWO{} representations predicted all displayed samples correctly, as expected. However, for correlated training data (Cram\'er’s $\text{V}=1$), i.e., clinical case distributions where the hospital is a confounder, it made obvious mispredictions: most notably, it mistook a morphologically unequivocal tumor patch for a normal patch based on its medical center origin (right). Figure~\ref{fig:mispredictions}b confirms that these mispredictions were a structural issue. On average, between 28\% (\matlas{}, \mvirchowTWO{}) and 94\% (\mphikonTWO{}) of normal patches from the RUMC medical center were incorrectly predicted as tumors. More importantly, between 19\% (\mvirchow{}) and 99\% (\mhipt{}) of tumor patches from the UMCU medical center were not recognized or misclassified. Notably, here, more robust foundation models (\mvirchowTWO{}, \matlas{}, \mconchONEFIVE{}, \mvirchow{}) yielded significantly lower false negative rates than less robust FMs (e.g., \mhoptimus{}, \mprovgigapath{}, \mmusk{}, \mphikonTWO{}).

We then applied the downstream models to predict all patches of unseen whole-slide images from the UMCU medical center (Figure~\ref{fig:mispredictions}c). Despite a large number of pre-training images and strong performance on standard benchmarks\footnote{\url{https://github.com/bioptimus/releases/tree/main/models/h-optimus/v0}}, the foundation models with lower robustness (shown for \muniTWO{}, \mphikonTWO{} as representatives) did not enable the identification of the tumor areas. With \mvirchowTWO{} as one of the most robust models, in contrast, recognition of larger and smaller tumor areas was still possible, even though the small tumor area (bottom slide) only received low prediction scores. Similarly, the \mconch{}-based downstream model still managed to highlight all tumor regions. Notice that \mconch{} achieved a high robustness rating, even though it was pre-trained on a significantly lower amount of whole-slide images and was mostly outperformed by the other three foundation models in regular biological benchmark tasks (see, e.g., \cite{ma2025pathbench}).

In summary, our results show that --- in the typical practical case where homogeneous training data across medical centers cannot be guaranteed --- FM-derived downstream models are prone to making severe mistakes that rule out their safe use in high-stakes clinical settings. A higher foundation model robustness, however, profoundly decreases the risk of such errors and thus increases practical utility and safety.

\subsubsection{Impact on clustering and retrieval}
\label{sec:clustering_retrieval}

Another key application of pathology foundation models is generating insights directly from the FM representations of diagnostic cases,
e.g., through clustering (grouping morphological patterns) or retrieval (search for similar samples) \cite{histo-xai-review, Dippel2024RudolfV, tizhoosh2024retrieval, alfasly2025retrievalval, vaidya2025pathobench}. In light of the presented robustness deficits in foundation models, we investigated how medical center signatures might impact these unsupervised algorithms.

\begin{figure}[t!]
    \centering
    \includegraphics[width=0.95\linewidth]{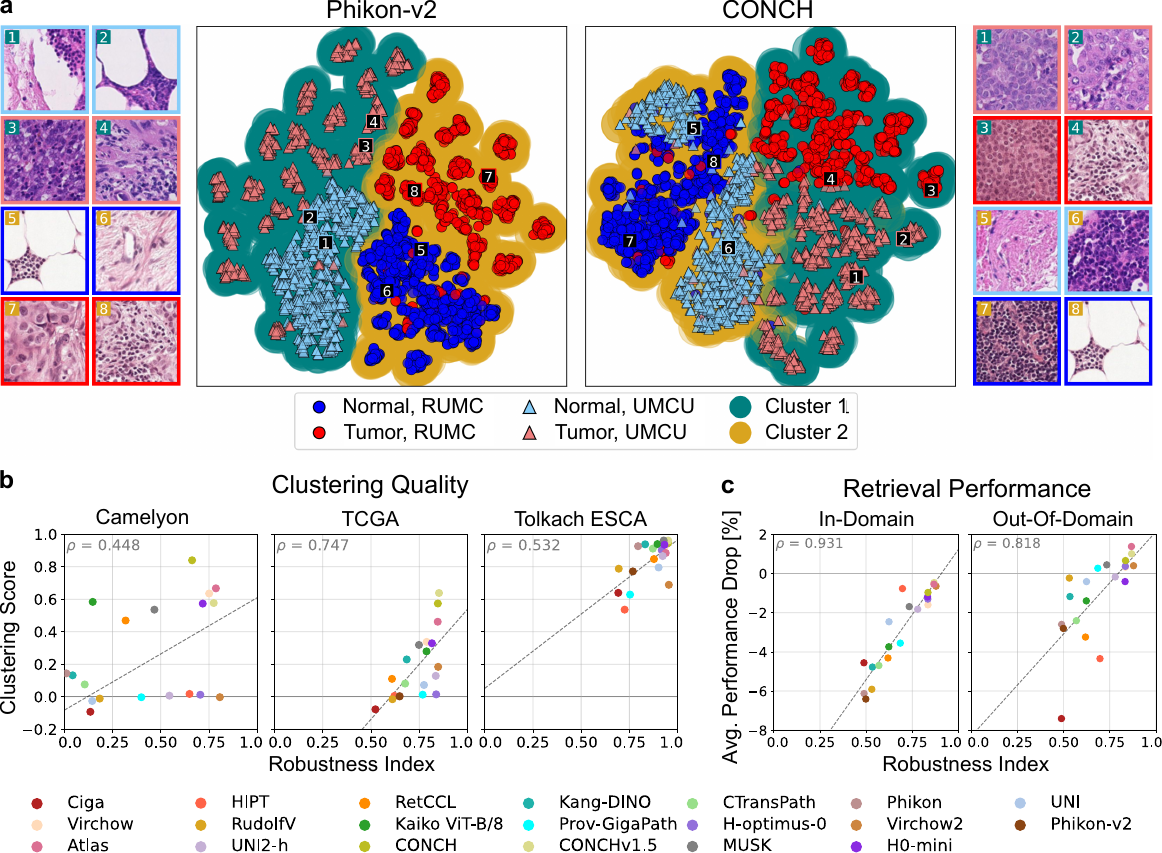}\caption{\textbf{Clustering and retrieval with non-robust FM representations.} \textbf{a} Comparison of how patches are clustered based on their similarity in the FM representation spaces of \mphikonTWO{} (left) and \mconch{} (right). The t-SNE visualizations of the representation spaces show the cluster assignments (green/yellow) alongside the biological class and medical center of each patch. Some example patches are illustrated for each cluster, together with their biological class and medical center encoded by the image frame. Clusters were found by $K$-means clustering without any prior information. In \mphikonTWO{}, patch representations clustered primarily by medical center rather than biological class, suggesting that the limited FM robustness hinders the generation of meaningful insights. In contrast, \mconch{} clusters mostly reflect morphological differences, successfully separating tumor from normal tissue across hospitals. \textbf{b}~Comparison of the robustness index and clustering score per dataset. The \textit{clustering score} quantifies the clustering quality ($1 =$ clusters are purely biological, $-1 =$ clusters are medical center-driven). Most FMs show low clustering scores, indicating center influence, while robust FMs generally yield better clustering, with data-specific variation. Moderate to strong correlations were observed (p-values: $0.0476$, $0.0004$, and $0.0181$ resp.). \textbf{c}~Correlation between the robustness index and the in-/out-of-domain average performance drops in heterogeneous patch retrieval databases. We observed strong correlations between FM robustness and the stability of patch retrieval performance (p-values: $0.00004$, $0.00008$).}
    \label{fig:robidx_clusidx_overview}
\end{figure}

Clustering is used, for example, to uncover novel disease subtypes or clinically relevant morphological patterns \cite{histo-xai-review, zhang2024histoclustering, chelebian2024depicter}. The underlying assumption is that samples with similar biological characteristics are close to each other in the FM representation space and will therefore form clusters. However, if medical center signatures influence the distances between embeddings, to-be-discovered clusters may be built around non-biological factors instead.
Figure~\ref{fig:robidx_clusidx_overview}a showcases this issue qualitatively. In \mphikonTWO{}, clustering is predominantly driven by medical center origin, with Cluster 1 composed mainly of UMCU patches and Cluster 2 of RUMC patches. As a result, the clusters are biologically heterogeneous and do not facilitate subtype discovery. Biologically similar patches from different centers are not grouped together, preventing the identification of biological commonalities across the cohort and medical centers. In contrast, \mconch{} representations yield biologically coherent clusters that combine patches of both medical centers but are separated by tumor status. This organization facilitates hierarchical analysis of intra-cluster biological similarities and inter-cluster biological differences.

To quantitatively assess and compare the quality of clusters across foundation models, we define a \textit{clustering score}. It favors biologically meaningful groupings while penalizing clusters driven by medical center, and ranges from -1 (pure medical center clusters) to 1 (perfect biological clusters), with 0 implying equal influence of both factors or random clustering. In contrast to the robustness index, which measures the influence of morphology and artifacts on the local neighborhood of each representation, the clustering score evaluates their impact on the global structure of the representation space (see Methods Section~\ref{sec:methods_clustering} for details).

Figure~\ref{fig:robidx_clusidx_overview}b highlights that, due to the impact of medical center signatures on FM representations, clustering often yields imperfect or poor results with clustering scores close to 0. Therefore, samples coming from the same medical center will frequently end up in the same cluster, even if their morphologies differ. Consequently, the utility of clustering in multi-center datasets, e.g., for identifying new disease subtypes or patterns, is substantially limited for non-robust FMs.
%
%
Earlier models, such as \mciga{}, \mhipt{}, \mctranspath{}, and \mphikon{}, tended to be more susceptible to center signatures, while more recent methods like \matlas{}, \mconchONEFIVE{}, and \mconch{} achieved the highest clustering scores across all datasets. Surprisingly, \mvirchowTWO{} produced considerably worse clustering scores, indicating that its clusters were heavily influenced by medical centers. Furthermore, the difficulty of clustering appeared to be dataset-dependent. In Tolkach ESCA, biological information was more salient, while in Camelyon, a more balanced influence of both medical center and biological features was observed. TCGA proved particularly challenging, especially LUSC-LUAD, for which none of the FMs produced representations suitable for accurate subtyping.

%
We furthermore observe a positive correlation $\rho$ between the robustness index and clustering score across datasets. This supports the notion that coherent local neighborhoods tend to reflect consistent global structures, suggesting that the robustness index may be predictive of the FMs' clustering quality in multi-center datasets. However, some models exhibited high robustness indices but clustering indices near zero, meaning that the local structure may reflect biological information, whereas the global clustering is influenced jointly by biological and confounding factors (this is further explored in Sup.~Note~\ref{sec:apx_upperbound_cs}).

We further assessed the efficacy of retrieving histologically similar images from a case database using the foundation model representations in the context of limited robustness (see Sup.~Note~\ref{sec:apx_retrieval_results} for details). This clinically relevant application has recently gained attention \cite{Dippel2024RudolfV, tizhoosh2024retrieval, alfasly2025retrievalval}.
%
%
The issues of non-robust FMs encountered for downstream models also re-occurred in this setting (Figure~\ref{fig:robidx_clusidx_overview}c and Sup.~Note~\ref{sec:apx_retrieval_results}). We observed worse performance and mispredictions when retrieval databases had heterogeneous clinical distributions across medical centers. This means qualitatively that retrieval for multi-site data is more likely to fail if the retrieval database cannot be guaranteed to have balanced data contributions from all medical centers. Quantitatively, however, the retrieval accuracies decreased more gently and did not drop as steeply as in the supervised downstream setting (Section~\ref{sec:downstream_experiments}), and were stable for out-of-domain patches when using some of the more robust foundation models (\matlas{}, \mconchONEFIVE{}, \mconch{}, \mmusk{}, \mvirchowTWO{}, \mvirchow{}, \mprovgigapath{}, \mhoptimus{}). Overall, more robust foundation models led to better retrieval results in heterogeneous databases (Figure~\ref{fig:robidx_clusidx_overview}c).

\subsection{Robustification of foundation models by adjusting FM representation spaces} \label{sec:robustification}

\begin{figure}[h!]
    \centering
    \includegraphics[width=0.95\linewidth]{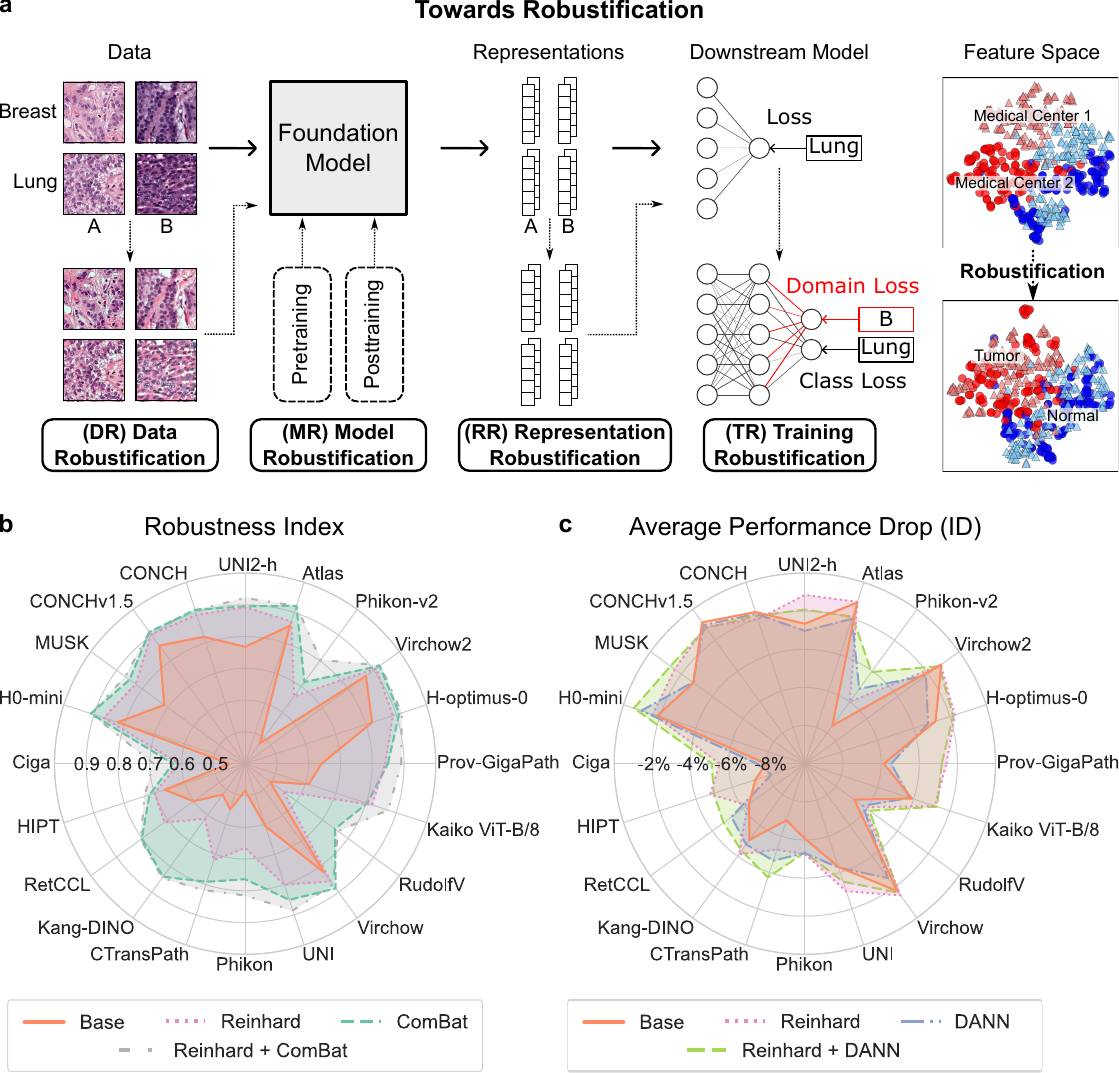}
    \caption{\textbf{Robustification of foundation model representations without FM retraining}. \textbf{a}~A~framework for FM robustness improvement. One option is to explicitly pre-train or post-train (i.e., finetune after initial pre-training) an FM to be more robust (MR: \textit{model robustification}). Here, we consider alternative approaches that do not require FM re-training: eliminating medical center signatures in image space (DR: \textit{data robustification}), in the FM representations (RR: \textit{representation robustification}), and during downstream model training (TR: \textit{training robustification}). Note that DR, RR, and TR have to be re-applied for each downstream dataset, while MR is only performed once. \textbf{b}~The effect of robustification on the robustness index. We report the mean robustness index over the Camelyon, TCGA~2x2, and Tolkach ESCA datasets after robustification as computed in Figure~\ref{fig:robustness_index}c (\textit{higher = better}). Reinhard stain normalization (DR) and ComBat batch correction (RR) considerably increased the robustness index across foundation models. TR was not applicable here as it requires downstream model training. \textbf{c}~The effect of robustification on Clever Hans learning of supervised downstream models. We report the average performance drops under spuriously correlated training data across Camelyon, TCGA~4x4, and Tolkach ESCA as computed in Figure~\ref{fig:sketch-downstream}c (\textit{higher / closer to zero = lower performance drop / better}). Reinhard stain normalization (DR) and domain-adversarial training (TR) improved the downstream model generalization performance for most foundation models. ComBat batch correction (RR) did not perform competitively.}
    \label{fig:sketch-mitigation}
\end{figure}
%

We will now discuss a framework to alleviate or correct the shortcomings of non-robust FMs. To address this, we explored approaches to robustify FM representation spaces for downstream analysis without requiring FM retraining, as the latter is generally costly and often practically impossible.

%
%
We studied three different approaches (Figure~\ref{fig:sketch-mitigation}a): (i) data robustification (\textbf{DR}), i.e., directly removing medical center signatures (e.g. staining or scanner artifacts) from the images, (ii) representation robustification (\textbf{RR}), i.e., removing signatures after feature extraction in the FM representations, and (iii) training robustification (\textbf{TR}), i.e., preventing downstream models from using the signatures as Clever Hans features for their predictions.

Specifically, as a typical DR approach, we inspected Reinhard stain normalization \cite{reinhard}, which harmonizes staining differences in image patches. For RR, we further adopted ComBat batch correction \cite{johnson2007combat, behdenna2023pycombat}, which fits an empirical Bayesian framework to identify and remove batch effects from high-dimensional vectors. Although originally developed for molecular data, it has recently shown potential for correcting histopathology representations \cite{murchan2024combat}. Finally, for TR, we assessed domain-adversarial neural network (DANN) training \cite{DBLP:journals/jmlr/GaninUAGLLML16}, which penalizes the use of medical center features in downstream models via an additional loss term (see Methods Section~\ref{sec:methods_mitigation} for details). 

\paragraph{Improved robustness index} We first re-computed the robustness index on the robustified FM representation spaces (Figure~\ref{fig:sketch-mitigation}b). Reinhard stain normalization considerably improved robustness for most foundation models, with relative increases of +16.2\% on average. Strikingly, ComBat enhanced it even further (+27.4\% on average), suggesting that the approach of representation robustification (RR) holds substantial potential for making FM representations more biologically meaningful. The most robust FM representations were achieved by combining both methods (DR+RR), yielding a robustness index of up to 0.92 (\mvirchowTWO{}, \matlas{}, \muniTWO{}). The largest jumps were observed for the initially less robust models, with a relative rise of up to 68.1\% (\mphikon{}), and \muniTWO{}, \mhoptimus{}, \mhmini{}, \mconch{}, and \mprovgigapath{} achieving robustness indices close to the best reported scores (0.89--0.92) after robustification.

\paragraph{Improved clustering} Similar effects can be observed for clustering, as described in Section~\ref{sec:clustering_retrieval}, on the robustified representations (detailed results in Sup.~Note~\ref{sec:apx_robustified_clustering}).
Robustification strategies led to an increase in clustering scores and thus clustering quality.
ComBat (RR) yielded the most consistent improvements, even transforming the lowest observed clustering score of \mphikonTWO{} on the Camelyon tumor detection dataset from $-0.99$ (indicating a clustering purely by medical center) to a significantly improved $+0.61$ (i.e., biologically more meaningful clusters). 
Robustification was particularly beneficial for the clustering quality in less robust FMs, whereas models that already exhibited high clustering scores without robustification saw only smaller gains.

\paragraph{Improved generalization} To understand to what extent the proposed robustification of FM representations can alleviate the negative consequences of non-robust foundation models, we re-assessed the generalization performance of downstream models as described in Section~\ref{sec:downstream_experiments} (see~Figure~\ref{fig:sketch-mitigation}c).
We find that on in-distribution test data, Reinhard stain normalization (DR) could reduce Clever Hans learning and improve the generalization performance for most FMs, on average by 1.11\%pt (max.\ +3.28\%pt). Notably, the approach was most effective for training data with strong spurious correlations (see Sup.~Note~\ref{sec:apx_robustification_downstream_results}).
Domain-adversarial training (TR) only increased performance for 12 out of 20 FMs, leading to a weaker positive effect of +0.23\%pt on average (max.\ +2.40\%pt). Reinhard and DANN could be combined to yield the highest increase of 1.30\%pt on average (max.\ +3.48\%pt), although using Reinhard stain normalization alone was still more effective for some FMs (e.g., \muniTWO{}, \mvirchow{}). After robustification, some originally less robust FMs (i.e., \mhmini{}, \muniTWO{}, \mmusk{}) enabled downstream model performance comparable to \mvirchowTWO{}, \matlas{}, and \mconchONEFIVE{}. Similar trends were observed for out-of-distribution test data (see Sup.~Note~\ref{sec:apx_robustification_downstream_results}).

Interestingly, despite substantially enhancing the robustness index scores, ComBat (RR) did not consistently lead to better downstream models; it even exacerbated the performance drops compared to the setting without robustification. More precisely, despite effectively reducing mispredictions for slightly heterogeneous training data, it likely removed important biological signals when they were strongly correlated with medical center information (see Sup.~Note~\ref{sec:apx_robustification_downstream_results}). This suggests that ComBat may only be effective if the data contribution of each medical center covers all biological characteristics present in the cohort to be analyzed.

We further find that no method \textit{entirely} eliminated the generalization performance drops, i.e.~supervised downstream models still learned aspects of medical center signatures as Clever Hans features instead of exclusively focusing on biological signals. A closer look into the FM representation spaces provided a potential explanation: medical center information is often entangled with biological features. Specifically, we found that biological and technical information was not encoded in separate linear directions of the FM representations; instead, the same linear directions often contained both types of signals (see Sup.~Note~\ref{sec:apx_feature-space-analysis} for details). Therefore, eliminating medical center signatures also risks damaging important biological signals.

In summary, representation robustification via ComBat has the potential to greatly improve FM robustness, but may also remove aspects of the biological signals when data distributions are heterogeneous across medical centers. Training robustification via DANN was found to be limited in effect and applicability. Data robustification via stain normalization brought consistent improvements, but cannot achieve complete robustification since it ignores technical factors other than staining.
Nonetheless, our results demonstrate that robustifying FM representations can alleviate performance drops without the need for retraining FMs.

\section{Discussion} \label{sec:discussion}

Foundation models have become a de facto standard across fields \cite{bommasani2021fms, radford2021clip, jia2021scaling, gpt4, azad2023mediafm_survey, zhou2024comprehensive, singhal2023large, Dippel2024RudolfV, kabylda2024molecular, kovacs2023evaluation, awais2025foundation, shaban2025foundation, tiu2022expert, zhang2025multimodal, theodoris2023geneformer, fu2025foundation} due to their broad application scope and their capacity to lower data requirements for downstream models. Ideally, their representations reflect the underlying (biomedical) problem well. Recently, however, several studies have demonstrated that the learned representations have the tendency to reflect local correlations well but can only suboptimally model long-range or semantic problem structure \cite{muttenthaler2024aligning}. Moreover,  representation learning can be subject to Clever Hans effects due to systematic failures occurring in unregularized unsupervised learning \cite{unsupervised-clever-hans}. 

In this work, we have contributed a further aspect to this discussion that has so far not received sufficient attention by the community, namely, the non-robustness of FMs that we show to cause critical failure in pathology FMs. 
We analyzed 20 current pathology FMs and observed that the non-robustness of FMs can lead to learned representations that entangle biologically meaningful (and desired) information with spurious confounders such as the medical center. We furthermore see that such confounders can give rise to critical failures when the pathology FMs are applied in diagnostic downstream tasks, clustering, or retrieval (see Section~\ref{sec:consequences}).

Our work has systematically contributed to a better understanding of pathology FMs by (i) establishing measures of robustness (see Section~\ref{sec:robustness_measure}), (ii) introducing benchmarks that allow to quantify (non-)robustness (see Section~\ref{sec:robustness_benchmarking}), (iii) demonstrating consequences of non-robustness in diagnostic tasks (see Section~\ref{sec:consequences}), and finally (iv) proposing a framework to robustify representations without the need to re-train FMs (see Section~\ref{sec:robustification}).

Let us put our findings into a broader perspective. First, our results show the importance of FM robustness, suggesting that robustness criteria should be included in future foundation model development as a core design principle. Integrating the proposed benchmark metrics into foundation model pre-training may lead to more robust pathology FMs that are better suited for clinical translation and research applications. Robust FMs are particularly desirable when only a small amount of data from multiple medical centers is available, e.g., for detecting or characterizing rare diseases. Second, our structured approach to analyzing and improving FMs may serve as a blueprint for other biomedical fields, as batch effects are prevalent in most biomedical data \cite{degrave2021ai,unsupervised-clever-hans, finak2016standardizing, goh2017batcheffects, vcuklina2021proteomics, zindler2020simulating}; subsampling balanced multi-site datasets, computing the robustness index, and measuring generalization performance drops are not conceptually limited to pathology. Thus, our framework could also contribute to improving the robustness of radiology or omics foundation models, among others.

An important debate has revolved around the question of whether making foundation models robust is desirable at all. In the context of managing biases in ML models based on sex or race, Weng et al.~\cite{weng2024bias} argued that foundation models should ideally contain as much information as their underlying data, as removal of supposedly irrelevant information could have unintended adverse effects. Extending this thought to medical center information, one could imagine cases where removing them leads to the elimination of relevant biological patterns of subpopulations that are overrepresented in a specific medical center, as center-specific biological information could be mistaken for technical artifacts. Instead, Weng et al.\ proposed to tackle biases through careful downstream model training and evaluation. However, we argue that collecting sufficient downstream training and evaluation data will often be impossible in medicine. For rare diseases, for example, the few available cases may need to be drawn from various medical centers with highly heterogeneous data characteristics. Here, robustified foundation models could have an enormous impact on downstream model efficacy and, therefore, on the quality of patient outcomes in clinical settings.

Adding to the debate, our findings indicate that robustification of FM representations may yield a disentanglement of biological and confounding features (such as medical center information), which gives rise to improvement in generalization. We stress that this aspect has so far been overlooked and not yet fully understood; thus, more technical efforts should go into this important direction.

Although the datasets and experiments presented in our benchmark are still somewhat limited in complexity, they were highly instructive in revealing limitations and differences between popular foundation models. Clearly, along with foundation models becoming increasingly robust in the future, the quest for more complex pathology prediction tasks to find more pronounced differences will need to continue.
Furthermore, the subsampling of data to have exactly one biological class and one medical center dimension, with carefully designed proportions, served the purpose of furthering our understanding of FM representations. More complex scenarios reflecting general clinical scenarios will be useful to gain more insights into FM robustness.


Finally, we focused our efforts on patch-level foundation models. Yet, practical applications often require slide-level predictions, which can be obtained via aggregation models (i.e., multiple instance learning) \cite{ilse2018attentionmil, campanella2019clinical, wagner2023transformer} or pre-trained slide representation models \cite{xu2024gigapath, wang2024slide, ding2024multimodal}. In both cases, though, patch representations build the basis of the slide representation computation, leading us to believe that the patch-level encoders are most crucial to evaluate for robustness. 
Nonetheless, robustness evaluation and comparison of slide-level foundation models form a welcome and logical extension for further work.

Paving the way towards more robust foundation models, future research should focus on furthering pre-training strategies --- particularly on how to favor alignment with clinical semantics while mitigating biases and ensuring transferability to downstream tasks.
Beyond refining the pre-training process, our study highlights a promising complementary research direction: post-hoc robustness improvements of FM representation spaces. Similarly to how modern LLM and vision training pipelines employ instruction tuning, reinforcement learning from human feedback (RLHF), or alignment as a crucial post-training step to mitigate harmful outputs \cite{ouyang2022training, bai2022constitutional, muttenthaler2024aligning}, future pathology foundation models may benefit from analogous post-training procedures to remove undesired non-biological features. This could offer a practical alternative to the extensive retraining of new pathology foundation models on the path to robust, trustworthy, and reliable AI models.

\section{Methods}\label{sec:methods}

\subsection{Dataset subsampling} \label{sec:methods_datasets}

We subsampled datasets from clinical sources that enable a comparison between biological signals and real-world medical center signatures in the representations of foundation models.
For this, we considered patch-level data where each patch represents a distinct biological class, e.g., a patch from the tumor area of a lung adenocarcinoma or a lung squamous cell carcinoma.
Additionally, we required that the patches originate from different medical centers.
A key idea was to make the subsampled datasets complete and balanced: ideally, every medical center contributes the same number of cases, slides, and patches per biological class.
Furthermore, we tried to ensure that systematic differences between medical centers are solely due to biologically irrelevant factors by sampling such that known medical and demographic characteristics are comparable across medical centers.

Following these principles, we subsampled four datasets from three public sources: (i) A subset from the \textbf{Camelyon} cohorts for tumor detection \cite{bejnordi2017camelyon, bandi2019camelyon17}, with two biological classes (normal vs.\ tumor) and two medical centers, plus three additional medical centers for out-of-domain (OOD) generalization. (ii) Two subsets from TCGA-UT~\cite{komura22tcga_uniform} for tumor type prediction. The first (\textbf{TCGA~4x4}) contains 4 biological classes (BRCA, COAD, LUAD, LUSC) and four medical centers (plus four OOD medical centers). The second (\textbf{TCGA~2x2}) comprises 94 class-class-center-center quartets, consisting of two biological classes and two medical centers each. (iii) One subset from the \textbf{Tolkach ESCA} resource~\cite{tolkach2023esca} for tissue compartment classification in oesophageal resections, with six biological classes (tumor, mucosa, muscularis propria, etc.) and three medical centers (plus one OOD cohort). Dataset statistics, example patches, and sampling details are provided in Sup.~Note~\ref{sec:apx_dataset_details}.

\subsection{Foundation models and feature extraction}
\label{sec:methods_fm}

In total, we evaluated 20 foundation models in our study that encompass a diverse range of architectures (from convolutional networks to vision transformers), utilize various pre-training objectives (including SSL objectives such as SimCLR, DINO, SRCL, iBOT, DINOv2 and image/text models such as \mconch{} and \mmusk{}), vary greatly in dataset size (from 6k to 3.1 million WSIs), and span multiple scales in model capacity (from 11 million to 1.1 billion parameters) (see Table~\ref{tab:models}).
For each SSL ViT model, we extracted the CLS and mean pooled patch token representations and concatenated them to a final representation, which was used throughout all analyses in the paper. For the image/text models, we used the recommended layers and settings on the respective Huggingface site. For CNN models (\mciga{}, \mretccl{}), we simply used the global average pooling representation at the end of the encoder.

\begin{table}[t]
    \centering
    \setlength{\tabcolsep}{5pt}
    \footnotesize
    \begin{tabular}{lccrcrr}
        \toprule
        Model & pre-training objective & \#pre-training WSIs & Architecture & \#Parameters   \\
        \midrule
        \mciga{} \cite{ciga2021self} & SimCLR & (400k patches) & ResNet-18 & 11M \\
        \mhipt{} \cite{chen2022scaling} & DINO & 10.7k & ViT-S/16 & 22M \\
        \mretccl{} \cite{wang2023retccl} & cluster-guided contrastive & 32k & ResNet-50 & 24M \\
        \mctranspath{} \cite{wang2022transformer} & SRCL & 32k & Swin-T &  28M \\
        \mkangdino \cite{kang2023benchmarking} & DINO & 36.7k & ViT-S/8 & 22M \\
        \mkaiko \cite{aben2024towardskaiko} & DINOv2 & 29k & ViT-B/8 & 86M \\
        \mphikon{} \cite{phikon} & iBOT & 6K & ViT-B & 86M \\
        \muni{} \cite{chen2024uni} & DINOv2 & 100K & ViT-L/16 & 303M \\
        \mprovgigapath{} \cite{xu2024gigapath} & DINOv2 & 171.2k & ViT-g/14 & 1100M \\
        \mvirchow{} \cite{vorontsov2024virchow_natmed} & DINOv2 & 1.5M & ViT-H & 632M \\
        \mvirchowTWO{} \cite{zimmermann2024virchow2} & DINOv2 & 3.1M & ViT-H & 632M \\
        \mhoptimus{} \tablefootnote{\url{https://github.com/bioptimus/releases/blob/main/models/h-optimus/v0/README.md}} & DINOv2 & $>$500K & ViT-g/14 & 1.1B \\
        \mrudolfv{} \cite{Dippel2024RudolfV} & DINOv2 & 134K & ViT-L/14 & 303M \\
        \mphikonTWO{} \cite{filiot2024phikon} & DINOv2 & 58.4K & ViT-L/16 & 303M \\
        \muniTWO{} \tablefootnote{\url{https://huggingface.co/MahmoodLab/UNI2-h}} & DINOv2 & 350K & ViT-H & 681M \\
        \matlas{} \cite{alber2025novel_atlas} & DINOv2 & 1.2M & ViT-H/14 & 632M \\
        \mhmini{} \cite{filiot2025distillingfoundationmodelsrobust} & distillation & $>$500K & ViT-B/14 & 86M \\
        \mconch{} \cite{lu2024visual} & iBOT + vision-language & 21.4K (+image/text pairs) & ViT-B/16 & 86M \\
        \mconchONEFIVE{} \tablefootnote{\url{https://huggingface.co/MahmoodLab/conchv1\_5}} & iBOT + vision-language & 100k (+image/text pairs)  & ViT-L/16 & 307M \\
        \mmusk{} \cite{xiang2025vision} & (MLM, MIM) + contrastive & 33k (+image/text pairs)  & BEiT3 & 675M \\
        \bottomrule
    \end{tabular}
    \caption{\textbf{Overview of all 20 histopathology foundation models used in this study.}}
    \label{tab:models}
\end{table}

\subsection{Representation space visualization and target prediction}
\label{sec:methods_figure1}

To visualize foundation model representation spaces, we used the default t-SNE implementation from the sklearn package\footnote{\url{https://scikit-learn.org/stable/modules/generated/sklearn.manifold.TSNE.html}} on extracted representation vectors from the Camelyon dataset for each of the four cancer/normal $\times$ RUMC/UMCU categories. For the computation of the t-SNE, we chose the perplexity as the optimal $k$ value derived during the robustness index calculation (see Sup.~Note~\ref{sec:apx_robustness_index_k_values}).
For measuring biological class and medical center prediction accuracies per foundation model, we trained linear probing models (aka.\ logistic regression), i.e., a simple neural network head without hidden layers, to predict either biological class or medical center (training details in Sup.~Note~\ref{sec:apx_training_details}). 
For this, we employed the subsampled PathoROB datasets of Camelyon, TCGA~4x4, and Tolkach ESCA. The data were split into approximately 0.6/0.1/0.3 train/validation/test on slide level. We report the test set accuracies averaged over 20 repetitions and the three datasets with 95\% confidence intervals $(\pm\ t_{0.975, 59} \cdot SE)$, corrected to remove common variance due to dataset (Masson \& Loftus \cite{Masson2003}).

\subsection{Robustness index} \label{sec:methods_robustness}
\label{ri_derivation}

The robustness index is inspired by $k$-nearest neighbor ($knn$) classification, one of the earliest, simplest, and most widely used machine learning methods for classifying samples based on feature vectors \cite{fix1951discriminatory, cover_hart_67}. $knn$ classification and related methods find the $k$ neighbors closest to a sample as determined by the distances between their feature vectors. We consider which neighbors have the same biological class and which have the same confounding class. This divides the neighbors of a sample into four groups; see Table~\ref{tab:same_other_table}.

\begin{table}[h!]
\centering
\begin{tabular}{|c|c|c|c|c|}
\hline
biological/confounding      & same confounding class & other confounding class  \\
\hline
same biological class    &   SS & SO  \\
\hline
other biological class    &  OS  & OO \\
\hline
\end{tabular}
\caption{Sample pairs are grouped into four categories, based on whether the samples have the Same (S) / Other (O) biological or confounding class.}
\label{tab:same_other_table}
\end{table}

The idea behind the robustness index is to express the degree to which models capture biological rather than confounding information. Neighbors with both the same biological and confounding class (SS) are irrelevant, however, as there is no way to determine whether the sample is close due to having the same biological or the same confounding class. The same goes for the OO class. We exclude the SS and OO combinations, therefore, and thus restrict the analysis to the neighbors that have \textit{either} the same biological class (SO), \textit{or} the same confounding class (OS).

For each sample in a given evaluation dataset $D$ containing $n$ samples, we obtain the $k$ nearest neighbors. From this set of $n\cdot k$ neighbors, we select the subsets of neighbors with either the {\bf S}ame biological class as the sample of which it is a neighbor and the {\bf O}ther confounding class ($SO_k$), or the {\bf O}ther biological class and {\bf S}ame confounding class ($OS_k$). The robustness index $\mathcal{R}_k$ is then defined as:

\begin{align*} 
\mathcal{R}_k(D) = \frac{|\,SO_k(D)\,|}{|\,SO_k(D)\,| \,+\, |\,OS_k(D)\,|}
\end{align*}
The robustness index expresses the degree to which biological rather than confounding features dominate the neighborhood of a sample in the representation space. The robustness index provides an easily interpretable metric that is shown in this work to capture a relevant dimension of model performance. A different question that could be posed is: how well does the foundation model generalize to Out Of Distribution (OOD) data? It has not escaped our notice that the four matrices computed above (SS, SO, OS, and OO) enable measuring this; the last subsection of Sup.~Note~\ref{sec:apx_robustness_index_computation} defines the \textit{Generalization Index} $\mathcal{G}$, which is not explored further in this work. For further results and details regarding the robustness index, see Sup.~Note~\ref{sec:apx_robustness_index_computation}.

\paragraph{Visual explanation: from category frequencies to robustness index}

\begin{figure}[htbp]
    \centering
    {\footnotesize
    \setlength{\tabcolsep}{2pt}
    \begin{tabular}{ccc}
                \includegraphics[width=0.35\linewidth]{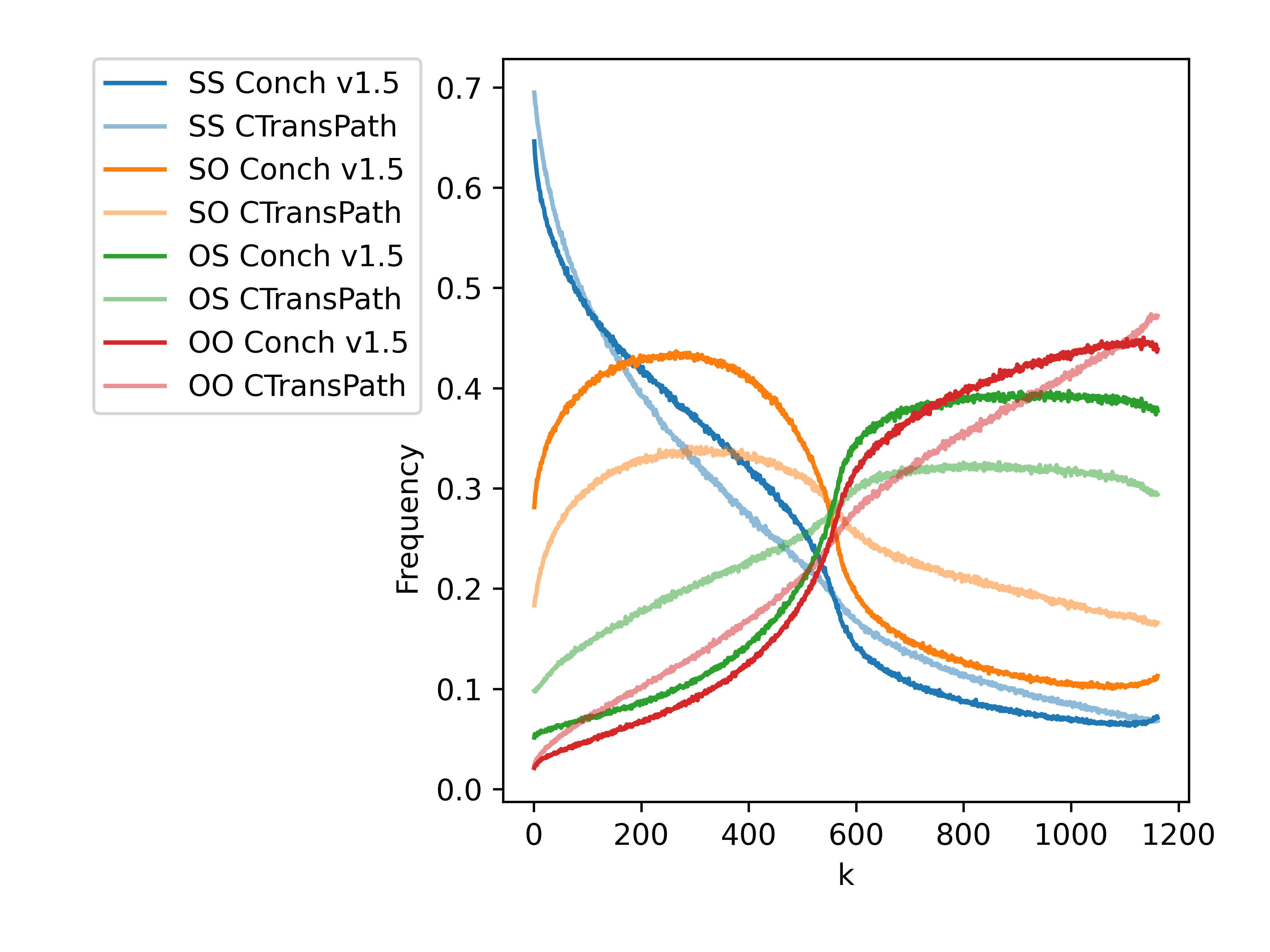} &
        \includegraphics[width=0.29\linewidth]{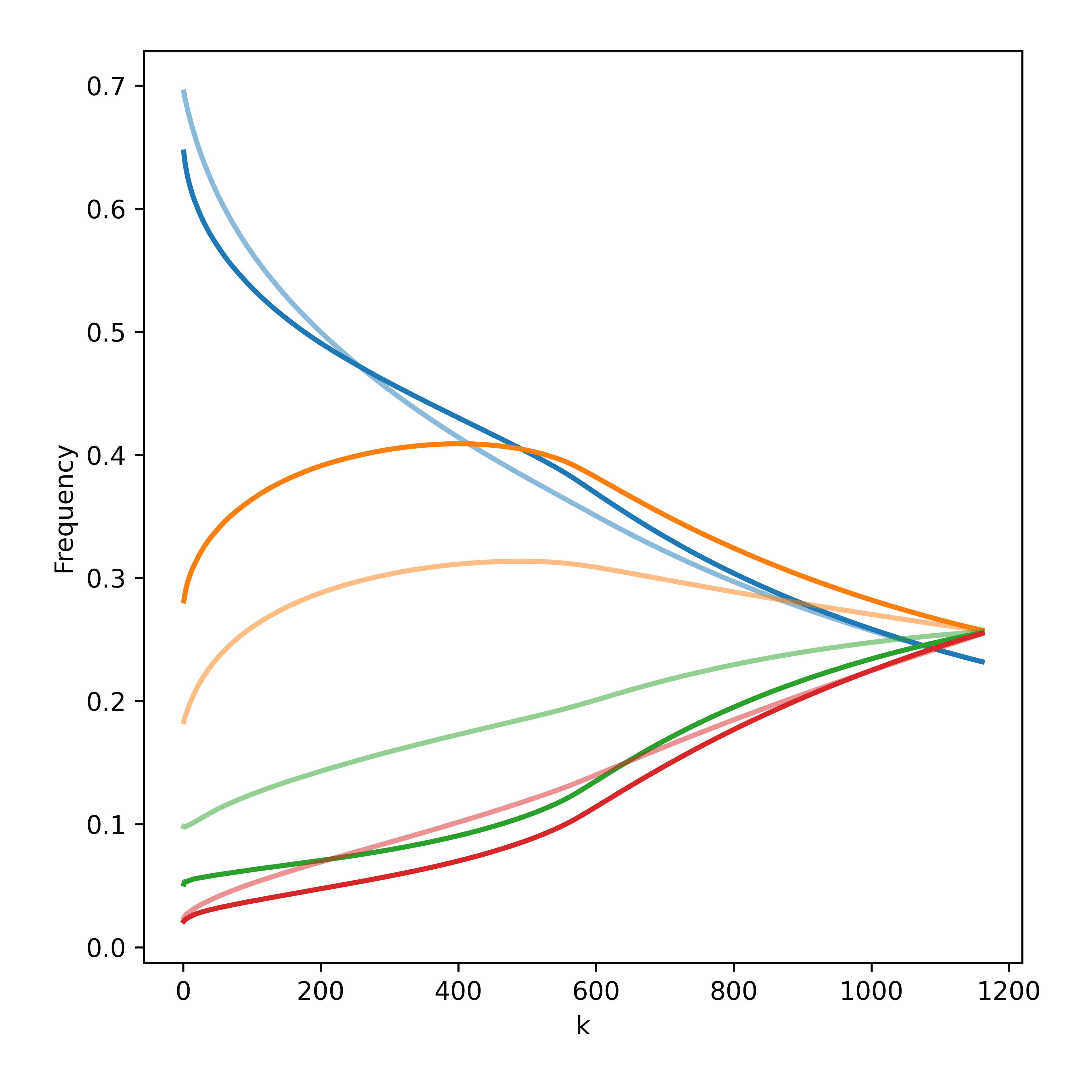} &
        \includegraphics[width=0.35\linewidth]{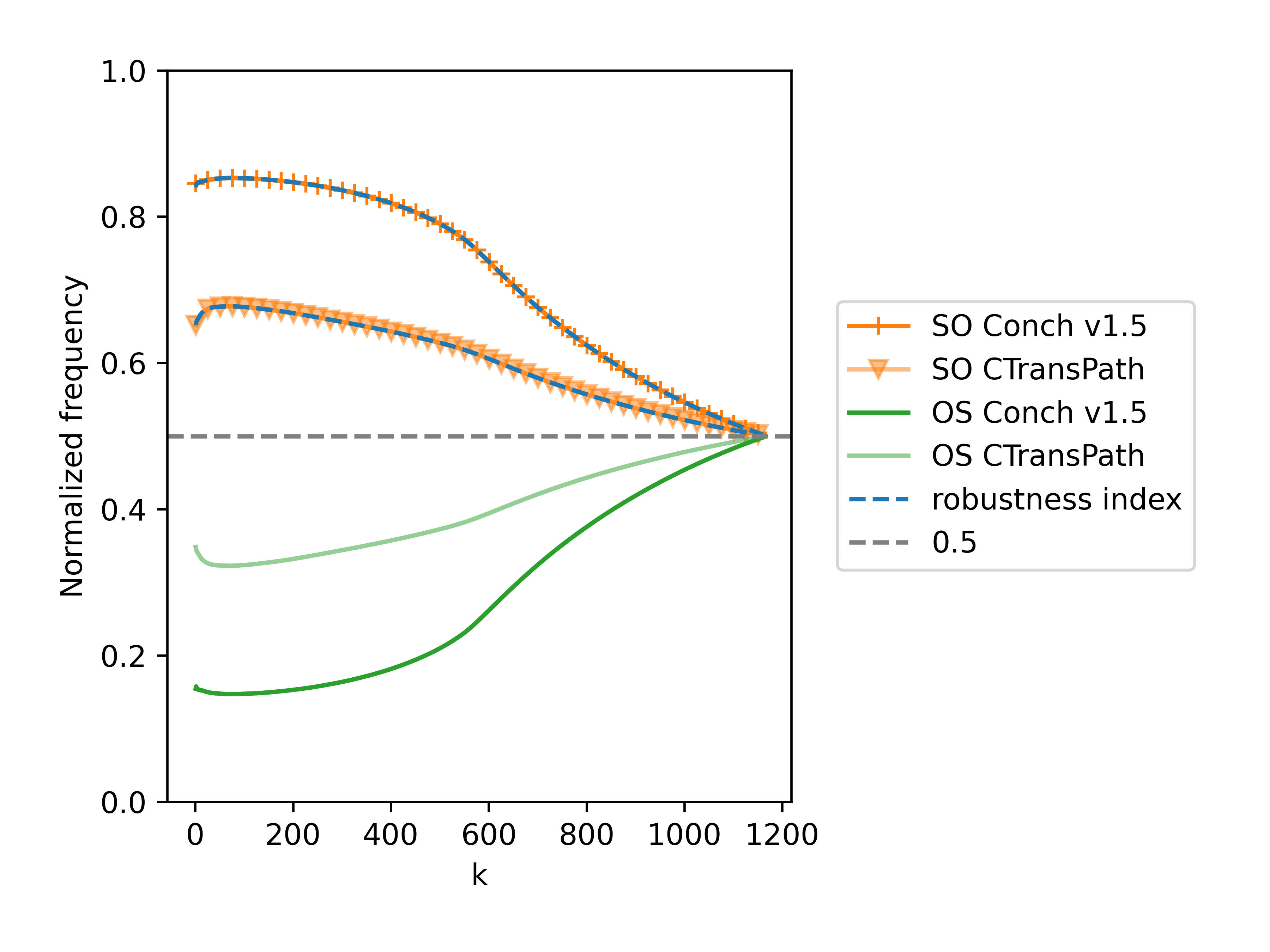} \\
        a.~Frequencies for neighbor $k$ & b.~Frequencies for $knn$  & c.~Normalized frequencies \\
    \end{tabular}
    }
    \caption{\textbf{Visual explanation of the robustness index calculation on TCGA~2x2.}
    \textbf{a.}~Frequencies of SS, SO, OS, and OO neighbors at distance $k$. \textbf{b.}~Frequencies over the $k$ nearest neigbhors. \textbf{c.}~Frequencies for the SO and OS categories, normalized to sum to one. The normalized SO frequency defines the robustness index. }
    \label{fig:rob-index-frequencies}
\end{figure}

To illustrate how the frequencies of these four categories determine robustness index values, we visualize the computation using real data. Figure \ref{fig:rob-index-frequencies} shows the robustness index calculation for \mctranspath{}, an earlier, often-cited pathology foundation model, and \mconchONEFIVE{}, a recent, more robust Vision-Language Model, on TCGA~2x2.

Fig.~\ref{fig:rob-index-frequencies}a shows the frequencies of all four categories (SS, SO, OS, OO) for both models, representing only the neighbors at each specific distance rank $k$. The frequencies of SS (blue) and OO neighbors (red) are similar for both models and do not play a role in the calculation. Fig.~\ref{fig:rob-index-frequencies}b aggregates this information over the $k$ nearest neighbors, resulting in more stable curves. Large differences are seen for the SO and OS lines in both a.~and b; the \mconchONEFIVE{} model places SO neighbors nearby much more often than \mctranspath{}, as shown by the higher orange line for $k$ values below the midpoint of 300, \textit{and} correctly places OS neighbors further away, as seen by the lower green line.

Fig.~\ref{fig:rob-index-frequencies}c focuses on the SO and OS lines, and normalizes these to sum to one. The resulting normalized SO curve equals the robustness index. The higher prevalence of SO neighbors and the lower frequency of OS neighbors combine to yield a substantially higher robustness index for \mconchONEFIVE{} across the full range of $k$ values.

\subsection{Supervised downstream model training and evaluation} \label{sec:methods_downstream}

\paragraph{Split generation}

We subsampled training splits from our PathoROB datasets (Camelyon, TCGA~4x4, Tolkach ESCA) in which the medical center is increasingly correlated with the target to be predicted. The first training split (Split~1, Cram\'er’s $\text{V}=0$) was fully balanced, i.e., every medical center evenly contributed the same number of patches per disease class. In Split~2, each medical center had one or more overrepresented disease classes, for which more data points were available than for the other classes, introducing a weak correlation between the medical center and the prediction target. This correlation was gradually increased until the final split (Cram\'er’s $\text{V}=1$). Here, each medical center contributes data for some but not for all other classes, resulting in a strong (or even perfect) correlation between the medical center and the biological target. In each split, the total number of training data points was the same overall, per disease class, and per medical center; only the dataset composition differed. For patches of each of the datasets, we ensured there was no slide overlap between the training and test sets. We held out a balanced in-distribution (ID) test set (different cases from the same medical centers) and a balanced out-of-distribution (OOD) data (different cases from different medical centers), allowing us to assess the generalization performance on data where the model cannot rely on correlations between medical centers and biological targets. Further details and the complete split statistics are provided in Sup.~Note~\ref{sec:apx_training_details} and Sup.~Figures~\ref{fig:camelyon_dataset_splits}--\ref{fig:tolkach_dataset_splits}.

\paragraph{Downstream model training and evaluation}

As a downstream model, we trained linear probing (aka.\ logistic regression), i.e., a simple neural network head without hidden layers to predict the biological classes (training details in Sup.~Note~\ref{sec:apx_training_details}). For evaluation, we calculated the accuracy on the ID and OOD test sets per dataset, model, and training split, and reported the average over the 20 repetitions with 95\% confidence intervals $(\pm\ t_{0.975, 19} \cdot SE)$.
For FM comparison, we aggregated the results into an \textit{average performance drop} (APD) per model: For a dataset $D$, model $M$, and repetition $r$ we define this as 
\begin{align*}
    \text{APD}_{D,M,r} = \frac{1}{\#splits - 1} \sum_{i = 2}^{\#splits} \frac{acc_{i} - acc_{1}}{acc_{1}}.
\end{align*}
An APD of 0\% corresponds to no generalization performance drops across splits, implying that the downstream model predictions did not rely on medical center information despite training data correlations incentivizing this. Increasingly negative APDs indicate steeper generalization performance drops, suggesting greater vulnerability to Clever Hans learning. We report the average APD over 20 repetitions and 3 datasets with 95\% confidence intervals $(\pm\ t_{0.975, 59} \cdot SE)$, corrected to remove common variance due to dataset (Masson \& Loftus \cite{Masson2003}).

\subsection{Clustering analysis} \label{sec:methods_clustering}

\paragraph{Measuring clustering quality}

While the robustness index captures the local neighborhood structure of each representation, the clustering score extends this analysis to a global level. It evaluates whether embedding vectors form distinct clusters, and to what extent they are driven by diagnostically relevant factors or misled by similarities in medical centers.
We used unsupervised $K$-means clustering with cosine distances, selecting the number of clusters $K$ by maximizing the silhouette score \cite{rousseeuw1987silhouettes}. This setup mimics real-world exploratory analyses where prior cluster information is unknown (e.g., discovering morphological subtypes).
The quality of clustering is evaluated by comparing the predicted cluster assignments $\hat{\mathcal{C}} = \lbrace \hat{C}_1, \ldots , \hat{C}_K \rbrace$ to the true biological $\mathcal{C}_{\text{bio}}$ and confounding labels $\mathcal{C}_{\text{mc}}$ (potentially of different sizes than $K$) using the adjusted Rand index (ARI)~\cite{hubert1985comparing}. The ARI measures the agreement between two clusterings primarily by the number of pairs of samples that are correctly clustered together or apart, accounting for chance. 
A perfect clustering is solely based on biological information and not influenced by the medical center origin. Hence, we define the clustering score as the difference between these ARIs: 
\begin{equation*}
    \text{clustering score} = \underbrace{\text{ARI}(\hat{\mathcal{C}}, \mathcal{C}_{\text{bio}})}_{\text{higher is better}} - \underbrace{\text{ARI}(\hat{\mathcal{C}}, \mathcal{C}_{\text{mc}})}_{\text{lower is better}}
\end{equation*}
The clustering score ranges within approximately $[-1,1]$, where values near zero suggest that the clustering is influenced by both biological and confounding information (or neither), positive values indicate a clustering aligned with medically relevant features, and negative values reflect a medical center-driven structure.
Sup.~Note~\ref{sec:apx_clustering_score_details} provides a more detailed description of the clustering score, and
Sup.~Note~\ref{sec:apx_upperbound_cs} discusses the clustering results with optimally chosen $K$, which reveals the effect of the silhouette score-based selection.

\paragraph{Clustering experiments}

Clustering experiments are conducted on balanced 2×2 configurations (two biological classes and two medical centers). For Camelyon and Tolkach ESCA, the same subsampled 2×2 datasets as in the robustness index evaluation are used, while for TCGA all 2×2 pairs derived from the 4×4 setting are considered. For each value of $K \in [2, 30]$, $K$-means clustering is performed, and the optimal $K$ is selected based on the silhouette score. The results of the final clustering, with the selected $K$, are compared with the true biological labels and the medical center origins using the proposed clustering score. An average clustering score and its standard deviation are estimated based on $50$ repetitions of the final clustering with different random initializations.

\subsection{Foundation model robustification without FM retraining} \label{sec:methods_mitigation}

\paragraph{Robustification methods}

While robustness improvement for standalone pathology models has been studied (see, e.g., \cite{jahanifar2025domain}), we view robustness improvement for foundation models as a major current open challenge for the field.

To remove medical center signatures in image space, we applied Reinhard \cite{reinhard} and Macenko \cite{macenko} stain normalization to each patch individually before extracting features from the foundation models. As a normalization target, we used the average statistics of 500 sampled patches from the TCGA-LUSC project. As we found that Reinhard stain normalization consistently achieved better robustification results than Macenko stain normalization, we only report the Reinhard results.

In molecular biology, data are represented in numerical vectors that describe cell or bulk expression profiles. Although potentially differently distributed, molecular expression vectors are structurally similar to FM feature vectors, motivating the application of ComBat~\cite{johnson2007combat}, one of the most popular batch correction methods in molecular biology. We used the PyComBat implementation\footnote{\url{https://pypi.org/project/inmoose/}}. For experiments with held-out test data, we first applied ComBat to the training data, and then used the corrected training data as a single reference batch to normalize the test sets without leaking information into the training set, as done by Murchan et al.~\cite{murchan2024combat}.

Domain-Adversarial Neural Networks (DANNs)~\cite{DBLP:journals/jmlr/GaninUAGLLML16} aim to learn a modified feature representation in which all domains (e.g.\ medical centers) are indistinguishable, while simultaneously optimizing for a prediction task.
Let $D = (x_i, y_i, c_i)_i$ be the dataset consisting of the feature vectors $x$, the biological target $y$, and the medical center $c$.
Further, let $\phi: x \mapsto x'$ be a learned projection function on top of the original representation, $f_{\mathsf{CL}}: \phi(x) \mapsto \hat{y}$ a classification head, and $f_{\mathsf{DA}}: \phi(x) \mapsto \hat{c}$ a domain discriminator on top of $\phi$.
The DANN objective can be defined as two competing loss parts:
\begin{align*}
    \mathcal{L}_{\mathsf{DANN}}(\hat{y},\hat{c}; y,c) &= \mathcal{L}_{\mathsf{CL}}(\hat{y},y) + \lambda \cdot \mathcal{L}_{\mathsf{DA}}(\hat{c},c),
    \label{eq:dann}
\end{align*}
where $\hat{y}, \hat{c}$ are the independent logits of the two separate heads, $f_{\mathsf{CL}}$ and $f_{\mathsf{DA}}$, on top of the learned feature space $\phi$.
The first term, $\mathcal{L}_{\mathsf{CL}}$, is used as a standard classification loss, whereas $\mathcal{L}_{\mathsf{DA}}$ is used to align the domains (both, e.g., using cross-entropy), and $\lambda$ is a weight to balance the loss terms. Training details and hyper-parameter choices are provided in Sup.~Note~\ref{sec:apx_training_details}.

Notice that ComBat and DANN require knowledge of the medical center origin of each patch, whereas Reinhard stain normalization does not. Further, DANN can only be applied when training a downstream model for a specific prediction task. Also note that the site stratification method proposed by Howard et al.\ \cite{Howard2021} cannot be usefully applied if training data are imbalanced --- balancing features of interest across folds is not possible if all samples of a biological class came exclusively from one medical center.

\paragraph{Experimental evaluation}

We re-ran the experiments from Section~\ref{sec:methods_downstream}, training supervised downstream models on top of robustified feature spaces (see Sup.~Note~\ref{sec:apx_training_details} for details). For each foundation model and robustness improvement method, we computed the average performance drops as described in Section~\ref{sec:methods_downstream} with respect to the averaged unnormalized baseline performance on the balanced Split 1 (Cram\'er’s $\text{V}=0$), i.e.,
\begin{align*}
\text{APD'}_{D,M,r} = \frac{1}{\#splits - 1} \sum_{i = 2}^{\#splits} \frac{acc_{i} - acc_{1, \text{unnormalized}}}{acc_{1, \text{unnormalized}}}
\end{align*}
for dataset $D$, model $M$, and repetition $r$. This enables a comparison of the robustification methods with the baseline setting: an average performance drop closer to zero than the baseline means that the method reduced the harm of Clever Hans learning. We report the averages per FM over all repetitions and datasets.

\subsection{Statistical information}
Due to the small sample size ($n = 20$), p-values of the Spearman rank-order correlation $\rho$ were estimated based on a two-sided paired permutation test with 50,000 permutations.

\section*{Additional information}

\paragraph{Data \& code availability}

All data and codes from the PathoROB benchmark will be made available in a public repository soon.

\paragraph{Acknowledgements}

This work was partly funded by the German Ministry for Education and Research (under refs 01IS14013A-E, 01GQ1115, 01GQ0850, 01IS18056A, 01IS18025A, 13GW0744D, and BIFOLD25B). Furthermore, K.-R.M. was partly supported by the Institute of Information and Communications Technology Planning and Evaluation (IITP) grants funded by the Korea government (MSIT) (no. 2019-0-00079, Artificial Intelligence Graduate School Program, Korea University, and no. 2022-0-00984, Development of Artificial Intelligence Technology for Personalized Plug-and-Play Explanation and Verification of Explanation).
We thank Florian C.F.\ Schulz and Augustin Krause for support with the foundation model representation extraction library.
We thank our colleagues Laure Ciernik, Alexander M{\"o}llers, and Stephan Tietz for their valuable feedback on earlier versions of this manuscript, which helped improve this work.
We would like to express our gratitude to Yuri Tolkach for sharing additional case metadata for the Tolkach ESCA dataset. 
The results shown here are in part based upon data generated by the TCGA Research Network: https://www.cancer.gov/tcga.

\paragraph{Author contributions}

Conceptualization and methodology: J.K., E.D.J., J.H., H.M., J.D., P.N., K.-R.M.
Development of the robustness index (Section 2.1): E.D.J.
Development of the downstream experiments (Section 2.3): J.K., J.H., H.M., J.D.
Data curation, code creation, and experiments: J.K., E.D.J., J.H., H.M., J.D., P.N.
Analysis of results: J.K., E.D.J., J.H., H.M., J.D., P.N.
Project administration: J.H.
Supervision: E.M., L.R., M.A., J.T., F.K., K.-R.M.
Writing: J.K., E.D.J., J.H., H.M., J.D., P.N., E.M., L.R., M.A., F.K., K.-R.M.

\newpage

\bibliographystyle{unsrtnat}
{\small \bibliography{bibliography}}


\clearpage

\appendix


{\centerline{\LARGE\bf Supplementary Notes}}
\setcounter{section}{0}    

\vspace{1em}

\noindent  These supplementary notes provide additional data, experiments, and analyses extending those presented in the main paper. The aim is to further support the claims of the main paper regarding the (limited) robustness of pathology foundation models (FMs), the consequences, and robustification strategies.

\vspace{0.5em}

\section{PathoROB dataset details} \label{sec:apx_dataset_details}

\subsection{Camelyon}

The CAMELYON16 dataset\footnote{\url{https://camelyon17.grand-challenge.org/Data}, License: Public Domain: CCO} \cite{bejnordi2017camelyon} comprises 400 sentinel lymph node slides from women with breast cancer, some of which contain precisely annotated regions of lymph node metastasis of different sizes. The dataset was issued by two hospitals in the Netherlands (RUMC, UMCU). The CAMELYON17 dataset \cite{bandi2019camelyon17} extends the CAMELYON16 dataset by 1000 slides from five medical centers (RUMC, UMCU, CWZ, RST, LPON), from which we only considered CZW, RST, and LPON as additional medical centers.

We extracted patches from the slides of $256 \times 256$ pixels without overlap at 20x magnification (0.5 microns per pixel). We identified and excluded background patches via Otsu's method \cite{otsu1975threshold} on slide thumbnails and applied a patch-level minimum standard deviation of 8. Each patch was labeled with the medical center from which its slide originated. We additionally labeled those patches fully inside the metastasis annotations as ``tumor'' and those fully outside as ``normal''.
%
For RUMC and UMCU, we sampled 300 normal patches each from 17 slides without metastases and 300 tumor patches each from 17 slides with metastases, making up our in-domain Camelyon dataset. For CWZ, RST, and LPON, where less data were available, we sampled 335 normal patches per slide from 5 slides without metastases and 327-335 tumor patches from 3-6 slides with metastases, respectively. These patches were used solely for out-of-domain evaluation. Table~\ref{tab:dataset_statistics_camelyon} details the resulting dataset statistics. Example patches per class and medical center are displayed in Figure~\ref{fig:dataset_samples}a.

\begin{table}[h!]
    \centering
    \scriptsize
    \begin{tabular}{lcc|ccc||c}
    \toprule
     & RUMC & UMCU & CWZ & RST & LPON & Total \\
    \midrule
    Normal & 5,100 (17) & 5,100 (17) & 335 (5) & 335 (5) & 335 (5) & 11,205 (49) \\
    Tumor & 5,100 (17) & 5,100 (17) & 327 (5) & 335 (6) & 335 (3) & 11,197 (48) \\
    \midrule \midrule
    Total & 10,200 (34) & 10,200 (34) & 662 (10) & 670 (11) & 670 (8) & 22,402 (97) \\
    \bottomrule
    \end{tabular}
    \caption{\textbf{Dataset statistics for Camelyon.} We report the number of patches together with the number of slides / cases in parentheses.}
    \label{tab:dataset_statistics_camelyon}
\end{table}

\subsection{TCGA}

\begin{table}[t!]
    \centering
    \scriptsize
    \begin{tabular}{lcc||c}
    \toprule
     & MCT 1 & MCT 2 & Total \\
    \midrule
    CLS 1 & 300 (10) & 300 (10) & 600 (20) \\
    CLS 2 & 300 (10) & 300 (10) & 600 (20) \\
    \midrule \midrule
    Total & 600 (20) & 600 (20) & 1,200 (40) \\
    \bottomrule
    \end{tabular}
    \caption{\textbf{Dataset statistics for TCGA~2x2 per 2x2 combination.} We report the number of patches together with the number of slides / cases in parentheses. CLS = tumor class, MCT = medical center. The total dataset includes 94 double pairs, totaling 112,800 patches. Note that the same patch can be present in multiple 2x2 combinations.}
    \label{tab:dataset_statistics_tcga2x2}
\end{table}

\begin{table}[t!]
    \centering
    \scriptsize
    \begin{tabular}{lcccc|cccc||c}
    \toprule
     & Asterand & Christiana & Roswell Park & Pittsburgh & Cureline & IGC & Poland & Johns Hopkins & Total \\
    \midrule
    BRCA & 360 (12) & 360 (12) & 360 (12) & 360 (12) & 300 (10) & 300 (10) & 0 (0) & 0 (0) & 2,040 (68) \\
    COAD & 360 (12) & 360 (12) & 360 (12) & 360 (12) & 0 (0) & 300 (10) & 300 (10) & 0 (0) & 2,040 (68) \\
    LUAD & 360 (12) & 360 (12) & 360 (12) & 360 (12) & 0 (0) & 0 (0) & 300 (10) & 300 (10) & 2,040 (68) \\
    LUSC & 360 (12) & 360 (12) & 360 (12) & 360 (12) & 300 (10) & 0 (0) & 0 (0) & 300 (10) & 2,040 (68) \\
    \midrule \midrule
    Total & 1,440 (48) & 1,440 (48) & 1,440 (48) & 1,440 (48) & 600 (20) & 600 (20) & 600 (20) & 600 (20) & 8,160 (272) \\
    \bottomrule
    \end{tabular}
    \caption{\textbf{Dataset statistics for TCGA~4x4.} We report the number of patches together with the number of slides / cases in parentheses. BRCA = breast invasive carcinoma; COAD = colon adenocarcinoma; LUAD = lung adenocarcinoma; LUSC = lung squamous cell carcinoma. Christiana = Christiana Healthcare; Pittsburgh = University of Pittsburgh; Poland = Greater Poland Cancer Center; IGC = International Genomics Consortium.}
    \label{tab:dataset_statistics_tcga4x4}
\end{table}

The TCGA-UT dataset\footnote{\url{https://zenodo.org/records/5889558}, License: Non-Commercial Use: CC-BY-NC-SA 4.0}~\cite{komura22tcga_uniform} provides tumor patches from 32 distinct cancer types within The Cancer Genome Atlas (TCGA). Each patch was extracted from an annotated tumor region, and each slide comes with multiple annotations created by trained pathologists. We used the highest available resolution level for the patches, corresponding to 0.5 $\mu$m/pixel or roughly 20x magnification. The patches have a size of 256x256 pixels.

To cover a large fraction of the dataset while maintaining class-center balancing, we extracted combinations of 2 cancer types and 2 medical centers (2x2 dataset). For each combination, we sampled 10 slides, from which we sampled 10 patches from 3 regions-of-interest, respectively. The resulting dataset statistics are shown in Table~\ref{tab:dataset_statistics_tcga2x2}. In total, we found 94 2x2 combinations for which sufficient data were available. The complete list is provided in Table~\ref{tab:tcga_2x2_dataset_details}. It covers combinations of similar cancer types (e.g.\ lung adenocarcinoma and lung squamous cell carcinoma) as well as unrelated tumor entities (e.g.\ lung adenocarcinoma and glioblastoma). We further composed a single dataset comprising 4 classes and 4 tissue source sites (4x4 dataset). In this case, we sampled 12 slides with 10 patches from 3 regions-of-interest, each, and added data from 4 further medical centers for out-of-domain assessment, resulting in the statistics of Table~\ref{tab:dataset_statistics_tcga4x4}. Example patches are provided in Figure~\ref{fig:dataset_samples}b.

\subsection{Tolkach ESCA}

\begin{table}[t!]
    \centering
    \scriptsize
    \begin{tabular}{lccc|c||c}
    \toprule
    & WNS & CHA & UKK & TCGA & Total \\
    \midrule
    TUMOR & 900 (9) & 900 (9) & 500 (5) & 500 (5) & 2,800 (28) \\
    REGR\_TU & 900 (9) & 900 (9) & 500 (5) & 0 & 2,300 (23) \\
    SH\_OES & 900 (9) & 900 (9) & 500 (5) & 500 (5) & 2,800 (28) \\
    SH\_MAG & 900 (9) & 900 (9) & 500 (5) & 500 (5) & 2,800 (28) \\
    MUSC\_PROP & 900 (9) & 900 (9) & 500 (5) & 500 (5) & 2,800 (28) \\
    ADVENT & 900 (9) & 900 (9) & 500 (5) & 500 (5) & 2,800 (28) \\
    \midrule \midrule
    Total & 5,400 (14) & 5,400 (31) & 3,000 (17) & 2,500 (15) & 16,300 (77) \\
    \bottomrule
    \end{tabular}
    \caption{\textbf{Dataset statistics for Tolkach ESCA.} We report the number of patches together with the number of slides / cases in parentheses. TUMOR = vital tumor tissue; REGR\_TU = regression areas; SH\_OES = oesophageal mucosa; SH\_MAG = gastric mucosa; MUSC\_PROP = muscularis propria; ADVENT = adventitial tissue. UKK = University Hospital Cologne; WNS = Landesklinikum Wiener Neustadt; CHA = University Hospital Berlin--Charité; TCGA = several participating institutions from the Cancer Genome Atlas, TCGA-ESCA project. Note that the same case may contribute patches to multiple biological classes from the same medical center.} \label{tab:dataset_statistics_tolkach}
\end{table}

The Tolkach ESCA dataset\footnote{\url{https://zenodo.org/records/7548828}, License: Custom, Non-Commercial Use}~\cite{tolkach2023esca} contains specimens from surgically resected oesophageal adenocarcinoma and adenocarcinoma of the oesophagogastric junction. The 320 slides from 126 patients were obtained from 4 medical centers, 3 of which (UKK, CHA, WNS) utilized the same tissue scanner. Further, they only contributed samples from patients who received neoadjuvant chemotherapy and had similar staging characteristics. In the remaining medical center (TCGA), different scanners were used, and no neoadjuvant treatment was given. Each slide was manually annotated and segmented into 11 tissue classes by board-certified pathologists, including tumor, regression tissue, and mucosa. Patches were extracted from these annotated regions at a resolution of approximately 0.75 microns per pixel and with a size of 256x256 pixels.

\begin{figure}[t!]
    \centering    \includegraphics[width=0.7\linewidth]{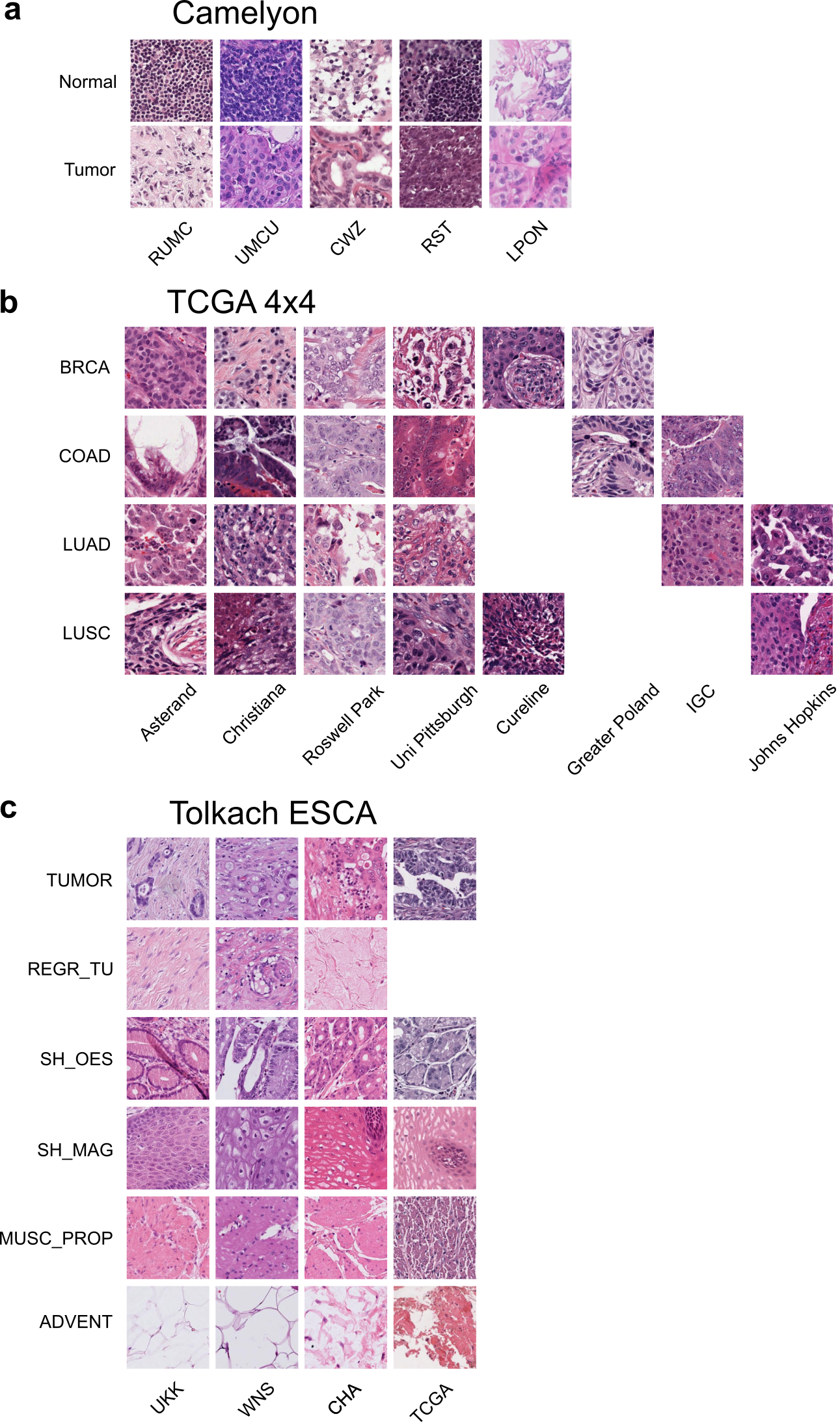}
    \caption{\textbf{Sample patches from the (\textbf{a}) Camelyon, (\textbf{b}) TCGA~4x4, and (\textbf{c}) Tolkach ESCA datasets.} The biological class is depicted on the vertical axis, and the medical center on the horizontal axis. Structural visual differences between medical centers are apparent but vary in strength and consistency.}
    \label{fig:dataset_samples}
\end{figure}

We selected the biological classes that have at least 900 (WNS, CHA) or 500 (UKK, TCGA) patches from 9 or 5 different cases in each medical center, and sampled 100 patches per case for each class-center combination. Notice that within one medical center, a case may contribute patches to multiple biological classes. The TCGA subset was only used for out-of-domain generalization considerations. The resulting dataset statistics are summarized in Table~\ref{tab:dataset_statistics_tolkach}, and sample patches are displayed in Figure~\ref{fig:dataset_samples}c.

\begin{table}[t!]
    \centering
    \tiny
    \resizebox{0.45\textwidth}{!}{
    \begin{tabular}{lll}
    \toprule
     & Disease Classes & Medical Centers \\
    \midrule
    1 & BLCA, BRCA & University of Pittsburgh, MD Anderson \\
    2 & BLCA, CESC & Asterand, Barretos Cancer Hospital \\
    3 & BLCA, COAD & Asterand, University of Pittsburgh \\
    4 & BLCA, ESCA & Asterand, Barretos Cancer Hospital \\
    5 & BLCA, HNSC & Asterand, Barretos Cancer Hospital \\
    6 & BLCA, LIHC & Asterand, ILSBio \\
    7 & BLCA, LUAD & Asterand, University of Pittsburgh \\
    8 & BLCA, LUSC & Asterand, University of Pittsburgh \\
    9 & BLCA, PRAD & University of Pittsburgh, University of California San Francisco \\
    10 & BLCA, SKCM & Asterand, University of Pittsburgh \\
    11 & BLCA, STAD & Barretos Cancer Hospital, ILSBio \\
    12 & BLCA, THCA & University of Pittsburgh, Memorial Sloan Kettering \\
    13 & BRCA, CESC & Asterand, International Genomics Consortium \\
    14 & BRCA, COAD & Asterand, Christiana Healthcare \\
    15 & BRCA, GBM & UCSF, Duke \\
    16 & BRCA, HNSC & Asterand, University of Miami \\
    17 & BRCA, KIRC & University of Pittsburgh, International Genomics Consortium \\
    18 & BRCA, KIRP & International Genomics Consortium, Roswell Park \\
    19 & BRCA, LGG & Duke, Cureline \\
    20 & BRCA, LIHC & Asterand, ILSBio \\
    21 & BRCA, LUAD & University of Pittsburgh, Christiana Healthcare \\
    22 & BRCA, LUSC & Cureline, Christiana Healthcare \\
    23 & BRCA, PRAD & International Genomics Consortium, Roswell Park \\
    24 & BRCA, READ & Greater Poland Cancer Center, Christiana Healthcare \\
    25 & BRCA, SKCM & Asterand, Cureline \\
    26 & BRCA, STAD & Asterand, International Genomics Consortium \\
    27 & BRCA, THCA & University of Pittsburgh, International Genomics Consortium \\
    28 & BRCA, UCEC & University of Pittsburgh, Duke \\
    29 & CESC, COAD & Asterand, International Genomics Consortium \\
    30 & CESC, ESCA & Asterand, Barretos Cancer Hospital \\
    31 & CESC, HNSC & Asterand, Barretos Cancer Hospital \\
    32 & CESC, LUAD & Asterand, International Genomics Consortium \\
    33 & CESC, LUSC & Asterand, International Genomics Consortium \\
    34 & CESC, STAD & Asterand, Barretos Cancer Hospital \\
    35 & CESC, THCA & Asterand, International Genomics Consortium \\
    36 & COAD, HNSC & University of Pittsburgh, International Genomics Consortium \\
    37 & COAD, KICH & MSKCC, Harvard \\
    38 & COAD, KIRC & University of Pittsburgh, MSKCC \\
    39 & COAD, KIRP & MSKCC, Roswell Park \\
    40 & COAD, LUAD & University of Pittsburgh, Roswell Park \\
    41 & COAD, LUSC & University of Pittsburgh, Roswell Park \\
    42 & COAD, PRAD & University of Pittsburgh, International Genomics Consortium \\
    43 & COAD, READ & Greater Poland Cancer Center, Christiana Healthcare \\
    44 & COAD, SKCM & Asterand, University of Pittsburgh \\
    45 & COAD, STAD & Greater Poland Cancer Center, International Genomics Consortium \\
    46 & COAD, THCA & Asterand, International Genomics Consortium \\
    47 & COAD, UCEC & University of Pittsburgh, MSKCC \\
    48 & ESCA, HNSC & Asterand, Barretos Cancer Hospital \\
    49 & ESCA, STAD & Asterand, Barretos Cancer Hospital \\
    50 & ESCA, THCA & Asterand, University Health Network \\
    51 & GBM, LGG & University of Florida, Duke \\
    52 & HNSC, KIRC & University of Pittsburgh, International Genomics Consortium \\
    53 & HNSC, KIRP & UNC, International Genomics Consortium \\
    54 & HNSC, LIHC & Asterand, UNC \\
    55 & HNSC, LUAD & UNC, International Genomics Consortium \\
    56 & HNSC, LUSC & Asterand, International Genomics Consortium \\
    57 & HNSC, PRAD & University of Pittsburgh, International Genomics Consortium \\
    58 & HNSC, SKCM & Asterand, University of Pittsburgh \\
    59 & HNSC, STAD & Asterand, Barretos Cancer Hospital \\
    60 & HNSC, THCA & Asterand, Johns Hopkins \\
    61 & KICH, KIRC & MSKCC, Harvard \\
    62 & KICH, UCEC & University of North Carolina, MSKCC \\
    63 & KIRC, KIRP & UNC, International Genomics Consortium \\
    64 & KIRC, LUAD & University of Pittsburgh, International Genomics Consortium \\
    65 & KIRC, LUSC & MSKCC, International Genomics Consortium \\
    66 & KIRC, PRAD & University of Pittsburgh, International Genomics Consortium \\
    67 & KIRC, THCA & University of Pittsburgh, International Genomics Consortium \\
    68 & KIRC, UCEC & University of Pittsburgh, MSKCC \\
    69 & KIRP, LIHC & Moffitt Cancer Center, UNC \\
    70 & KIRP, LUAD & UNC, International Genomics Consortium \\
    71 & KIRP, LUSC & International Genomics Consortium, Roswell Park \\
    72 & KIRP, PAAD & Moffitt Cancer Center, International Genomics Consortium \\
    73 & KIRP, PRAD & University of California San Francisco, International Genomics Consortium \\
    74 & LGG, LUSC & Mayo Clinic - Rochester, Cureline \\
    75 & LGG, UCEC & MSKCC, Duke \\
    76 & LIHC, LUAD & Asterand, UNC \\
    77 & LIHC, LUSC & Asterand, Mayo Clinic - Rochester \\
    78 & LIHC, PAAD & Moffitt Cancer Center, Alberta Health Services \\
    79 & LIHC, STAD & Asterand, ILSBio \\
    80 & LUAD, LUSC & Roswell Park, Johns Hopkins \\
    81 & LUAD, PRAD & International Genomics Consortium, Roswell Park \\
    82 & LUAD, SKCM & Asterand, University of Pittsburgh \\
    83 & LUAD, STAD & Asterand, International Genomics Consortium \\
    84 & LUAD, THCA & International Genomics Consortium, Johns Hopkins \\
    85 & LUSC, PRAD & International Genomics Consortium, Roswell Park \\
    86 & LUSC, SKCM & Asterand, University of Pittsburgh \\
    87 & LUSC, STAD & Asterand, International Genomics Consortium \\
    88 & LUSC, THCA & Asterand, University of Pittsburgh \\
    89 & LUSC, UCEC & University of Pittsburgh, MSKCC \\
    90 & PRAD, THCA & University of Pittsburgh, International Genomics Consortium \\
    91 & PRAD, UCEC & Washington University, University of Pittsburgh \\
    92 & SKCM, THCA & Asterand, University of Pittsburgh \\
    93 & STAD, THCA & Asterand, International Genomics Consortium \\
    94 & THCA, UCEC & University of North Carolina, University of Pittsburgh \\
    \bottomrule
    \end{tabular}}
    \caption{\textbf{Disease class and medical center combinations in the TCGA~2x2 dataset.}}
    \label{tab:tcga_2x2_dataset_details}
\end{table}

\FloatBarrier
\section{Robustness index computation and further results} \label{sec:apx_robustness_index_computation}

\noindent This section provides details regarding the implementation and calculation of the robustness index.

\subsection{Data pre-processing: l2 normalization}
Each representation vector was normalized using l2 normalization before computing the robustness index. By virtue of this choice, applying the Euclidean distance to the normalized embeddings, as used here, will result in the same neighbor ranking as using the cosine distance. 

\subsection{Selecting $k$ values} \label{sec:apx_robustness_index_k_values}

The robustness index metric is defined for all values of $k$. To compare models, it's preferable to obtain a single scalar metric; below, we discuss how this selection is performed. 

In this work, we choose to analyze the exact set of neighbors used by a $knn$ classifier, so that the robustness index value represents the robustness of the $knn$ classifier. We therefore base our choice of $k$ on values for $k$ that would be used in practice when applying $knn$ classification. We note that considerations for choosing $k$ will depend on the context and application of a foundation model; for example, if the nearest neighbors are to be used for reference search rather than prediction, then the number of desired similar samples could guide the choice of $k$.

When a $knn$ model is applied for a prediction task, $k$ will in current practice typically be chosen so as to facilitate high prediction performance\footnote{Our work suggests this should not be the only consideration; since the choice of $k$ influences the robustness of the model, we argue it's advisable to select $k$ such that a favorable tradeoff between prediction performance and robustness is attained}. Therefore, per model, we first determine the value of $k$ that maximizes the prediction performance of the biological class. Once all per-model optimal $k$ values have been determined, the median of these is used as the common choice of $k$ for a particular dataset.

The following median optimal $k$ values are obtained:
\begin{itemize}
\item TCGA Uniform: 61
\item Camelyon16: 11
\item Tolkach ESCA: 46
\end{itemize}

\paragraph{Leave one-case-out cross-validation for $knn$ classification}
To evaluate $knn$ prediction performance in the above procedure to identify the optimal value of $k$, we first determine all $k$ nearest neighbors for each sample in the dataset once for a high number of $k$. Next, for each sample, we remove all neighbors that come from the same case, as using neighbors from the same case would mean that case is split across training and evaluation. This evaluation can be viewed as leave-one-case-out cross-validation; each case is used in evaluation once, using all remaining cases for training. It can equivalently be viewed as $n$-fold case-level cross-validation, where $n$ is maximized to equal the total number of cases.

This yields the largest possible training set for each evaluation sample, and thus the most accurate, lowest variance result. It also implies that the procedure is deterministic, and has a single true outcome; i.e., there is no variation in training/evaluation splits, and thus no variance in the resulting outcome. Given that the dataset can be viewed as a sample from a larger, unknown full distribution, we estimate the variance in the resulting robustness index values using bootstrapping.

\paragraph{Efficient optimization of $k$ for $knn$ classification}

As noted above, we first determine all $k$ neighbors for a high value of $k$ once. Next, we iteratively decrease $k$, compute the corresponding predictions, and track the resulting prediction accuracies, as measured using balanced accuracy. Since the sets of neighbors for decreasing numbers of $k$ are strict subsets of the neighbors for larger values of $k$, we do not need to recompute the set of neighbors, but can simply take increasingly small subsets.

\subsection{Efficient computation of the robustness index}
To efficiently compute the robustness index for a range of k values, analogous to the above, a single knn model with the highest k value of the range can be fit. After retrieving the set of nearest neighbors from the fitted model (e.g.~using the {\textrm sklearn kneighbors} function) and retrieving binary matrices indicating, for each combination of sample x neighbor, whether the biological and confounding class are the same, the robustness index can be obtained by taking the cumulative sum of the column sums for SO and OS, and normalizing these vectors to sum to one, after which the normalized cumulative SO vector represents the robustness index.

\paragraph{Two dataset scenario}
The robustness index can also be computed for an evaluation set $D_{eval}$ in a two-dataset setup where neighbors are selected from a different dataset $D'$, provided that both datasets share the same set of biological and confounding classes.

\paragraph{Dataset details}
The details of the datasets used for robustness index calculation are as follows:
\begin{itemize}
\item Camelyon: We used the patches from the RUMC and UMCU medical centers (see Table~\ref{tab:dataset_statistics_camelyon}), yielding 20,400 patches.
\item TCGA~2x2: We used the full TCGA~2x2 dataset as specified in Table~\ref{tab:dataset_statistics_tcga2x2}, yielding a dataset of 4 * 300 * 94 = 112,800 patches in total.
\item Tolkach ESCA: We used the selection of the Tolkach ESCA dataset, described in Table~\ref{tab:dataset_statistics_tolkach}, with 5 randomly selected cases per combination of biological and confounding class, 100 patches per case, and excluding TCGA (as this contains multiple centers), yielding 9,000 patches in total. \footnote{The case ID for this dataset was kindly provided by Yuri Tolkach upon request.}
\end{itemize}

\paragraph{TCGA~2x2: paired version of the robustness index}

The robustness index, as defined in this article, can be applied to any dataset or pair of datasets for which both biological and confounding class information is available. To analyze whether biological or confounding information is more strongly represented, ideally, a balanced dataset should be used that has a similar number of biological and confounding classes, and where the number of samples for each combination of biological and confounding is similar (see Sup.~Note~\ref{sec:apx_dataset_details}). An example is the TCGA-2k dataset described in \cite{dejong_pathology_fm_robustness}, which has five biological and confounding classes chosen such that for 20 of the 25 combinations of these, an equal number of samples can be selected.

The requirement of similar distributions of biological and confounding classes is a limiting factor. To overcome this, when many biological and confounding classes are available, as is the case in TCGA, a paired variant of the robustness index can be calculated as follows:
\begin{itemize}
\item Double pairs of two biological classes (TCGA projects) x two confounding classes (tissue source sites) are selected. The TCGA~2x2 subset in the PathoROB benchmark contains 94 such double pairs.
\item The robustness index statistics (i.e.~the SO and OS frequencies) are calculated separately per double pair, and then aggregated over all 94 double pairs.
\end{itemize}

\subsection{Trading off performance vs robustness}
We argue that prediction performance and robustness are both important objectives that must be included in the evaluation of medical foundation models. Both objectives are to be optimized.

In the context of $knn$ classification, selecting a value for $k$ implies making a tradeoff between prediction performance and robustness. Our results show that for many models, robustness can easily be improved compared to the selected optimal $k$ value by selecting a different value for $k$. We encourage practitioners to carefully consider what tradeoff between robustness and prediction performance is advisable, and note that for medical applications, robustness is of the utmost importance, given that the lives of patients may be at stake.

\begin{figure}
    \centering
    \includegraphics[width=0.75\linewidth]{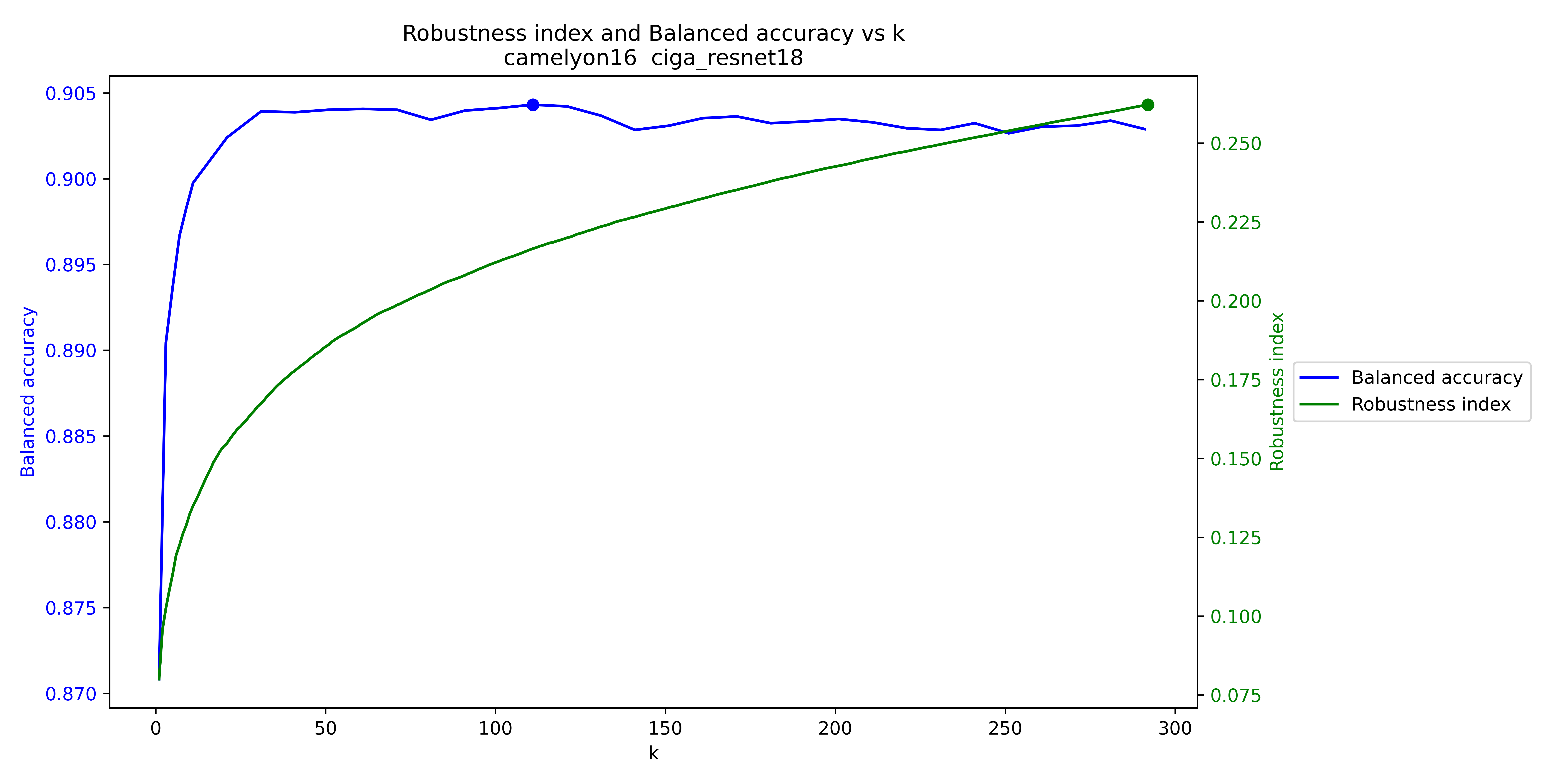}
    \caption{\textbf{Trading off performance vs robustness}. by choosing different values for $k$, different tradeoffs can be obtained.}
    \label{fig:tradeoff-performance-robustness}
\end{figure}

Figure \ref{fig:tradeoff-performance-robustness} shows how the choice of $k$ affects the prediction performance and robustness. While maximal prediction performance is obtained for $k$ values around 100, substantially higher robustness values can be achieved by choosing larger values of $k$, with relatively limited impact on prediction performance.

\subsection{Standard deviation estimation}
\label{bootstrapping}
For a given dataset, the robustness index calculation is deterministic, and so there is no variance in the value. We can view the dataset as having been sampled from the hypothetical larger distribution of all possible datasets that could have been obtained in the same manner from the same population, and the robustness index values for such different instantiations do have a variance.

Bootstrapping \cite{efron1992bootstrap} approximates this variance by treating the original dataset as a proxy for the population, generating multiple datasets via resampling, and computing the statistic's variability across these resamples. Given the $knn$ neighbor lists for all samples in the dataset, we randomly sample a dataset of the same size with replacement. The resulting dataset will contain some samples multiple times and will not include all samples. We calculate the robustness index over the sampled set. This procedure is repeated $n=1000$ times, yielding 1000 robustness index values. The estimated standard deviation of these robustness values is computed by taking the standard deviation over the computed robustness index values.

\subsection{Results per dataset}
\label{sec:apx_ri_resultsperset}
This section provides overview graphs with:
\begin{itemize}
    \item Top left: the frequencies of SO (solid lines) and OS (dashed lines) for neighbor $k$
    \item Top right: the same information, aggregated over the $k$ nearest neighbors
    \item Bottom left: the effect of k on prediction accuracy; the dot shows the optimal choice of k per model. The median of these optimal k values is chosen as the common choice of $k$.
    \item Bottom right: robustness index for all models; the value at the median $k$ value is used to compare models.
\end{itemize}

\begin{figure}[h]
    \centering
    \begin{minipage}{\textwidth}
        \centering
        \begin{minipage}[t]{0.48\textwidth}
            \centering
            \includegraphics[width=\linewidth]{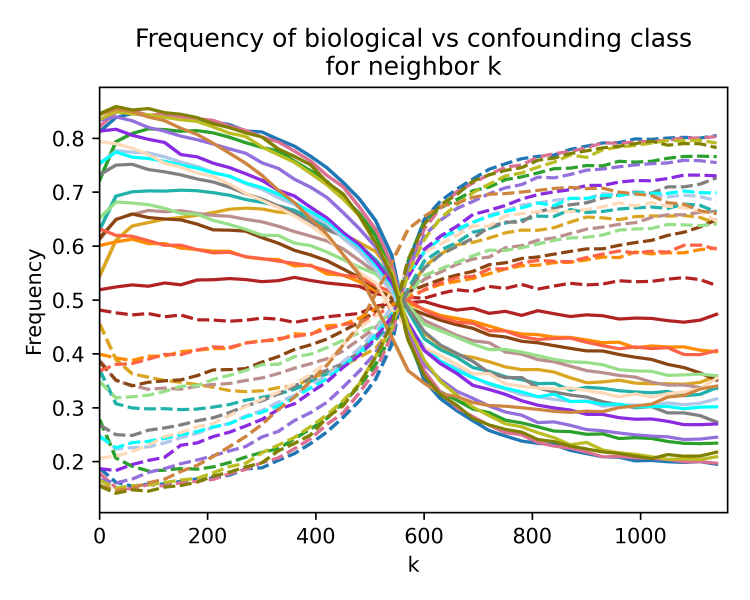}            
            \includegraphics[width=\linewidth]{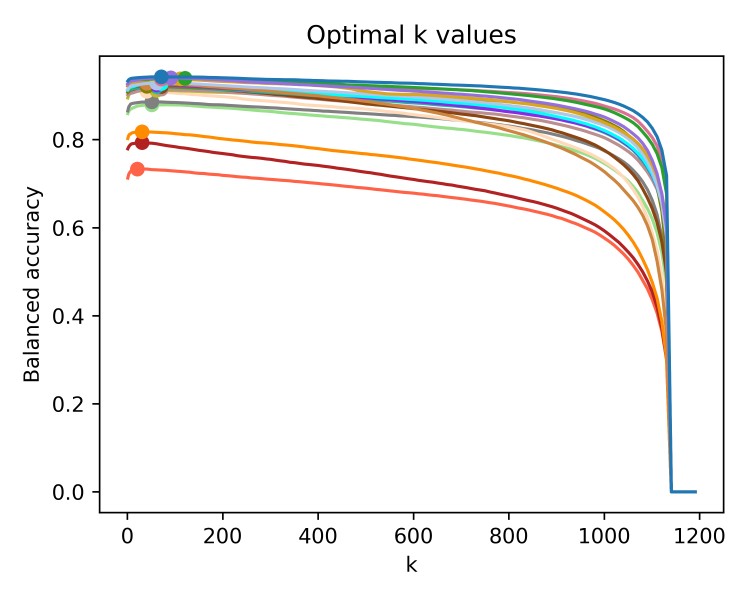}
        \end{minipage}%
        \hfill
        \begin{minipage}[t]{0.48\textwidth}
            \centering
            \includegraphics[width=\linewidth]{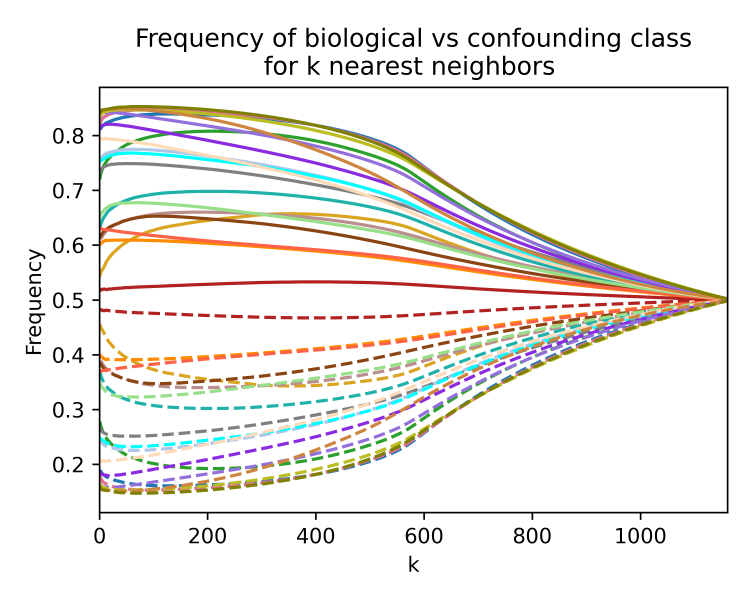}           
            \includegraphics[width=\linewidth]{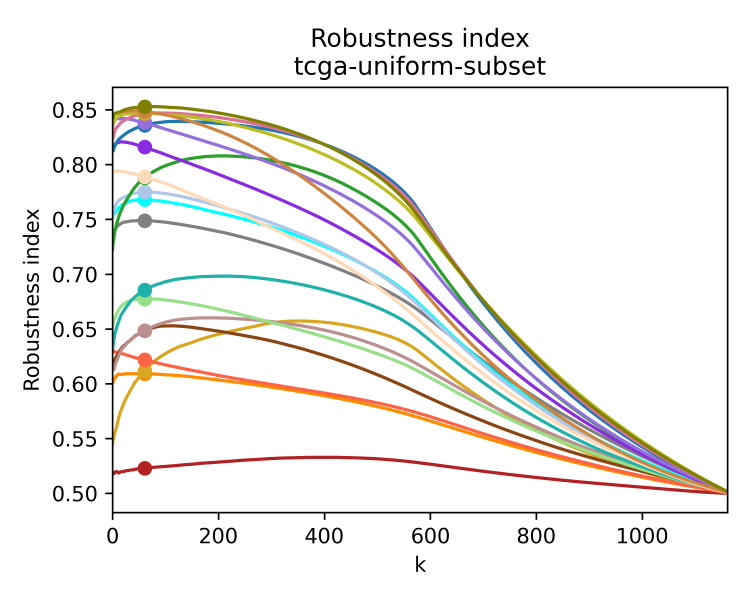}
        \end{minipage}
    \end{minipage}
        
    \begin{minipage}{\textwidth}
        \centering
        \includegraphics[width=0.8\linewidth]{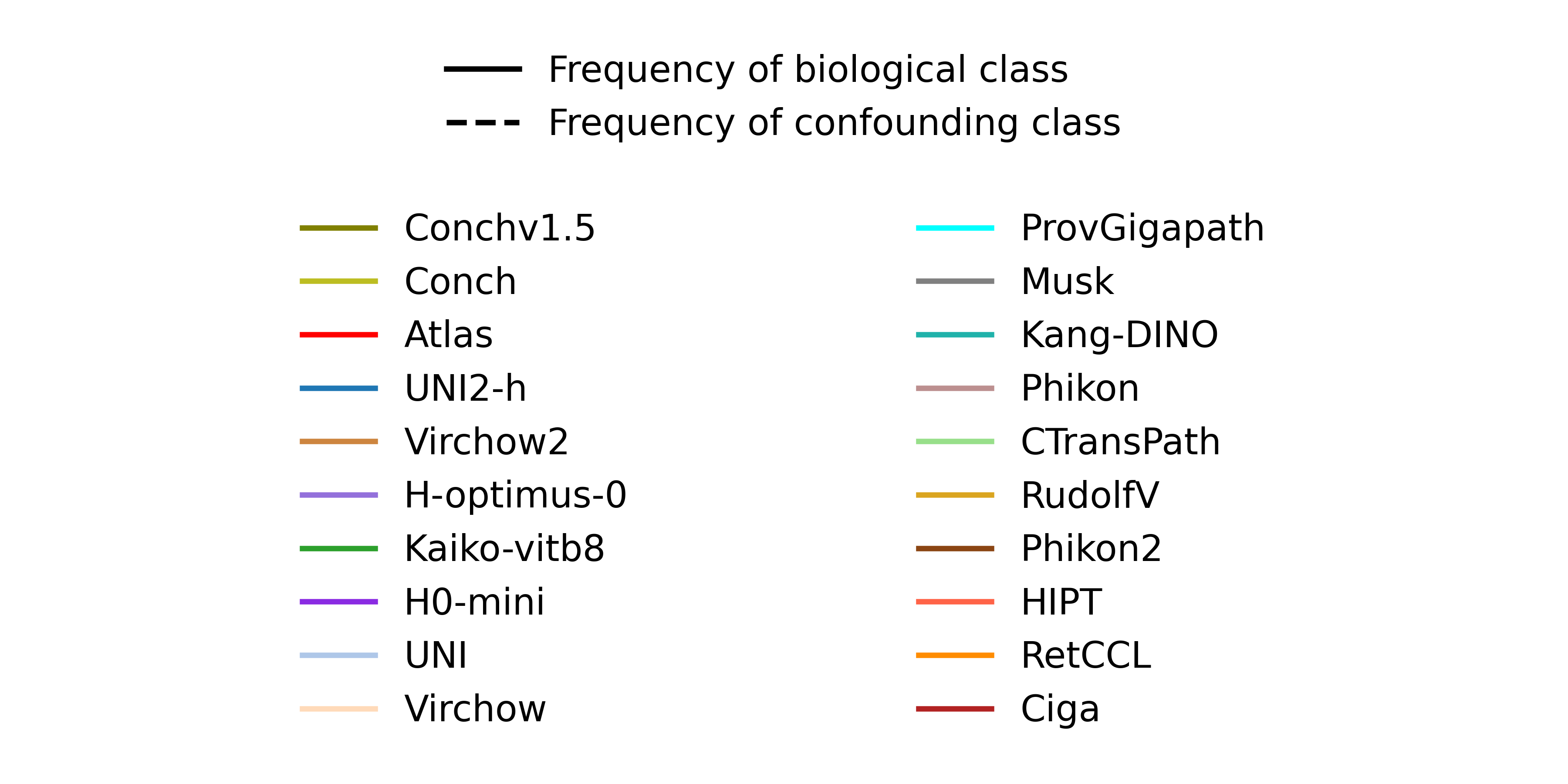}
    \end{minipage}
    \caption{\textbf{Robustness index results on TCGA~2x2 for all models}. See text.}
   \label{fig:rob-index-tcga}
\end{figure}

Figure \ref{fig:rob-index-tcga} shows the robustness index results on TCGA~2x2 for all models. As the starting point (top left), the frequency with which the {\it k}th-nearest neighbor (e.g. {\it k}=3 represent the third nearest neighbor) has the {\bf S}ame biological class (e.g.~cancer type) but the {\bf O}ther confounding class (e.g.~medical center) is measured (combination {\bf SO}; the solid lines show this for the {\it k} = 1 to 600 (the total number of SO + OS neighbors for one of the 94 TCGA project combinations). The complement of this, i.e.~the fraction of neighbors at distance rank {\it k} that have the {\bf O}ther biological class but the {\bf S}ame confounding class (combination {\bf OS}) is represented by the dashed lines. As the analysis is restricted to the {\bf SO} and {\bf OS} combinations, these two lines sum to one, and the total number of neighbors considered is restricted to an exclusive upper bound of 600 of the 4*300=1200 for each TCGA project combination. The top right show the same information in aggregated form, reflecting these frequencies of all nearest neighbors from distance rank 1 up to the specified value of {\it k}, as used in {\it knn} classifiers. The solid lines at the top equal the robustness index. To select a value for {\it k}, the bottom left shows the performance for predicting the biological class as a function of {\it k}; dots mark the optimal value of {\it k} for each model. The median of these optimal {\it k} values, 61, is used in the bottom right zoomed-in robustness index plot to obtain the scalar robustness index value.

\begin{figure}[h]
    \centering
    \begin{minipage}{\textwidth}
        \centering
        \begin{minipage}[t]{0.48\textwidth}
            \centering
            \includegraphics[width=\linewidth]{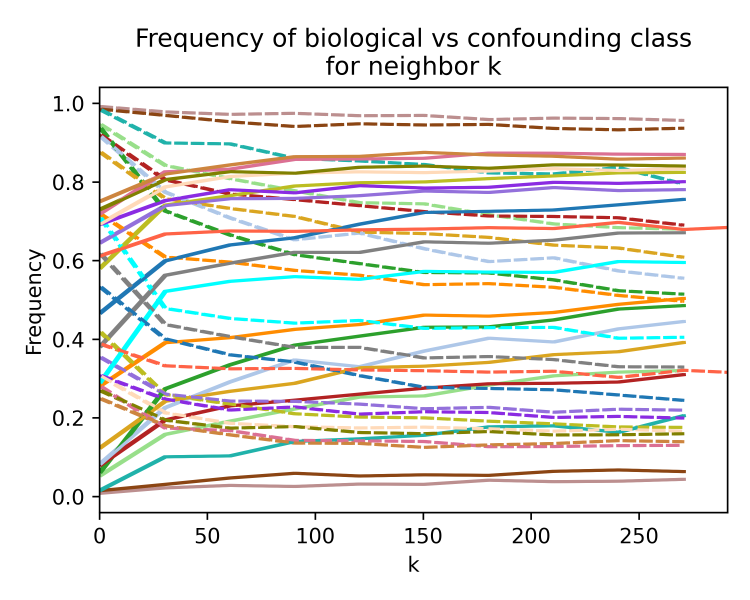}            
            \includegraphics[width=\linewidth]{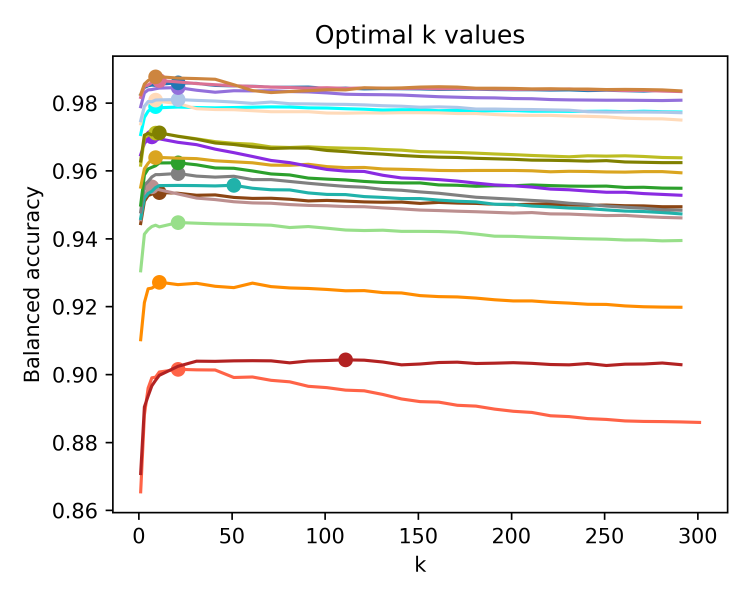}
        \end{minipage}%
        \hfill
        \begin{minipage}[t]{0.48\textwidth}
            \centering
            \includegraphics[width=\linewidth]{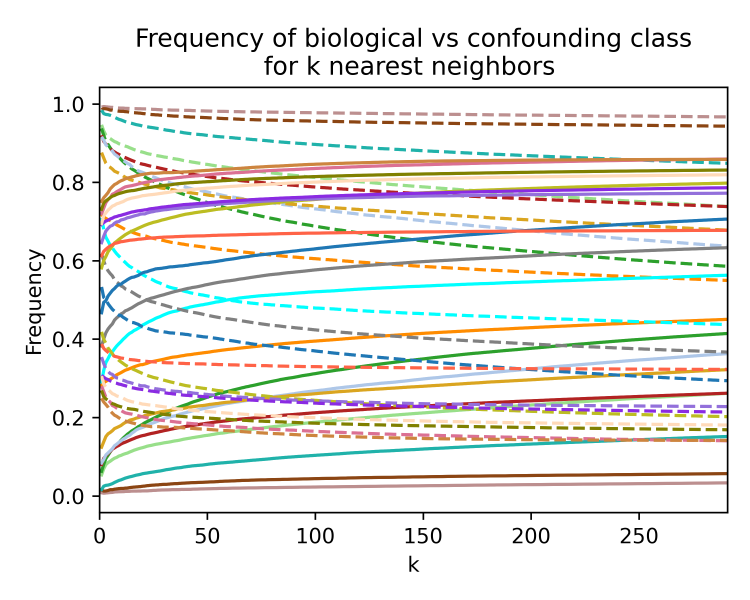}           
            \includegraphics[width=\linewidth]{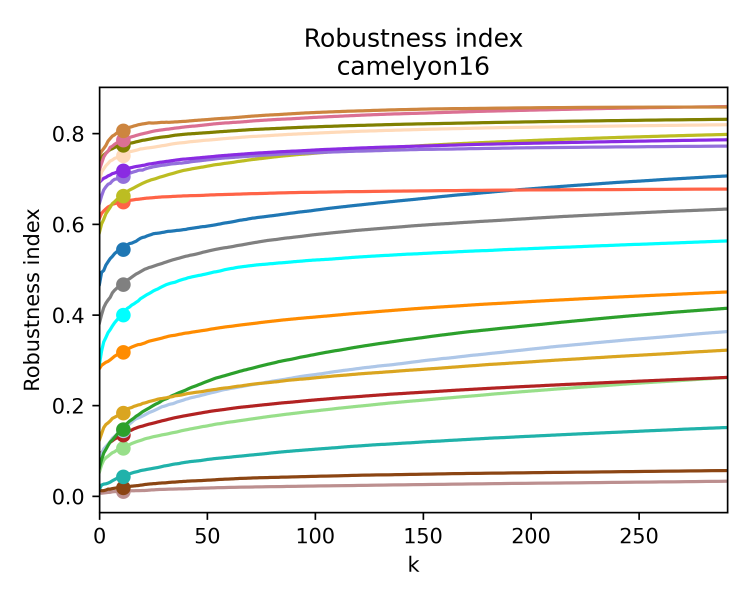}
        \end{minipage}
    \end{minipage}
        
    \begin{minipage}{\textwidth}
        \centering
        \includegraphics[width=0.8\linewidth]{figures/robustness_index/standalone-legend.png}
    \end{minipage}
    \caption{\textbf{Robustness index results on Camelyon}.}
   \label{fig:rob-index-camelyon16}
\end{figure}

\begin{figure}[h]
    \centering
    \begin{minipage}{\textwidth}
        \centering
        \begin{minipage}[t]{0.48\textwidth}
            \centering
            \includegraphics[width=\linewidth]{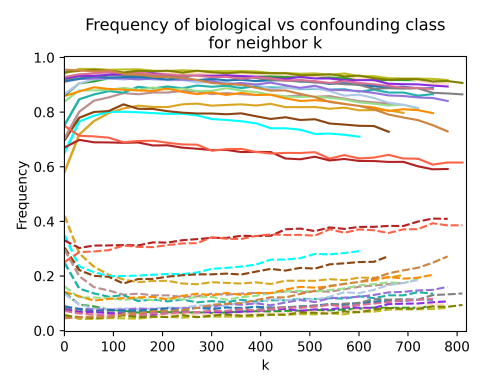}            
            \includegraphics[width=\linewidth]{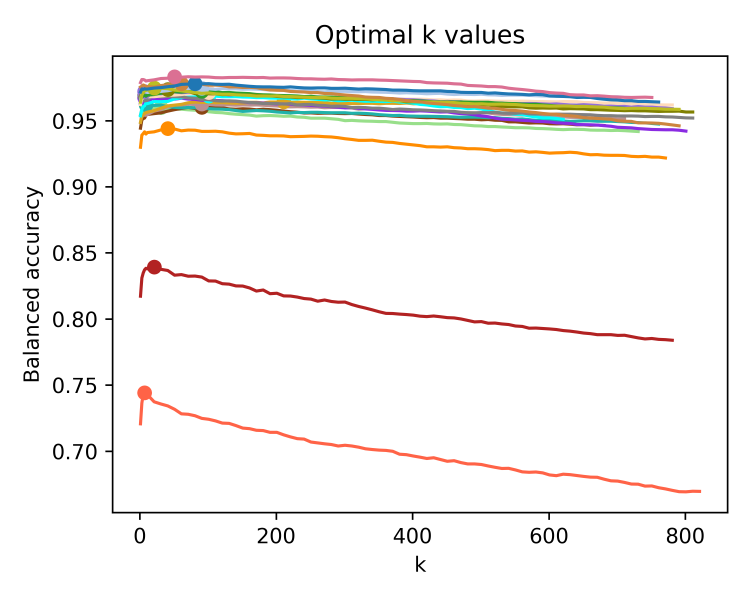}
        \end{minipage}%
        \hfill
        \begin{minipage}[t]{0.48\textwidth}
            \centering
            \includegraphics[width=\linewidth]{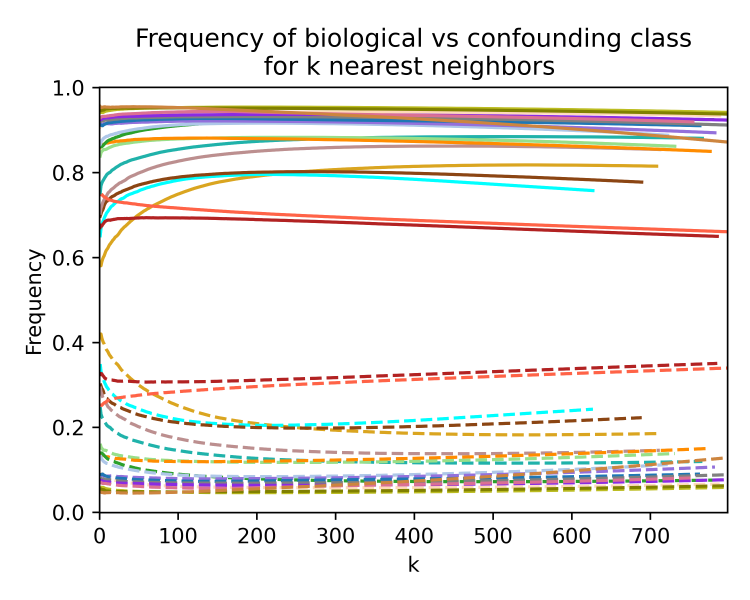}           
            \includegraphics[width=\linewidth]{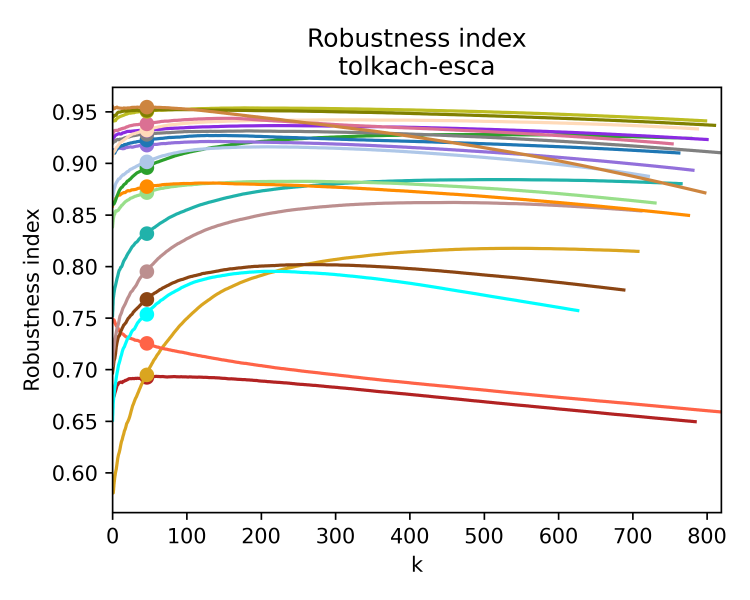}
        \end{minipage}
    \end{minipage}
        
    \begin{minipage}{\textwidth}
        \centering
        \includegraphics[width=0.8\linewidth]{figures/robustness_index/standalone-legend.png}
    \end{minipage}
    \caption{\textbf{Robustness index results on Tolkach ESCA}.}
   \label{fig:rob-index-tolkach}
\end{figure}

Figures~\ref{fig:rob-index-camelyon16}~and~\ref{fig:rob-index-tolkach} on the following pages show the same graphs for the Camelyon and Tolkach ESCA datasets.

\FloatBarrier
\subsection{Robustness index: numerical results}
This subsection summarizes the robustness index results for all processings. For the default processing (no robustification method applied), we include the results of bootstrapping, see above under Section \ref{bootstrapping}; the second and third numerical columns report the mean and standard deviation of the resulting 1000 robustness index values.

\setlength{\tabcolsep}{3pt}
\begin{table}[h!]
\centering
\begin{tabular}{|l|c|c|c|c|c|c|c|c|}
\hline
Processing & Default & Default & Default & Reinhard & Combat & Reinhard \\
 &  &  &  &  &  & Combat \\
\hline
&  &  &  &  &  & \\
Model & RI & RI mean & std dev & RI & RI & RI \\
\hline 
\matlas{} & 0.846 & 0.846 & 0.002 & 0.849 & 0.868 & 0.870  \\
\mciga{} & 0.523 & 0.523 & 0.002 & 0.569 & 0.623 & 0.630  \\
\mconch{} & 0.847 & 0.847 & 0.001 & 0.843 & 0.834 & 0.827  \\
\mconchONEFIVE{} & 0.853 & 0.853 & 0.001 & 0.861 & 0.843 & 0.839  \\
\mhoptimus{} & 0.838 & 0.838 & 0.002 & 0.843 & 0.87 & 0.866  \\
\mhmini{} & 0.816 & 0.816 & 0.002 & 0.824 & 0.853 & 0.850  \\
\mhipt{} & 0.622 & 0.622 & 0.001 & 0.637 & 0.628 & 0.636  \\
\mkaiko{} & 0.788 & 0.788 & 0.002 & 0.799 & 0.825 & 0.830  \\
\mkangdino{} & 0.685 & 0.685 & 0.002 & 0.692 & 0.82 & 0.818  \\
\mphikon{} & 0.648 & 0.648 & 0.002 & 0.7 & 0.782 & 0.782  \\
\mphikonTWO{} & 0.648 & 0.648 & 0.002 & 0.683 & 0.784 & 0.778  \\
\mprovgigapath{} & 0.768 & 0.768 & 0.002 & 0.804 & 0.833 & 0.839  \\
\mretccl{} & 0.609 & 0.609 & 0.002 & 0.622 & 0.706 & 0.691  \\
\mrudolfv{} & 0.611 & 0.611 & 0.003 & 0.62 & 0.748 & 0.750  \\
\mctranspath{} & 0.677 & 0.677 & 0.002 & 0.696 & 0.758 & 0.746  \\
\muni{} & 0.775 & 0.775 & 0.002 & 0.783 & 0.844 & 0.831  \\
\muniTWO{} & 0.836 & 0.836 & 0.002 & 0.86 & 0.861 & 0.870  \\
\mmusk{} & 0.749 & 0.749 & 0.002 & 0.765 & 0.778 & 0.779  \\
\mvirchow{} & 0.789 & 0.789 & 0.002 & 0.742 & 0.8 & 0.731  \\
\mvirchowTWO{} & 0.848 & 0.848 & 0.001 & 0.852 & 0.859 & 0.856  \\
\hline
\end{tabular}
\caption{\textbf{Robustness index results for different processings on TCGA~2x2}. The first column shows the robustness index computed over the full dataset with leave-one-case-out cross-validation for the default processing (no robustification method applied). The second and third columns show the mean and standard deviation of the robustness index computed using bootstrapping. The remaining columns show the robustness index for the different robustification methods compared.}
\end{table}

\begin{table}[h!]
\centering
\begin{tabular}{|l|c|c|c|c|c|c|}
\hline
Processing & Default & Default & Default & Reinhard & Combat & Reinhard \\
 &  &  &  &  &  & Combat \\
\hline
Model & RI & RI mean & std dev & RI & RI & RI \\
\hline
\matlas{} & 0.785 & 0.786 & 0.008  &0.823  &0.927    &0.926  \\
\mciga{} & 0.135 & 0.135 & 0.004  &0.569  &0.458    &0.654  \\
\mconch{} & 0.662 & 0.662 & 0.007  &0.889  &0.921    &0.931  \\
\mconchONEFIVE{} & 0.774 & 0.774 & 0.006  &0.892  &0.925   &0.930  \\
\mhoptimus{} & 0.705 & 0.705 & 0.009  &0.890  &0.907  &0.931  \\
\mhmini{} & 0.718 & 0.717 & 0.007  &0.840  &0.916   &0.918  \\
\mhipt{} & 0.649 & 0.649 & 0.004  &0.734  &0.755   &0.790  \\
\mkaiko{} & 0.147 & 0.147 & 0.007 &0.742  &0.631   &0.835  \\
\mkangdino{} & 0.043 & 0.043 & 0.004  &0.290  &0.758   &0.776  \\
\mphikon{} & 0.011 & 0.011 & 0.002  &0.462  &0.575   &0.740  \\
\mphikonTWO{} & 0.019 & 0.019 & 0.002  &0.484  &0.446   &0.724  \\
\mprovgigapath{} & 0.399 & 0.399 & 0.012  &0.831  &0.800   &0.896  \\
\mretccl{} & 0.318 & 0.318 & 0.006  &0.665  &0.757   &0.801  \\
\mrudolfv{} & 0.184 & 0.183 & 0.008  &0.232  &0.616   &0.634  \\
\mctranspath{} & 0.106 & 0.106 & 0.005  &0.589  &0.668   &0.772  \\
\muni{} & 0.145 & 0.145 & 0.009  &0.712  &0.751   &0.875  \\
\muniTWO{} & 0.544 & 0.545 & 0.013  &0.872  &0.868  &0.931  \\
\mmusk{} & 0.467 & 0.467 & 0.008  &0.782  &0.812   &0.866  \\
\mvirchow{} & 0.751 & 0.750 & 0.008  &0.891  &0.896  &0.918  \\
\mvirchowTWO{} & 0.806 & 0.806 & 0.007  &0.923  &0.941  &0.953  \\
\hline
\end{tabular}
\caption{\textbf{Robustness index results for different processings on Camelyon}. The first column shows the robustness index computed over the full dataset with leave-one-case-out cross-validation for the default processing (no mitigation method applied). The second and third columns show the mean and standard deviation of the robustness index computed using bootstrapping. The remaining columns show the robustness index for the different mitigation methods compared. }
\end{table}

\begin{table}[h!]
\centering
\begin{tabular}{|l|c|c|c|c|c|c|}
\hline
Processing & Default & Default & Default & Reinhard & Combat & Reinhard \\
 &  &  &  &  &  & Combat \\
\hline
Model & RI & RI mean & std dev & RI & RI & RI \\
\hline
\matlas{} & 0.938 & 0.939 & 0.003  &0.940  &0.972  &0.971  \\
\mciga{} & 0.693 & 0.693 & 0.004 &0.685  &0.801   &0.781  \\
\mconch{} & 0.951 & 0.951 & 0.002 &0.945  &0.972  &0.965  \\
\mconchONEFIVE{} & 0.951 & 0.951 & 0.002 &0.956  &0.969  &0.969  \\
\mhoptimus{} & 0.918 & 0.918 & 0.003 &0.935  &0.953  &0.958  \\
\mhmini{} & 0.932 & 0.932 & 0.003 &0.940  &0.963   &0.962  \\
\mhipt{} & 0.726 & 0.725 & 0.004 &0.728  &0.765  &0.774  \\
\mkaiko{} & 0.896 & 0.896 & 0.004 &0.920  &0.955  &0.954  \\
\mkangdino{} & 0.832 & 0.832 & 0.005 &0.863  &0.941   &0.939  \\
\mphikon{} & 0.795 & 0.795 & 0.006  &0.835  &0.934  &0.922  \\
\mphikonTWO{} & 0.768 & 0.768 & 0.006 &0.820  &0.923  &0.910  \\
\mprovgigapath{} & 0.754 & 0.754 & 0.007  &0.887  &0.915  &0.939  \\
\mretccl{} & 0.878 & 0.877 & 0.003  &0.853  &0.943  &0.925  \\
\mrudolfv{} & 0.695 & 0.695 & 0.007 &0.794  &0.883   &0.899  \\
\mctranspath{} & 0.872 & 0.872 & 0.004 &0.872  &0.947  &0.937  \\
\muni{} & 0.902 & 0.901 & 0.004 &0.907  &0.957   &0.951  \\
\muniTWO{} & 0.923 & 0.923 & 0.003 &0.944  &0.957  &0.963  \\
\mmusk{} & 0.928 & 0.928 & 0.002  &0.931  &0.958  &0.954  \\
\mvirchow{} & 0.932 & 0.932 & 0.003 &0.940  &0.959  &0.959  \\
\mvirchowTWO{} & 0.955 & 0.955 & 0.002 &0.955  &0.970  &0.967  \\
\hline
\end{tabular}
\caption{\textbf{Robustness index results for different processings on Tolkach ESCA}. The first column shows the robustness index computed over the full dataset with leave-one-case-out cross-validation for the default processing (no mitigation method applied). The second and third columns show the mean and standard deviation of the robustness index computed using bootstrapping. The remaining columns show the robustness index for the different mitigation methods compared. }
\end{table}

\setlength{\tabcolsep}{6pt} 
\FloatBarrier

\subsection{Robustness index per biological and confounding category}
The main results measure the degree to which biological features dominate confounding ones in the embedding space in total, computed over the full dataset. To see to what extent the robustness of a foundation model varies across different biological or confounding categories, we computed the robustness index per class for all biological and confounding categories.

\begin{figure}[h]
    \centering
    \begin{tabular}{cc}
    \includegraphics[width=0.5\linewidth]{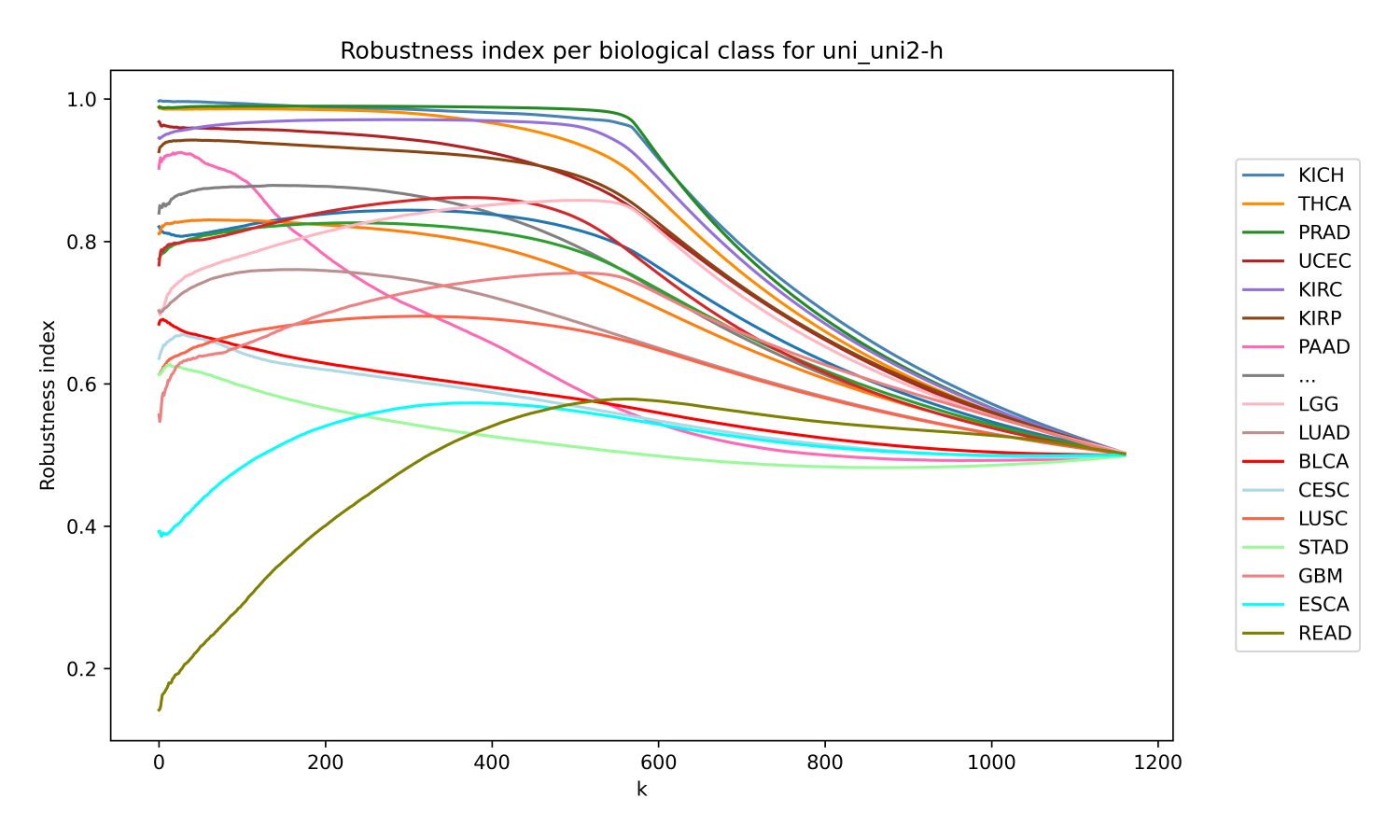}
    \includegraphics[width=0.5\linewidth]{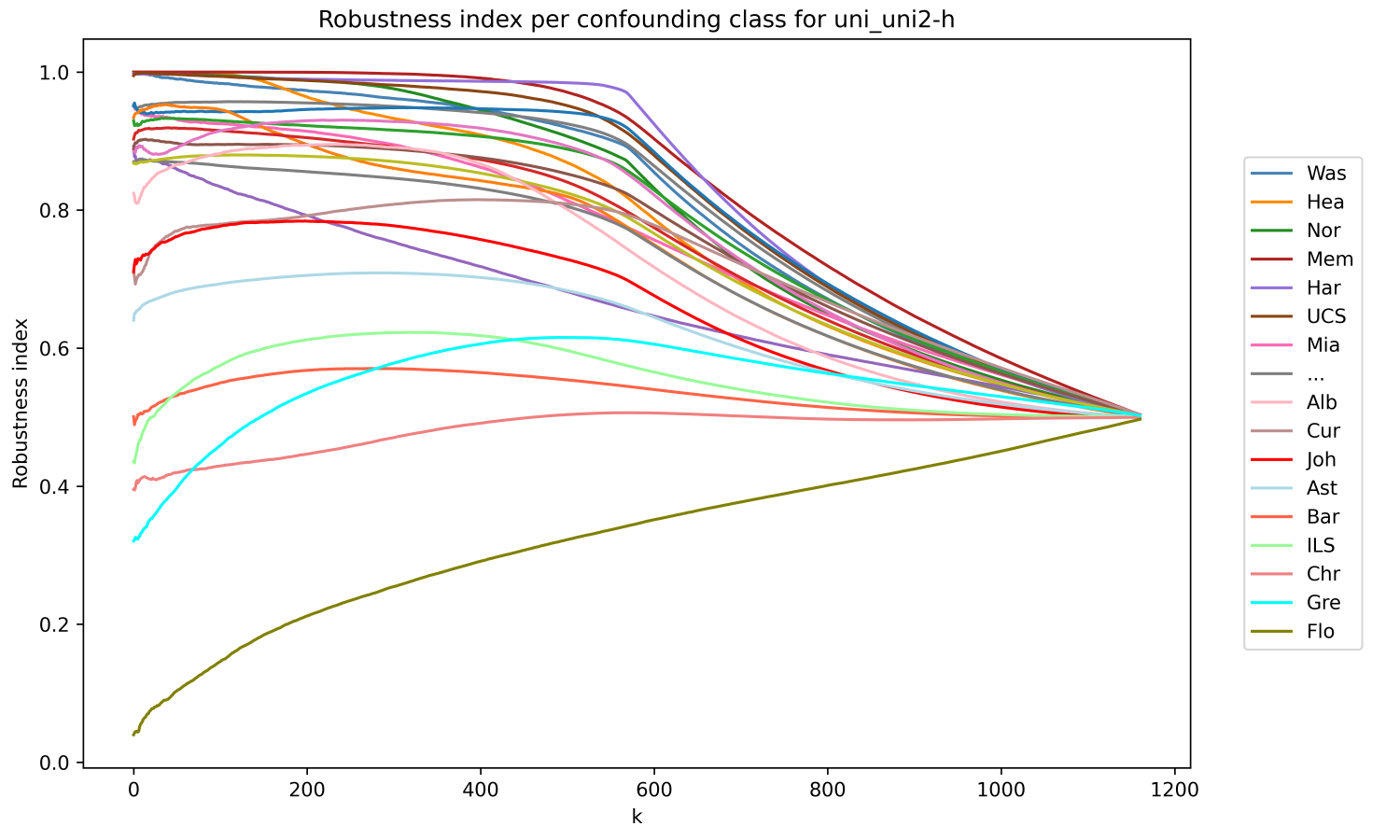}
    \end{tabular}
    \caption{\textbf{Robustness index per biological class (left) and per confounding class (right) for \muniTWO{}}. This breakdown shows that robustness varies substantially across different categories, for both biological and confounding categories.}
    \label{fig:robustness_per_category}
\end{figure}

Figure \ref{fig:robustness_per_category} shows the results for \muniTWO{}. For both biological and confounding classes, robustness varies substantially across different categories.

\FloatBarrier

\subsection{Potential variants of the robustness index and room for robustness improvement}
The current work motivates the choice for focusing on the SO and OS combinations in Section~\ref{sec:methods_robustness}, and demonstrates that this choice provides an informative metric. 
When viewing the curves in Figure 7 in the main text, it is seen that the largest differences between the two models are observed when comparing the two SO lines (orange) with each other, or the two OS lines (green). By contrast, the SS lines (blue) and the OO lines (red) differ less between the models.

\paragraph{Generalization index}
A different type of comparison we can make is within one model. For both models, there is a sharp discrepancy between the blue and orange lines, representing the frequency of the correct biological class among the nearest neighbors for same-medical-center samples (SS, blue) and other-medical-center samples (SO, orange). If the model were insensitive to medical center differences, these lines should look similar. They clearly do not; the orange lines start much lower than the blue lines, showing that both models are far less successful at identifying same-biological-class samples from other medical centers than from the same medical center as the sample. 

This difference is highly concerning, and even occurs for \mconchONEFIVE{}, one of the most robust models evaluated in this work. This suggests there is still a long way to go for current foundation models to become more robust. It also suggests that comparing the SO and SS lines can provide an additional useful metric of robustness. 

When the confounding classes are medical centers or datasets, along the same lines, we can measure {\em generalization performance} on unseen data, and compare this to performance on seen data, i.e.~data from the same medical center as the evaluation sample. To do so, we can compute the relation between Out Of Distribution (OOD) prediction performance and In Distribution (ID) performance. Rather than measuring accuracy (fraction of correct classifications), however, which reflects binary classification decisions per sample, we can again consider the fraction of neighbors of each sample representing the correct biological class; this provides a more fine-grained and precise metric. 

OOD performance is then reflected by $\frac{SO}{SO+OO}$, and ID performance by $\frac{SS}{SS+OS}$. For a fully robust model, OOD performance should equal ID performance, as the model performs equally well for unseen centers. Thus, the ratio between OOD performance and ID performance provides a metric of generalization to unseen data, which we name the {\em Generalization Index}, $\mathcal G$:
$$\mathcal G=\frac{\text{OOD~performance}}{\text{ID~performance}} = \frac{\frac{SO}{SO+OO}}  {\frac{SS}{SS+OS}} =  \frac{SO}{SO+OO} \cdot \frac{SS+OS}{SS} = \frac{SS \cdot SO + SO \cdot OS}{SS \cdot SO + SS \cdot OO}$$

Finally, we note that the current work has evaluated the robustness of current pathology foundation models on patch-level tasks. Evaluating slide-level foundation models on robustness would form a valuable follow-up study.

\FloatBarrier
\section{Robust foundation models prevent entangled features}
\label{sec:apx_downstream_entanglement}

\subsection{Spurious correlation in downstream datasets}

Multi-site histopathology datasets are typically imbalanced across contributing centers; for example, a disease class may be overrepresented at one medical site and underrepresented at others. This is particularly pronounced for rare diseases or tumor subtypes. Such imbalances, coupled with center-specific technical artifacts, induce \textit{spurious correlations} between biological labels and the site signatures: a deceptive relationship. While these correlations reflect co-occurrence, they do not correspond to a medically meaningful coherence or causation \cite{Deangeli2022, Guerrero2024}. The risk of spurious correlation increases particularly in small datasets from a limited number of medical centers. Without deliberate data curation and balancing, it is unlikely that the distributions of biological classes, confounding factors, or their combinations will be uniform across sites.

\subsection{Spurious correlation can hamper generalizability}
Spurious correlation can significantly impair the generalizability of downstream models trained on representations of non-robust foundation models. An unrobust foundation model might encode both biologically meaningful and confounding information, thereby potentially propagating spurious correlation from the input to its representation space; its representations hold biological and confounding features, each correlated with the prediction target. A feature refers to a dimension or a one-dimensional direction in the representation space. Although the representations may contain sufficient information to learn the correct association between the biological features and the target label, the downstream model might just as well base its predictions on the confounding feature instead. Crucially, the model has no inherent way of distinguishing or disentangling whether the correlation between a given feature and the target is medically meaningful or merely spurious. As a result, downstream models trained on spuriously correlated representations may inadvertently leverage these medical center-specific signals as \textit{Clever Hans} (or shortcut) features. Although these Clever Hans features may enhance apparent model accuracy during development, they compromise generalization when such spurious correlations are absent at deployment \cite{clever-hans, geirhos2020shortcut, hermannfoundations, unsupervised-clever-hans, Howard2021, mahmood-demographic, brown2023detecting}.

In contrast to the abundance of unlabeled datasets used for pre-training foundation models, only limited labeled datasets are typically available for training on a downstream task, effectively increasing the risk of having spurious correlations in downstream inputs. A non-robust FM that encodes confounding variables within its representations then inherently snowballs this spurious correlation, possibly resulting in Clever Hans learning in downstream tasks. Conversely, the representations of a robust FM are free of such confounding information. Robustness in FMs effectively precludes the risk of Clever Hans learning even under strongly biased data --- after all, downstream models cannot exploit signals that are not encoded.

\subsection{Polysemy aggravates robustification}
The problem of Clever Hans learning is further exacerbated by \textit{polysemy}. Polysemantic features (cf.~\cite{elhage2022superposition}) that encode a mixture of biologically meaningful and confounding information make it substantially more difficult for the downstream model to ascertain whether a given relationship is medically meaningful or the result of spurious correlation. Polysemy also presents a major challenge for robustification: if a feature encodes both types of information, removing it to eliminate confounding signals will also discard relevant biological information, thereby reducing the overall utility of the representation. This issue is also discussed in Sup.~Note~\ref{sec:apx_feature-space-analysis}.

\subsection{Illustrative example:  unrobust foundation models enable Clever Hans learning}
\begin{table}[h!]
\centering
\begin{tabular}{*{8}{b{0.09\textwidth}}}
\toprule
\textbf{\rotatebox[origin=lb]{90}{\begin{tabular}[c]{@{}l@{}}Biological\\ target\end{tabular}}} & 
\multicolumn{1}{l}{\textbf{\rotatebox[origin=lb]{90}{\begin{tabular}[c]{@{}l@{}}Medical\\ center\end{tabular}}}} & 
\multicolumn{1}{l}{\textbf{\rotatebox[origin=lb]{90}{\begin{tabular}[c]{@{}l@{}}Feature 1\\ morphology\end{tabular}}}} & 
\multicolumn{1}{l}{\textbf{\rotatebox[origin=lb]{90}{\begin{tabular}[c]{@{}l@{}}Prediction\\ based on f1\end{tabular}}}} & 
\multicolumn{1}{l}{\textbf{\rotatebox[origin=lb]{90}{\begin{tabular}[c]{@{}l@{}}Feature 2\\ staining\end{tabular}}}} & 
\multicolumn{1}{l}{\textbf{\rotatebox[origin=lb]{90}{\begin{tabular}[c]{@{}l@{}}Prediction\\ based on f2\end{tabular}}}} & 
\multicolumn{1}{l}{\textbf{\rotatebox[origin=lb]{90}{\begin{tabular}[c]{@{}l@{}}Feature 3\\ polysemantic\end{tabular}}}} & 
\multicolumn{1}{l}{\textbf{\rotatebox[origin=lb]{90}{\begin{tabular}[c]{@{}l@{}}Prediction\\ based on f3\end{tabular}}}} \\
\midrule
\rowcolor[HTML]{C0C0C0} 
\multicolumn{8}{l}{\cellcolor[HTML]{C0C0C0}\textbf{Training set}} \\
normal & A & 0 & \textcolor{green}{\checkmark} & 0 & \textcolor{green}{\checkmark} & 0 & \textcolor{green}{\checkmark} \\
\rowcolor[HTML]{EFEFEF} 
normal & A & 0 & \textcolor{green}{\checkmark} & 0 & \textcolor{green}{\checkmark} & 0 & \textcolor{green}{\checkmark} \\
normal & A & 0 & \textcolor{green}{\checkmark} & 0 & \textcolor{green}{\checkmark} & 0 & \textcolor{green}{\checkmark} \\
\rowcolor[HTML]{EFEFEF} 
normal & A & 0 & \textcolor{green}{\checkmark} & 0 & \textcolor{green}{\checkmark} & 0 & \textcolor{green}{\checkmark} \\
tumor & B & 1 & \textcolor{green}{\checkmark} & 1 & \textcolor{green}{\checkmark} & 1 & \textcolor{green}{\checkmark} \\
\rowcolor[HTML]{EFEFEF} 
tumor & B & 1 & \textcolor{green}{\checkmark} & 1 & \textcolor{green}{\checkmark} & 1 & \textcolor{green}{\checkmark} \\
tumor & B & 1 & \textcolor{green}{\checkmark} & 1 & \textcolor{green}{\checkmark} & 1 & \textcolor{green}{\checkmark} \\
\rowcolor[HTML]{EFEFEF} 
tumor & B & 1 & \textcolor{green}{\checkmark} & 1 & \textcolor{green}{\checkmark} & 1 & \textcolor{green}{\checkmark} \\
\rowcolor[HTML]{C0C0C0} 
\multicolumn{8}{l}{\cellcolor[HTML]{C0C0C0}\textbf{Test set}} \\
normal & A & 0 & \textcolor{green}{\checkmark} & 0 & \textcolor{green}{\checkmark} & 0 & \textcolor{green}{\checkmark} \\
\rowcolor[HTML]{EFEFEF} 
normal & B & 0 & \textcolor{green}{\checkmark} & 1 & \textcolor{red}{$\times$} & 0.5 & \textcolor{red}{?} \\
tumor & A & 1 & \textcolor{green}{\checkmark} & 0 & \textcolor{red}{$\times$} & 0.5 & \textcolor{red}{?} \\
\rowcolor[HTML]{EFEFEF} 
tumor & B & 1 & \textcolor{green}{\checkmark} & 1 & \textcolor{green}{\checkmark} & 1 & \textcolor{green}{\checkmark}\\
\bottomrule
\end{tabular}
\caption{\textbf{Effect of foundation model robustness on downstream model learning with spuriously correlated training data.} Non-robust FMs encode both biological (f1) and technical (confounding) features (f2, f3) in their representations, enabling downstream models to exploit spurious correlations between medical center signatures and prediction targets. Robust FMs minimize confounding features in their representations, thereby enforcing downstream models to learn from biological signals alone.}
\label{tab:toy_example}
\end{table}

To illustrate this point, Table~\ref{tab:toy_example} presents a synthetic example resembling the final data split described in Section 2.3.1. A small dataset is derived from two centers, A and B. The downstream task is to classify tumor status. In the training set, however, the center of origin is perfectly (spuriously) correlated with the tumor label.
All samples from center A are normal, and all from center B are tumorous. Thus, the model could learn to associate tumor status with center identity rather than true morphological features. In the test set, this spurious correlation is absent: samples from center B are normal, and those from center A are tumorous.

We compare the possible prediction outcomes of a downstream model trained on the representations of a non-robust foundation model. An unrobust foundation model might encode three features: Feature 1 captures biologically relevant information (e.g., tissue morphology) that accurately reflects tumor status. A downstream model relying on this feature learns the correct, generalizable association between input and label, and performs reliably on the test set.
In contrast, feature 2 encodes a confounding feature which corresponds to a site-specific artifact (e.g., staining) and is spuriously correlated with the biological target in the training set. A downstream model that relies on this feature performs well during training but fails to generalize, leading to misclassification once the spurious correlation is absent.
Feature 3 is polysemantic, encoding both morphological and staining information. On the test set, it partially activates in response to either the tumor status or the medical center artifact. As a result, a downstream model trained on this feature exhibits ambiguous behavior, potentially leading to misclassification and poor generalization.

Crucially, all three features yield identical prediction outcomes on the training set. Consequently, a downstream model has no inherent way of distinguishing which feature encodes biologically meaningful information and which relies on spurious correlations. It may therefore base its predictions on any of the available features, including those that do not generalize. In contrast, a fully robust foundation model encodes only the biologically relevant information (feature 1), thereby ensuring that any downstream model learns the true, medically meaningful relationship between input and target.

\subsection{Connection to real-data experiments}

Section~\ref{sec:downstream_experiments} illustrated these issues using real clinical data rather than a synthetic example. In this experiment, the degree of spurious correlation between the biological target and the medical center of origin was systematically varied in the training data. This design allowed us to assess the consequences of spurious correlations and the corresponding risk of Clever Hans learning. Notably, in all splits, the downstream model had both the capacity (evident from its strong performance on the first split) and the necessary information (even for the skewed splits, the training data held information about \textit{all} biological classes) to learn the true association between the representations and the biological target. 

However, as the level of spurious correlation between the medical center and the biological target increased, the risk of Clever Hans learning became more pronounced. Although the FMs continued to encode all relevant biological information, the influence of confounding features became increasingly apparent. In these settings, downstream models were more likely to base their predictions on these confounding signals, leading to degraded performance across nearly all FMs. Only the most robust FMs maintained high accuracy across all splits, which is also reflected in a high correlation between the robustness index and the average performance drop.

At first glance, these results may appear to contradict the high medical center classification accuracies shown in Figure~\ref{fig:sketch-overview}d: even the most robust FMs enabled a reliable prediction of the medical center from their representations. However, a feature space analysis (described in detail in Sup.~Note~\ref{sec:apx_feature-space-analysis}) clarified this apparent discrepancy. While confounding information is encoded by all foundation models, non-robust FMs tend to encode it more strongly and blatantly (linearly, in directions of largest variance). This makes such features more readily accessible to downstream models, which tend to rely on easily extractable signals \cite{hermannfoundations, geirhos2018imagenet}. Furthermore, with an increasing level of spurious correlation between the medical center and the biological target, the apparent predictivity of the confounding feature is heightened, ultimately reaching the same alignment with the target as the biological feature. Thus, unrobust FMs pose a severe risk of Clever Hans learning because they accessibly encode confounding features that have a fallacious predictive value in spuriously correlated datasets.
While perfect robustness of foundation models might not be achievable (yet), ensuring --- possibly through robustification --- that biological features dominate the representation space reduces the risk of Clever Hans learning in downstream models. 

\subsection{Summary}
To summarize, the experiment described in Section~\ref{sec:downstream_experiments} highlighted a fundamental issue: non-robust foundation models increase the likelihood of Clever Hans learning and poor generalization, especially in high-stakes clinical tasks. Reducing or eliminating confounding features from FM representation spaces is therefore essential for maximizing robustness and ensuring reliable performance in downstream applications.

\FloatBarrier
\section{Downstream model training} \label{sec:apx_training_details}
\subsection{General training details}

We trained various downstream models from FM representations of (small) labeled training data cohorts to evaluate FMs from different perspectives. This is a common way of assessing foundation models, and, in addition, an effective way of deriving special-purpose models for medical applications (see e.g.\ \cite{chen2020simple, vorontsov2023virchow, chen2024uni}).

As a downstream model, we used linear probing (aka.\ logistic regression), i.e., training a simple neural network head without hidden layers to predict the classes of interest, which is a common choice in the field \cite{chen2020simple, vorontsov2023virchow, chen2024uni}. 

Using the sklearn implementation\footnote{\url{https://scikit-learn.org/stable/modules/generated/sklearn.linear\_model.LogisticRegression.html}}, we selected the optimal regularization parameter $C$ based on validation performance, choosing from 15 logarithmically spaced values in the range $[10^{-8}, 10^4]$. To account for variations due to splitting, we repeated all linear probing trainings 20 times by resampling the training, validation, and test data from the respective PathoROB dataset. Notice that the number and composition of samples in training, validation, and test sets always remained constant across repetitions.

\subsection{Training with spuriously correlated data}

\begin{figure}[h!]
    \centering
    \begin{tikzpicture}[scale=0.5]
        \tiny
        \node[above] at (9.5,2) {Training Sets};
        \node[left] at (0,0.5) {RUMC};
        \node[left] at (0,-0.5) {UMCU};
        \draw (0,1) rectangle (2,0) node[pos=.5] {2100 (7)};
        \draw (2,1) rectangle (4,0) node[pos=.5] {2100 (7)};
        \draw (0,0) rectangle (2,-1) node[pos=.5] {2100 (7)};
        \draw (2,0) rectangle (4,-1) node[pos=.5] {2100 (7)};
        \node[above] at (2,1) {Split 1 ($V=0.00$)};
        
        \draw (5,1) rectangle (7,0) node[pos=.5] {1800 (6)};
        \draw (7,1) rectangle (9,0) node[pos=.5] {2400 (8)};
        \draw (5,0) rectangle (7,-1) node[pos=.5] {2400 (8)};
        \draw (7,0) rectangle (9,-1) node[pos=.5] {1800 (6)};
        \node[above] at (7,1) {Split 2 ($V=0.14$)};
        
        \draw (10,1) rectangle (12,0) node[pos=.5] {1500 (5)};
        \draw (12,1) rectangle (14,0) node[pos=.5] {2700 (9)};
        \draw (10,0) rectangle (12,-1) node[pos=.5] {2700 (9)};
        \draw (12,0) rectangle (14,-1) node[pos=.5] {1500 (5)};
        \node[above] at (12,1) {Split 3 ($V=0.29$)};
        
        \draw (15,1) rectangle (17,0) node[pos=.5] {1200 (4)};
        \draw (17,1) rectangle (19,0) node[pos=.5] {3000 (10)};
        \draw (15,0) rectangle (17,-1) node[pos=.5] {3000 (10)};
        \draw (17,0) rectangle (19,-1) node[pos=.5] {1200 (4)};
        \node[above] at (17,1) {Split 4 ($V=0.43$)};
        
        \node[left] at (0,-2.5) {RUMC};
        \node[left] at (0,-3.5) {UMCU};
        \draw (0,-2) rectangle (2,-3) node[pos=.5] {900 (3)};
        \draw (2,-2) rectangle (4,-3) node[pos=.5] {3300 (11)};
        \draw (0,-3) rectangle (2,-4) node[pos=.5] {3300 (11)};
        \draw (2,-3) rectangle (4,-4) node[pos=.5] {900 (3)};
        \node[above] at (2,-2) {Split 5 ($V=0.57$)};
        \node[below] at (1,-4) {Normal};
        \node[below] at (3,-4) {Tumor};
        
        \draw (5,-2) rectangle (7,-3) node[pos=.5] {600 (2)};
        \draw (7,-2) rectangle (9,-3) node[pos=.5] {3600 (12)};
        \draw (5,-3) rectangle (7,-4) node[pos=.5] {3600 (12)};
        \draw (7,-3) rectangle (9,-4) node[pos=.5] {600 (2)};
        \node[above] at (7,-2) {Split 6 ($V=0.71$)};
        \node[below] at (6,-4) {Normal};
        \node[below] at (8,-4) {Tumor};
        
        \draw (10,-2) rectangle (12,-3) node[pos=.5] {300 (1)};
        \draw (12,-2) rectangle (14,-3) node[pos=.5] {3900 (13)};
        \draw (10,-3) rectangle (12,-4) node[pos=.5] {3900 (13)};
        \draw (12,-3) rectangle (14,-4) node[pos=.5] {300 (1)};
        \node[above] at (12,-2) {Split 7 ($V=0.86$)};
        \node[below] at (11,-4) {Normal};
        \node[below] at (13,-4) {Tumor};

        \draw (15,-2) rectangle (17,-3) node[pos=.5] {};
        \draw (17,-2) rectangle (19,-3) node[pos=.5] {4200 (14)};
        \draw (15,-3) rectangle (17,-4) node[pos=.5] {4200 (14)};
        \draw (17,-3) rectangle (19,-4) node[pos=.5] {};
        \node[above] at (17,-2) {Split 8 ($V=1.00$)};
        \node[below] at (16,-4) {Normal};
        \node[below] at (18,-4) {Tumor};
    \end{tikzpicture}
    
    \vspace{0.05cm}
    
    \begin{tikzpicture}[scale=0.5]
        \tiny
        \node[above] at (2,1.5) {In-Domain Test Set};
        \draw (0,1) rectangle (2,0) node[pos=.5] {600 (2)};
        \draw (2,1) rectangle (4,0) node[pos=.5] {600 (2)};
        \draw (0,0) rectangle (2,-1) node[pos=.5] {600 (2)};
        \draw (2,0) rectangle (4,-1) node[pos=.5] {600 (2)};
        \node[left] at (0,0.5) {RUMC};
        \node[left] at (0,-0.5) {UMCU};
        \node[below] at (1,-1) {Normal};
        \node[below] at (3,-1) {Tumor};
        
        \begin{scope}[xshift=10cm]
            \node[above] at (2,1.5) {Out-of-Domain Test Set};
            \draw (0,1) rectangle (2,0) node[pos=.5] {335 (5)};
            \draw (2,1) rectangle (4,0) node[pos=.5] {327 (5)};
            \draw (0,0) rectangle (2,-1) node[pos=.5] {335 (5)};
            \draw (2,0) rectangle (4,-1) node[pos=.5] {335 (6)};
            \draw (0,-1) rectangle (2,-2) node[pos=.5] {335 (5)};
            \draw (2,-1) rectangle (4,-2) node[pos=.5] {335 (3)};
            \node[left] at (0,0.5) {CWZ};
            \node[left] at (0,-0.5) {RST};
            \node[left] at (0,-1.5) {LPON};
            \node[below] at (1,-2) {Normal};
            \node[below] at (3,-2) {Tumor};
        \end{scope}
    \end{tikzpicture}
    \caption{\textbf{Training and test data compositions, including the number of patches (slides), for the Camelyon downstream task.} For each training split, we report Cramer's V between the medical center and biological class as a measure of correlation between the two variables. Each split's validation set maintains the same correlation and balance as the training set. For each of the four training categories, the validation set is composed of patches coming from one single held-out validation slide, while, in total, rows and columns add up to 300 patches.}
    \label{fig:camelyon_dataset_splits}
\end{figure}

\begin{figure}[t!]
    \centering
    \begin{tikzpicture}[scale=0.5]
        \tiny
        \node[above] at (12.25,6) {Training Sets};

        \node[left] at (0,4.5) {AST};
        \node[left] at (0,3.5) {CH};
        \node[left] at (0,2.5) {RP};
        \node[left] at (0,1.5) {UP};
        \draw (0,5) rectangle (2,4) node[pos=.5] {60 (2)};
        \draw (2,5) rectangle (4,4) node[pos=.5] {60 (2)};
        \draw (4,5) rectangle (6,4) node[pos=.5] {60 (2)};
        \draw (6,5) rectangle (8,4) node[pos=.5] {60 (2)};

        \draw (0,4) rectangle (2,3) node[pos=.5] {60 (2)};
        \draw (2,4) rectangle (4,3) node[pos=.5] {60 (2)};
        \draw (4,4) rectangle (6,3) node[pos=.5] {60 (2)};
        \draw (6,4) rectangle (8,3) node[pos=.5] {60 (2)};

        \draw (0,3) rectangle (2,2) node[pos=.5] {60 (2)};
        \draw (2,3) rectangle (4,2) node[pos=.5] {60 (2)};
        \draw (4,3) rectangle (6,2) node[pos=.5] {60 (2)};
        \draw (6,3) rectangle (8,2) node[pos=.5] {60 (2)};

        \draw (0,2) rectangle (2,1) node[pos=.5] {60 (2)};
        \draw (2,2) rectangle (4,1) node[pos=.5] {60 (2)};
        \draw (4,2) rectangle (6,1) node[pos=.5] {60 (2)};
        \draw (6,2) rectangle (8,1) node[pos=.5] {60 (2)};

        \node[above] at (4,5) {Split 1 ($V=0.00$)};

        \begin{scope}[xshift=8.25cm]
            \draw (0,5) rectangle (2,4) node[pos=.5] {90 (3)};
            \draw (2,5) rectangle (4,4) node[pos=.5] {60 (2)};
            \draw (4,5) rectangle (6,4) node[pos=.5] {60 (2)};
            \draw (6,5) rectangle (8,4) node[pos=.5] {30 (1)};
    
            \draw (0,4) rectangle (2,3) node[pos=.5] {60 (2)};
            \draw (2,4) rectangle (4,3) node[pos=.5] {90 (3)};
            \draw (4,4) rectangle (6,3) node[pos=.5] {30 (1)};
            \draw (6,4) rectangle (8,3) node[pos=.5] {60 (2)};

            \draw (0,3) rectangle (2,2) node[pos=.5] {60 (2)};
            \draw (2,3) rectangle (4,2) node[pos=.5] {30 (1)};
            \draw (4,3) rectangle (6,2) node[pos=.5] {90 (3)};
            \draw (6,3) rectangle (8,2) node[pos=.5] {60 (2)};

            \draw (0,2) rectangle (2,1) node[pos=.5] {30 (1)};
            \draw (2,2) rectangle (4,1) node[pos=.5] {60 (2)};
            \draw (4,2) rectangle (6,1) node[pos=.5] {60 (2)};
            \draw (6,2) rectangle (8,1) node[pos=.5] {90 (3)};
    
            \node[above] at (4,5) {Split 2 ($V=0.20$)};
        \end{scope}

        \begin{scope}[xshift=16.5cm]
            \draw (0,5) rectangle (2,4) node[pos=.5] {120 (4)};
            \draw (2,5) rectangle (4,4) node[pos=.5] {30 (1)};
            \draw (4,5) rectangle (6,4) node[pos=.5] {30 (1)};
            \draw (6,5) rectangle (8,4) node[pos=.5] {60 (2)};
    
            \draw (0,4) rectangle (2,3) node[pos=.5] {30 (1)};
            \draw (2,4) rectangle (4,3) node[pos=.5] {120 (4)};
            \draw (4,4) rectangle (6,3) node[pos=.5] {60 (2)};
            \draw (6,4) rectangle (8,3) node[pos=.5] {30 (1)};

            \draw (0,3) rectangle (2,2) node[pos=.5] {30 (1)};
            \draw (2,3) rectangle (4,2) node[pos=.5] {60 (2)};
            \draw (4,3) rectangle (6,2) node[pos=.5] {120 (4)};
            \draw (6,3) rectangle (8,2) node[pos=.5] {30 (1)};

            \draw (0,2) rectangle (2,1) node[pos=.5] {60 (2)};
            \draw (2,2) rectangle (4,1) node[pos=.5] {30 (1)};
            \draw (4,2) rectangle (6,1) node[pos=.5] {30 (1)};
            \draw (6,2) rectangle (8,1) node[pos=.5] {120 (4)};
    
            \node[above] at (4,5) {Split 3 ($V=0.35$)};
        \end{scope}

        \begin{scope}[yshift=-5cm]
            \node[left] at (0,4.5) {AST};
            \node[left] at (0,3.5) {CH};
            \node[left] at (0,2.5) {RP};
            \node[left] at (0,1.5) {UP};
            \draw (0,5) rectangle (2,4) node[pos=.5] {150 (5)};
            \draw (2,5) rectangle (4,4) node[pos=.5] {30 (1)};
            \draw (4,5) rectangle (6,4) node[pos=.5] {30 (1)};
            \draw (6,5) rectangle (8,4) node[pos=.5] {30 (1)};
    
            \draw (0,4) rectangle (2,3) node[pos=.5] {30 (1)};
            \draw (2,4) rectangle (4,3) node[pos=.5] {150 (5)};
            \draw (4,4) rectangle (6,3) node[pos=.5] {30 (1)};
            \draw (6,4) rectangle (8,3) node[pos=.5] {30 (1)};

            \draw (0,3) rectangle (2,2) node[pos=.5] {30 (1)};
            \draw (2,3) rectangle (4,2) node[pos=.5] {30 (1)};
            \draw (4,3) rectangle (6,2) node[pos=.5] {150 (5)};
            \draw (6,3) rectangle (8,2) node[pos=.5] {30 (1)};

            \draw (0,2) rectangle (2,1) node[pos=.5] {30 (1)};
            \draw (2,2) rectangle (4,1) node[pos=.5] {30 (1)};
            \draw (4,2) rectangle (6,1) node[pos=.5] {30 (1)};
            \draw (6,2) rectangle (8,1) node[pos=.5] {150 (5)};
    
            \node[above] at (4,5) {Split 4 ($V=0.50$)};
        \end{scope}

        \begin{scope}[xshift=8.25cm, yshift=-5cm]
            \draw (0,5) rectangle (2,4) node[pos=.5] {180 (6)};
            \draw (2,5) rectangle (4,4) node[pos=.5] {30 (1)};
            \draw (4,5) rectangle (6,4) node[pos=.5] {30 (1)};
            \draw (6,5) rectangle (8,4) node[pos=.5] {};
    
            \draw (0,4) rectangle (2,3) node[pos=.5] {30 (1)};
            \draw (2,4) rectangle (4,3) node[pos=.5] {180 (6)};
            \draw (4,4) rectangle (6,3) node[pos=.5] {};
            \draw (6,4) rectangle (8,3) node[pos=.5] {30 (1)};

            \draw (0,3) rectangle (2,2) node[pos=.5] {30 (1)};
            \draw (2,3) rectangle (4,2) node[pos=.5] {};
            \draw (4,3) rectangle (6,2) node[pos=.5] {180 (6)};
            \draw (6,3) rectangle (8,2) node[pos=.5] {30 (1)};

            \draw (0,2) rectangle (2,1) node[pos=.5] {};
            \draw (2,2) rectangle (4,1) node[pos=.5] {30 (1)};
            \draw (4,2) rectangle (6,1) node[pos=.5] {30 (1)};
            \draw (6,2) rectangle (8,1) node[pos=.5] {180 (6)};

            \node[below] at (1,1) {BRCA};
            \node[below] at (3,1) {COAD};
            \node[below] at (5,1) {LUAD};
            \node[below] at (7,1) {LUSC};
    
            \node[above] at (4,5) {Split 5 ($V=0.68$)};
        \end{scope}

        \begin{scope}[xshift=16.5cm, yshift=-5cm]
            \draw (0,5) rectangle (2,4) node[pos=.5] {210 (7)};
            \draw (2,5) rectangle (4,4) node[pos=.5] {};
            \draw (4,5) rectangle (6,4) node[pos=.5] {};
            \draw (6,5) rectangle (8,4) node[pos=.5] {30 (1)};
    
            \draw (0,4) rectangle (2,3) node[pos=.5] {};
            \draw (2,4) rectangle (4,3) node[pos=.5] {210 (7)};
            \draw (4,4) rectangle (6,3) node[pos=.5] {30 (1)};
            \draw (6,4) rectangle (8,3) node[pos=.5] {};

            \draw (0,3) rectangle (2,2) node[pos=.5] {};
            \draw (2,3) rectangle (4,2) node[pos=.5] {30 (1)};
            \draw (4,3) rectangle (6,2) node[pos=.5] {210 (7)};
            \draw (6,3) rectangle (8,2) node[pos=.5] {};

            \draw (0,2) rectangle (2,1) node[pos=.5] {30 (1)};
            \draw (2,2) rectangle (4,1) node[pos=.5] {};
            \draw (4,2) rectangle (6,1) node[pos=.5] {};
            \draw (6,2) rectangle (8,1) node[pos=.5] {210 (7)};

            \node[below] at (1,1) {BRCA};
            \node[below] at (3,1) {COAD};
            \node[below] at (5,1) {LUAD};
            \node[below] at (7,1) {LUSC};
    
            \node[above] at (4,5) {Split 6 ($V=0.84$)};
        \end{scope}

        \begin{scope}[yshift=-10cm]
            \node[left] at (0,4.5) {AST};
            \node[left] at (0,3.5) {CH};
            \node[left] at (0,2.5) {RP};
            \node[left] at (0,1.5) {UP};
            \draw (0,5) rectangle (2,4) node[pos=.5] {240 (8)};
            \draw (2,5) rectangle (4,4) node[pos=.5] {};
            \draw (4,5) rectangle (6,4) node[pos=.5] {};
            \draw (6,5) rectangle (8,4) node[pos=.5] {};
    
            \draw (0,4) rectangle (2,3) node[pos=.5] {};
            \draw (2,4) rectangle (4,3) node[pos=.5] {240 (8)};
            \draw (4,4) rectangle (6,3) node[pos=.5] {};
            \draw (6,4) rectangle (8,3) node[pos=.5] {};

            \draw (0,3) rectangle (2,2) node[pos=.5] {};
            \draw (2,3) rectangle (4,2) node[pos=.5] {};
            \draw (4,3) rectangle (6,2) node[pos=.5] {240 (8)};
            \draw (6,3) rectangle (8,2) node[pos=.5] {};

            \draw (0,2) rectangle (2,1) node[pos=.5] {};
            \draw (2,2) rectangle (4,1) node[pos=.5] {};
            \draw (4,2) rectangle (6,1) node[pos=.5] {};
            \draw (6,2) rectangle (8,1) node[pos=.5] {240 (8)};

            \node[below] at (1,1) {BRCA};
            \node[below] at (3,1) {COAD};
            \node[below] at (5,1) {LUAD};
            \node[below] at (7,1) {LUSC};
    
            \node[above] at (4,5) {Split 7 ($V=1.00$)};
        \end{scope}
    \end{tikzpicture}

    \vspace{0.05cm}

    \begin{tikzpicture}[scale=0.5]
        \tiny
        \node[above] at (4,5.5) {In-Domain Test Set};
        \node[left] at (0,4.5) {AST};
        \node[left] at (0,3.5) {CH};
        \node[left] at (0,2.5) {RP};
        \node[left] at (0,1.5) {UP};
        \draw (0,5) rectangle (2,4) node[pos=.5] {90 (3)};
        \draw (2,5) rectangle (4,4) node[pos=.5] {90 (3)};
        \draw (4,5) rectangle (6,4) node[pos=.5] {90 (3)};
        \draw (6,5) rectangle (8,4) node[pos=.5] {90 (3)};

        \draw (0,4) rectangle (2,3) node[pos=.5] {90 (3)};
        \draw (2,4) rectangle (4,3) node[pos=.5] {90 (3)};
        \draw (4,4) rectangle (6,3) node[pos=.5] {90 (3)};
        \draw (6,4) rectangle (8,3) node[pos=.5] {90 (3)};

        \draw (0,3) rectangle (2,2) node[pos=.5] {90 (3)};
        \draw (2,3) rectangle (4,2) node[pos=.5] {90 (3)};
        \draw (4,3) rectangle (6,2) node[pos=.5] {90 (3)};
        \draw (6,3) rectangle (8,2) node[pos=.5] {90 (3)};

        \draw (0,2) rectangle (2,1) node[pos=.5] {90 (3)};
        \draw (2,2) rectangle (4,1) node[pos=.5] {90 (3)};
        \draw (4,2) rectangle (6,1) node[pos=.5] {90 (3)};
        \draw (6,2) rectangle (8,1) node[pos=.5] {90 (3)};

        \node[below] at (1,1) {BRCA};
        \node[below] at (3,1) {COAD};
        \node[below] at (5,1) {LUAD};
        \node[below] at (7,1) {LUSC};

        \begin{scope}[xshift=12cm]
            \node[above] at (4,5.5) {Out-of-Domain Test Set};
            \node[left] at (0,4.5) {CU};
            \node[left] at (0,3.5) {GPCC};
            \node[left] at (0,2.5) {IGC};
            \node[left] at (0,1.5) {JH};
            \draw (0,5) rectangle (2,4) node[pos=.5] {300 (10)};
            \draw (2,5) rectangle (4,4) node[pos=.5] {};
            \draw (4,5) rectangle (6,4) node[pos=.5] {};
            \draw (6,5) rectangle (8,4) node[pos=.5] {300 (10)};
    
            \draw (0,4) rectangle (2,3) node[pos=.5] {300 (10)};
            \draw (2,4) rectangle (4,3) node[pos=.5] {300 (10)};
            \draw (4,4) rectangle (6,3) node[pos=.5] {};
            \draw (6,4) rectangle (8,3) node[pos=.5] {};

            \draw (0,3) rectangle (2,2) node[pos=.5] {};
            \draw (2,3) rectangle (4,2) node[pos=.5] {300 (10)};
            \draw (4,3) rectangle (6,2) node[pos=.5] {300 (10)};
            \draw (6,3) rectangle (8,2) node[pos=.5] {};

            \draw (0,2) rectangle (2,1) node[pos=.5] {};
            \draw (2,2) rectangle (4,1) node[pos=.5] {};
            \draw (4,2) rectangle (6,1) node[pos=.5] {300 (10)};
            \draw (6,2) rectangle (8,1) node[pos=.5] {300 (10)};

            \node[below] at (1,1) {BRCA};
            \node[below] at (3,1) {COAD};
            \node[below] at (5,1) {LUAD};
            \node[below] at (7,1) {LUSC};
        \end{scope}
    \end{tikzpicture}

    \caption{\textbf{Training and test data compositions, including the number of patches (slides), for the TCGA downstream task based on the TCGA~4x4 dataset.} For each training split, we report Cramer's V between the medical center and biological class as a measure of correlation between the two variables. Each split's validation set maintains the same correlation and balance as the training set. For each of the 16 training categories, the validation set is composed of patches coming from one single held-out validation slide, while, in total, rows and columns add up to 30 patches.}
    \label{fig:tcga_dataset_splits}
\end{figure}

\begin{figure}[t!]
    \centering
    \begin{tikzpicture}[scale=0.5]
        \tiny
        \node[above] at (12.25,4) {Training Sets};

        \node[left] at (0,2.5) {WNS};
        \node[left] at (0,1.5) {CHA};
        \draw (0,3) rectangle (2,2) node[pos=.5] {300};
        \draw (2,3) rectangle (4,2) node[pos=.5] {300};
        \draw (4,3) rectangle (6,2) node[pos=.5] {300};
        \draw (6,3) rectangle (8,2) node[pos=.5] {300};
        \draw (8,3) rectangle (10,2) node[pos=.5] {300};
        \draw (10,3) rectangle (12,2) node[pos=.5] {300};

        \draw (0,2) rectangle (2,1) node[pos=.5] {300};
        \draw (2,2) rectangle (4,1) node[pos=.5] {300};
        \draw (4,2) rectangle (6,1) node[pos=.5] {300};
        \draw (6,2) rectangle (8,1) node[pos=.5] {300};
        \draw (8,2) rectangle (10,1) node[pos=.5] {300};
        \draw (10,2) rectangle (12,1) node[pos=.5] {300};

        \node[above] at (6,3) {Split 1 ($V=0.00$)};

        \begin{scope}[xshift=12.5cm]
            \draw (0,3) rectangle (2,2) node[pos=.5] {200};
            \draw (2,3) rectangle (4,2) node[pos=.5] {200};
            \draw (4,3) rectangle (6,2) node[pos=.5] {200};
            \draw (6,3) rectangle (8,2) node[pos=.5] {400};
            \draw (8,3) rectangle (10,2) node[pos=.5] {400};
            \draw (10,3) rectangle (12,2) node[pos=.5] {400};

            \draw (0,2) rectangle (2,1) node[pos=.5] {400};
            \draw (2,2) rectangle (4,1) node[pos=.5] {400};
            \draw (4,2) rectangle (6,1) node[pos=.5] {400};
            \draw (6,2) rectangle (8,1) node[pos=.5] {200};
            \draw (8,2) rectangle (10,1) node[pos=.5] {200};
            \draw (10,2) rectangle (12,1) node[pos=.5] {200};
            \node[above] at (6,3) {Split 2 ($V=0.33$)};
        \end{scope}

        \begin{scope}[yshift=-3cm]
            \node[left] at (0,2.5) {WNS};
            \node[left] at (0,1.5) {CHA};
            \draw (0,3) rectangle (2,2) node[pos=.5] {100};
            \draw (2,3) rectangle (4,2) node[pos=.5] {100};
            \draw (4,3) rectangle (6,2) node[pos=.5] {100};
            \draw (6,3) rectangle (8,2) node[pos=.5] {500};
            \draw (8,3) rectangle (10,2) node[pos=.5] {500};
            \draw (10,3) rectangle (12,2) node[pos=.5] {500};

            \draw (0,2) rectangle (2,1) node[pos=.5] {500};
            \draw (2,2) rectangle (4,1) node[pos=.5] {500};
            \draw (4,2) rectangle (6,1) node[pos=.5] {500};
            \draw (6,2) rectangle (8,1) node[pos=.5] {100};
            \draw (8,2) rectangle (10,1) node[pos=.5] {100};
            \draw (10,2) rectangle (12,1) node[pos=.5] {100};
            \node[below] at (1,1) {TU};
            \node[below] at (3,1) {MP};
            \node[below] at (5,1) {SO};
            \node[below] at (7,1) {SM};
            \node[below] at (9,1) {RT};
            \node[below] at (11,1) {AD};
            \node[above] at (6,3) {Split 3 ($V=0.67$)};
        \end{scope}

        \begin{scope}[yshift=-3cm, xshift=12.5cm]
            \draw (0,3) rectangle (2,2) node[pos=.5] {};
            \draw (2,3) rectangle (4,2) node[pos=.5] {};
            \draw (4,3) rectangle (6,2) node[pos=.5] {};
            \draw (6,3) rectangle (8,2) node[pos=.5] {600};
            \draw (8,3) rectangle (10,2) node[pos=.5] {600};
            \draw (10,3) rectangle (12,2) node[pos=.5] {600};

            \draw (0,2) rectangle (2,1) node[pos=.5] {600};
            \draw (2,2) rectangle (4,1) node[pos=.5] {600};
            \draw (4,2) rectangle (6,1) node[pos=.5] {600};
            \draw (6,2) rectangle (8,1) node[pos=.5] {};
            \draw (8,2) rectangle (10,1) node[pos=.5] {};
            \draw (10,2) rectangle (12,1) node[pos=.5] {};
            \node[below] at (1,1) {TU};
            \node[below] at (3,1) {MP};
            \node[below] at (5,1) {SO};
            \node[below] at (7,1) {SM};
            \node[below] at (9,1) {RT};
            \node[below] at (11,1) {AD};
            \node[above] at (6,3) {Split 4 ($V=1.00$)};
        \end{scope}
    \end{tikzpicture}

    \vspace{0.05cm}

    \begin{tikzpicture}[scale=0.5]
        \tiny
        \node[above] at (6,3.5) {In-Domain Test Set};
        \node[left] at (0,2.5) {WNS};
        \node[left] at (0,1.5) {CHA};
        \draw (0,3) rectangle (2,2) node[pos=.5] {200};
        \draw (2,3) rectangle (4,2) node[pos=.5] {200};
        \draw (4,3) rectangle (6,2) node[pos=.5] {200};
        \draw (6,3) rectangle (8,2) node[pos=.5] {200};
        \draw (8,3) rectangle (10,2) node[pos=.5] {200};
        \draw (10,3) rectangle (12,2) node[pos=.5] {200};

        \draw (0,2) rectangle (2,1) node[pos=.5] {200};
        \draw (2,2) rectangle (4,1) node[pos=.5] {200};
        \draw (4,2) rectangle (6,1) node[pos=.5] {200};
        \draw (6,2) rectangle (8,1) node[pos=.5] {200};
        \draw (8,2) rectangle (10,1) node[pos=.5] {200};
        \draw (10,2) rectangle (12,1) node[pos=.5] {200};

        \begin{scope}[yshift=-3.5cm]
            \node[above] at (6,3.5) {Out-of-Domain Test Set};
            \node[left] at (0,2.5) {UKK};
            \node[left] at (0,1.5) {TCGA};
            \draw (0,3) rectangle (2,2) node[pos=.5] {500};
            \draw (2,3) rectangle (4,2) node[pos=.5] {500};
            \draw (4,3) rectangle (6,2) node[pos=.5] {500};
            \draw (6,3) rectangle (8,2) node[pos=.5] {500};
            \draw (8,3) rectangle (10,2) node[pos=.5] {500};
            \draw (10,3) rectangle (12,2) node[pos=.5] {500};

            \draw (0,2) rectangle (2,1) node[pos=.5] {500};
            \draw (2,2) rectangle (4,1) node[pos=.5] {500};
            \draw (4,2) rectangle (6,1) node[pos=.5] {500};
            \draw (6,2) rectangle (8,1) node[pos=.5] {500};
            \draw (8,2) rectangle (10,1) node[pos=.5] {};
            \draw (10,2) rectangle (12,1) node[pos=.5] {500};
            \node[below] at (1,1) {TU};
            \node[below] at (3,1) {MP};
            \node[below] at (5,1) {SO};
            \node[below] at (7,1) {SM};
            \node[below] at (9,1) {RT};
            \node[below] at (11,1) {AD};
        \end{scope}
    \end{tikzpicture}

    \caption{\textbf{Training and test data compositions, including the number of patches, for the Tolkach ESCA downstream task.} For each training split, we report Cramer's V between the medical center and biological class as a measure of correlation between the two variables. Each split's validation set maintains the same correlation and balance as the training set. For each of the 12 training categories, the validation set is composed of held-out validation patches, while, in total, the columns add up to 100 patches.}
    \label{fig:tolkach_dataset_splits}
\end{figure}

For assessing Clever Hans learning in supervised downstream models, we trained linear probing models on training data with increasing spurious correlations between the medical center and the biological class (see Section~\ref{sec:downstream_experiments}, Section~\ref{sec:methods_downstream}). The detailed statistics of the training, validation, and test splits are depicted in Figures~\ref{fig:camelyon_dataset_splits},~\ref{fig:tcga_dataset_splits},~\ref{fig:tolkach_dataset_splits}. As a validation set, we took a small amount of data, maintaining the same correlation as the training set -- further specified in the captions of Figures~\ref{fig:camelyon_dataset_splits},~\ref{fig:tcga_dataset_splits},~\ref{fig:tolkach_dataset_splits}.

To ensure a fair estimation of the generalization performance drop when increasing correlation, we always used the same data points across training splits (within a single repetition). That is, if a patch was present in one training split, it was also present in every other training split, with at least the same amount of data for the respective class-center combination. To estimate the stability of the experiment across repetitions, we repeated the split sampling procedure 20 times. That is, we randomly filled the numbers depicted in (Figures~\ref{fig:camelyon_dataset_splits},~\ref{fig:tcga_dataset_splits},~\ref{fig:tolkach_dataset_splits}) with different patches sampled from the PathoROB datasets.

\subsection{Domain-adversarial training}

We assessed domain-adversarial neural network (DANN)~\cite{DBLP:journals/jmlr/GaninUAGLLML16} training as a method for robustifying FM representation spaces.
Recall from the main text that the DANN objective can be defined through two competing loss parts:
\begin{align*}
    \mathcal{L}_{\mathsf{DANN}}(\hat{y},\hat{c}; y,c) &= \mathcal{L}_{\mathsf{CL}}(\hat{y},y) + \lambda \cdot \mathcal{L}_{\mathsf{DA}}(\hat{c},c) \, ,
\end{align*}
where the domain discriminator part $\mathcal{L}_{\mathsf{DA}}$ makes use of a \textit{gradient reversal} layer~\cite{DBLP:journals/jmlr/GaninUAGLLML16}.
While originally developed in the context of domain adaptation (cf.~\cite{DBLP:journals/jmlr/GaninUAGLLML16}), we revisit it here for post-hoc removing medical center signatures from FMs. The technical difference from the latter is that we do not consider unlabeled domains; instead, we assume complete access to the biological class and confounding information for all domains. This means our goal is not to propagate label information to an unsupervised domain to perform classification, but instead we aim to learn a feature space that is free from non-biological features. To summarize, our approach consists of two steps: We first try to learn an unbiased feature representation of the original training data using DANN.
Then, we trained a new linear probe (identical to the one used in the experiments without mitigation) on the training data embedded in the new feature space.

For the DANN architecture, we found that a one-layer feature extractor with ReLU, which reduces the original feature space by half, combined with a linear classification head and a shallow non-linear domain discriminator, results in the best overall model performances.
The idea of halving the original feature space is to compress the downstream accessible information, hoping to discard confounding signals in favor of biological signals.
We trained DANN for 20 epochs and chose the loss-balancing weight $\lambda$ to increase linearly from zero to approach one over the epochs (i.e.\ in each epoch we slightly increase the impact of the domain discriminator loss while the weight for the classification loss term remains the same).

\FloatBarrier
\section{Clustering analysis}
\label{sec:apx_clustering_score_details}

\subsection{Clustering score}

The clustering score evaluates the robustness of the global organization within the representation space. Unlike the robustness index, which assesses the composition of each embedding’s nearest neighbors and, thus, provides a local perspective, the clustering score assesses the composition of entire groups of embeddings, offering a global view. Similar to the robustness index, it quantifies the extent to which the spatial arrangement of the representations is influenced by biologically meaningful factors as opposed to confounding variables.

To examine the global structure, we performed unsupervised $K$-means clustering using cosine distances. The number of clusters $K$ was not predefined but selected by maximizing the silhouette score \cite{rousseeuw1987silhouettes}. The silhouette score quantifies cluster cohesion (within cluster distances) compared to cluster separation (distances to closest, distinct cluster), thereby serving as a proxy for the cluster quality in the absence of ground-truth labels. 
Let $\hat{\mathcal{C}} = \lbrace \hat{C}_1, \ldots, \hat{C}_K\rbrace$ denote the predicted partition (clustering) of a data set $\mathcal{D}$ such that $\bigcup_{k} \hat{C}_k = \mathcal{D}$, and $d(i,j)$ is the distance between samples $i \in \hat{C}_I$ and $j \in \hat{C}_J$, then
\begin{align*}
    \text{silhouette score}(\hat{\mathcal{C}}) = \frac{1}{\vert\mathcal{D}\vert} \sum_{i\in \mathcal{D}} s(i),& \quad 
    \overbrace{s(i) = \frac{a(i)-b(i)}{\max{ \lbrace a(i), b(i) \rbrace }}}^{\text{silhouette of sample }i}, \\
    \underbrace{a(i) = \frac{1}{\vert \hat{C}_I \vert -1}\sum_{i\neq i' \in \hat{C}_I} d(i,i')}_{\text{average inter-cluster distances}},& \quad 
    \underbrace{b(i) = \min_{\hat{C}_I\neq \hat{C}_J\subseteq \hat{\mathcal{C}}}{\frac{1}{\vert \hat{C}_J \vert}\sum_{j \in \hat{C}_J}d(i,j)}}_{\text{average intra-cluster distances}}.
\end{align*}
This fully unsupervised setup deliberately reflects realistic conditions in clinical knowledge discovery and novelty detection, where prior cluster information is typically unavailable. See Sup.~Note~\ref{sec:apx_upperbound_cs} for an ablation study that eliminates the effect of the silhouette score-based selection of $K$ on the clustering performance.

The quality of the clustering is evaluated by comparing the predicted cluster assignments $\hat{\mathcal{C}}$ to the known biological $\mathcal{C}_{\text{bio}}$ and confounding labels $\mathcal{C}_{\text{mc}}$ using the adjusted Rand index (ARI)~\cite{hubert1985comparing}. The ARI measures the agreement between two clusterings mainly by the number of pairs of samples correctly clustered together or apart, accounting for chance. It is symmetric, invariant to label permutations, and bounded in $[-0.5,1]$, with $1$ indicating perfect agreement, $0$ random labeling, and negative values suggesting poor, independent alignment:

\begin{equation*}
    \text{ARI}(\hat{\mathcal{C}}, \mathcal{C}) = \frac{\text{RI}-\mathbb{E}(\text{RI})}{\max(\text{RI})-\mathbb{E}(\text{RI})}, \quad \underbrace{\text{RI} = \frac{n+m}{\vert \mathcal{D}\vert (\vert \mathcal{D}\vert-1)/2 }}_{\text{Rand index}}
\end{equation*}
\begin{align*}
    n : &\text{ number of pairs } (i,j) \text{ that are in the same cluster in } \hat{\mathcal{C}} \\
    &\text{ and in the same cluster in } \mathcal{C}\\
    m : &\text{ number of pairs } (i,j) \text{ that are in different clusters in } \hat{\mathcal{C}} \\
    &\text{ and in different clusters in } \mathcal{C}\\
    \vert \mathcal{D}\vert (\vert \mathcal{D}\vert-1)/2 : &\text{ total number of pairs } (i,j)\\
 \mathbb{E}(\text{RI}) : &\text{ expectation under random cluster assignments}.
\end{align*}
To evaluate the influence of confounding factors, we compute the ARI twice: once comparing the predicted clusters to the biological class labels (e.g., diagnosis, tissue type), and once to the confounding labels (e.g., medical center origin). The clustering score is then defined as the difference between these two ARI values
\begin{equation*}
    \text{clustering score} = \overbrace{\text{ARI}(\hat{\mathcal{C}}, \mathcal{C}_{\text{bio}})}^{\text{higher is better}} - \overbrace{\text{ARI}(\hat{\mathcal{C}}, \mathcal{C}_{\text{mc}})}^{\text{lower is better}}
\end{equation*}
The clustering score inherits ARI’s desirable properties --- chance adjustment, symmetry, and boundedness (in $[-1.5, 1.5]$).
It ranges from approximately $[-1,1]$, 
where values near zero suggest the clustering is influenced by both biological and confounding information (or neither), positive values indicate clustering aligned with medically relevant features, and negative values reflect confounding-driven structure.

\subsection{Implementation details}

To ensure comparability between $\text{ARI}(\hat{\mathcal{C}}, \mathcal{C}_{\text{bio}})$ and $\text{ARI}(\hat{\mathcal{C}}, \mathcal{C}_{\text{mc}})$, an equal number of biological and confounding classes was selected, with a balanced number of representations per class. 
Specifically, the clustering experiments were conducted on the paired settings Camelyon~2x2, Tolkach~2x2, and all 2x2 class pairings from TCGA~4x4 (for reducing the complexity). 
Prior to clustering, all embeddings were normalized to unit $\ell_2$-norm, so that clustering with Euclidean distances effectively corresponded to clustering based on cosine distances. 
We chose the optimal number of clusters $K$ among $[2,30]$ by maximizing the silhouette score and finally quantifying the quality of the resulting $K$-means clustering by the proposed clustering score. 

Because $K$-means clustering is sensitive to the random cluster initialization, the best solution among consecutive random runs is generally selected based on the lowest inertia. For finding $K$, the clustering is repeated for $20$ and the final clustering only for $5$ random initializations. Additionally, the estimation of an average clustering score and its standard deviation was enabled by repeating the final clustering for $50$ trials (with each $5$ initializations) with different random seeds. Thereby, the dependency of the $K$-means clustering on its random cluster initialization is diminished, which is reflected by very low standard deviations of the clustering score.

\FloatBarrier
\section{Analysis of the FM representation space} \label{sec:apx_feature_space_analysis}

\subsection{FM representation space visualizations} \label{sec:apx_feature_space_visus}

\begin{figure}[t!]
    \centering    \includegraphics[width=0.9\linewidth]{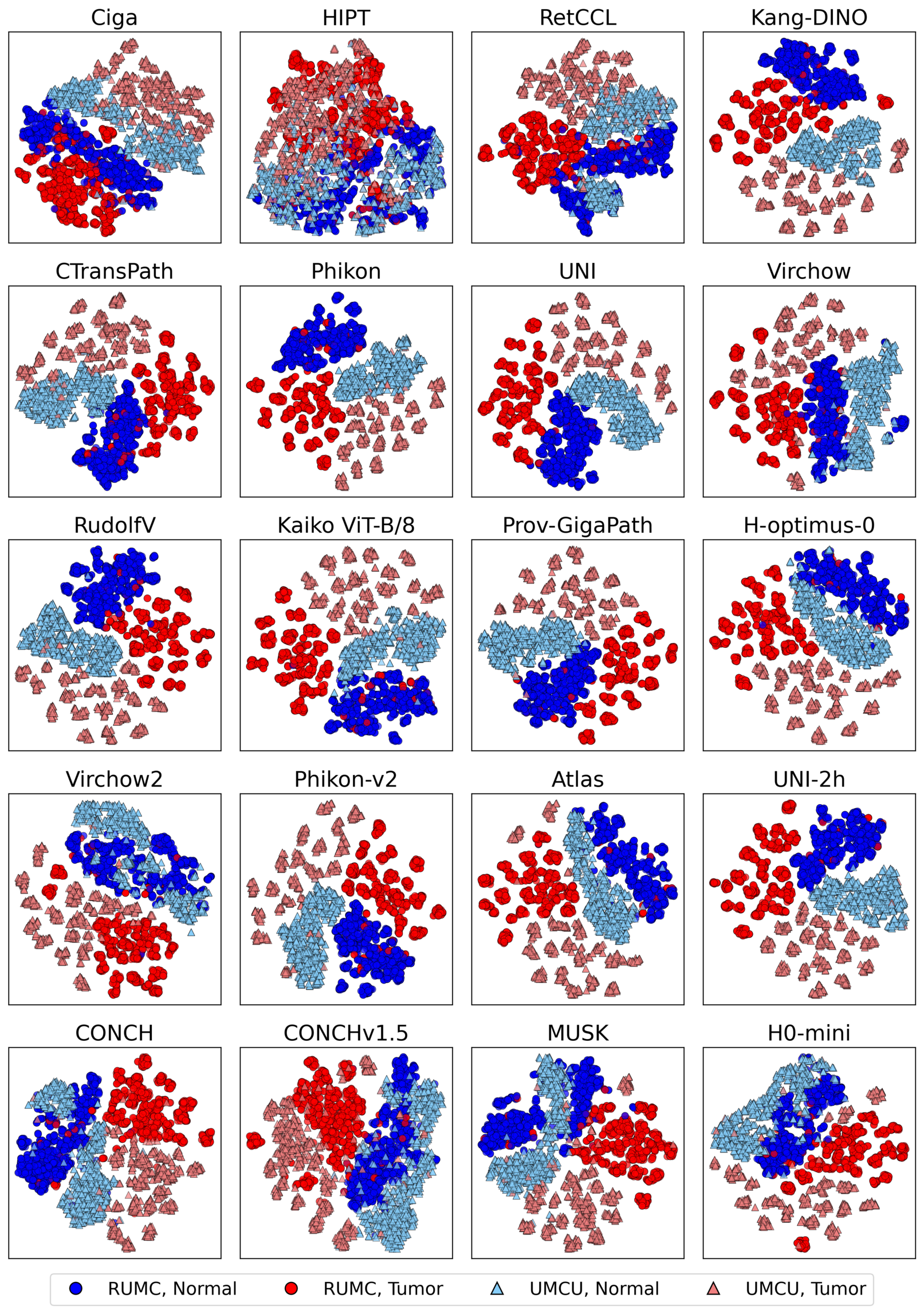}
    \caption{\textbf{t-SNE visualization of the representation spaces per FM on Camelyon.} Groups of embeddings are separated based on their medical center origin (dark circles / bright triangles) and tumor status (blue / red).}
    \label{fig:tsne}
\end{figure}

\begin{figure}[t!]
    \centering    \includegraphics[width=0.9\linewidth]{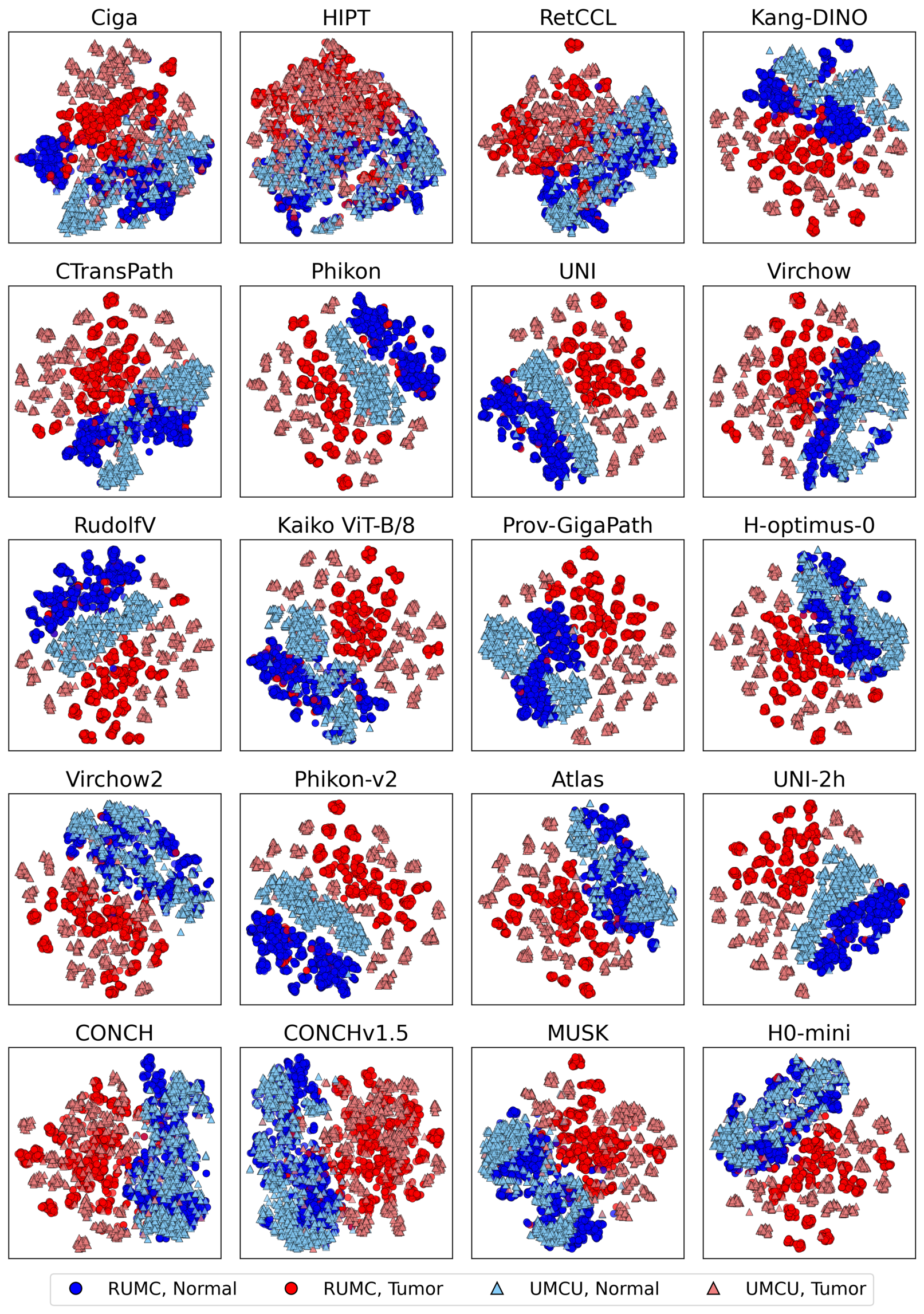}
    \caption{\textbf{t-SNE visualization of the representation spaces after robustification by ComBat (RR) per FM on Camelyon.} Groups of embeddings are mostly separated based on their tumor status (blue / red) while overlapping by their medical center origin (dark circles / bright triangles).}
    \label{fig:tsne_combat}
\end{figure}

Figure~\ref{fig:tsne} shows additional t-SNE plots computed on Camelyon representation vectors of all foundation models. The underlying patches were sampled from the RUMC and UMCU medical centers. For the computation of the t-SNE, we chose the perplexity as the optimal $k$ value derived during the robustness index calculation (see Sup.~Note\ref{sec:apx_robustness_index_computation}). Notably, in older and less robust models (e.g. \mciga{}, \mretccl{}), tumor and normal patches tend to cluster more closely within individual medical centers. In contrast, newer and more robust models like \mhmini{} and \mconchONEFIVE{} show more overlap of centers within the same biological subclass. However, most models exhibit a recognizable clustering pattern corresponding to the four cancer/normal $\times$ RUMC/UMCU subclasses, suggesting that the learned feature spaces are at least partially structured by medical center information at some hierarchical level.

In comparison, Figure~\ref{fig:tsne_combat} visualizes the embedding spaces after robustification using ComBat batch correction (RR) \cite{johnson2007combat, behdenna2023pycombat}. It appears that robustification successfully leads to stronger overlap between embeddings of the same morphology, especially if there was already correct organization based on the tumor status prior to robustification, e.g., see \mvirchowTWO{}, \mconch{} variants, \mhmini{}. In some cases, robustification altered the organization such that it now reflects the biology rather than the medical center origin, e.g., \mretccl{}, \muni{} variants, \mciga{}, \mctranspath{}, \mkaiko{}, and so forth. However, ComBat also caused stronger overlaps between groups of embeddings of different medical status, e.g., the \mconch{} variants, \mrudolfv{}, \mkaiko{}, \mretccl{}.

While t-SNE captures and visualizes local neighborhood structures, longer distances are not properly reflected. Due to the small choice of the perplexity, an embedding is only represented in terms of its closest neighbors, most likely corresponding to patches of the same slide. Relations between sub-groups, e.g., different WSIs, are not properly represented. Explaining the calculated robustness indices or clustering scores in terms of the t-SNE representations is limited. Yet, some trends can be observed. Overlaps between normal and tumorous embeddings have a negative impact on the robustness index and the clustering score. Non-isotropic clusters, e.g., a banana-shaped group of tumors around the normal embeddings for \muni{}, violate the assumption of spherical-shaped, isotropic clusters made by the $K$-means clustering.

\subsection{Analysis of FM representation spaces with PCA}
\label{sec:apx_feature-space-analysis}


To better understand \textit{how} medical center and biologically relevant information are encoded in the representation space, we performed multiple analyses based on Principal Component Analysis (PCA). PCA reduces the embeddings to linear directions of greatest variance (principal components, PCs), i.e., to those directions in which the embeddings vary the most.

\begin{figure}[t!]
    \centering
    \includegraphics[width=0.52\linewidth]{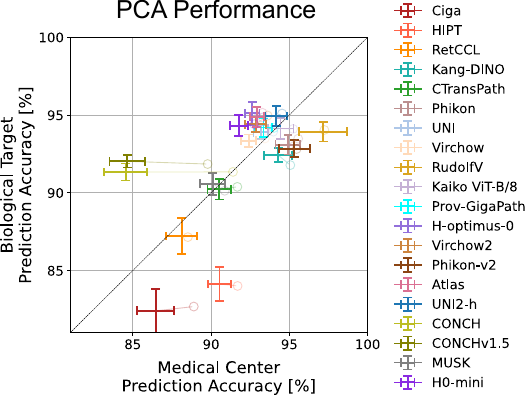}
    \caption{\textbf{Classification accuracy for predicting the medical center and biological class before and after reducing the dimensionality of the representations using PCA.} The representation spaces were reduced to their first 10\% of principal components with PCA applied individually per dataset and FM. We report mean prediction accuracies with 95\% confidence intervals over 20 repetitions and three held-out test sets (Camelyon, TCGA~4x4, Tolkach ESCA), corrected to remove common variance due to dataset (Masson~\&~Loftus~\cite{Masson2003}). Only minor differences in accuracy were observed, indicating that medically relevant and medical center information is strongly encoded in the representation space.}
    \label{fig:pca_top10}
\end{figure}

First, we predicted the medical center and biological class from the reduced representations. We used the same linear probing model training setup as described in Sup.~Note~\ref{sec:apx_training_details} but on the PCA-reduced embeddings derived using the sklearn implementation\footnote{\url{https://scikit-learn.org/stable/modules/generated/sklearn.decomposition.PCA.html}}. We considered representations projected to their first 10\% of principal components. The results are shown in Figure~\ref{fig:pca_top10}. The biological target and the medical center classification performance using only the top $10\%$ of the principal components remained comparable to that using the full-dimensional embeddings, implying that medically relevant and irrelevant medical center information are both strongly encoded in the linear directions of the representation space, and are readily available and accessible by downstream models. 
Given that downstream models tend to exploit easily accessible linear features (as a Clever Hans feature or shortcut), even if less predictive \cite{hermannfoundations, geirhos2018imagenet}, this explains both the strong supervised downstream model performance on balanced data and their vulnerability to spurious correlations (cf.~Figure~\ref{fig:sketch-downstream}, see also Sup.~Note~\ref{sec:apx_downstream_entanglement} for a thorough discussion). 

Second, we analyzed the class-separability of the biological class or the medical center within the individual PCs by measuring AUROC scores for the biological target and medical center classification. This is to assess how these signals are encoded in the embedding space, how prevalent they are in the most important directions, and whether they are polysemantic.

Specifically, we first projected the embeddings on each PC individually, which gave one-dimensional representations. Then, we analyzed how well these one-dimensional representations can be linearly separated with respect to either their medical center of origin or their biological class. For this, we measured their separability via a multi-class one-vs-one area under the receiver operating characteristic curve (AUROC) using the one-dimensional projections as scores and the medical center or biological classes as labels. To ensure that biological and medical center AUROCs are comparable, we used the paired datasets Camelyon~2x2, TCGA~4x4, and a reduced Tolkach~2x2, with the medical centers WNS and CHA and the biological classes SH\_OES and SH\_MAG. The AUROCs range from $0.5$, indicating no separation is possible, to $1$, with perfect separability (PC fully explains the class separability and target). They naturally handle class imbalances, guarantee comparability across models as they are always in $[0.5, 1]$, and are nonparametric since it does not rely on a particular choice of the classification algorithm. PCs for which both signals have an AUROC exceeding $0.6$, which indicates that both signals are linearly pronounced, are considered \textit{polysemantic} (cf.~\cite{elhage2022superposition}). Polysemantic features encode both biological and confounding information.
The depth of the encoding, i.e., the number of PCs for which a significant AUROC $> 0.6$ is observed, reflects the complexity of the data representation in the embedding space (loosely comparable to the dimensionality of the subspace that encodes relevant information). 

\begin{figure}[h!]
    \centering    \includegraphics[width=0.78\linewidth]{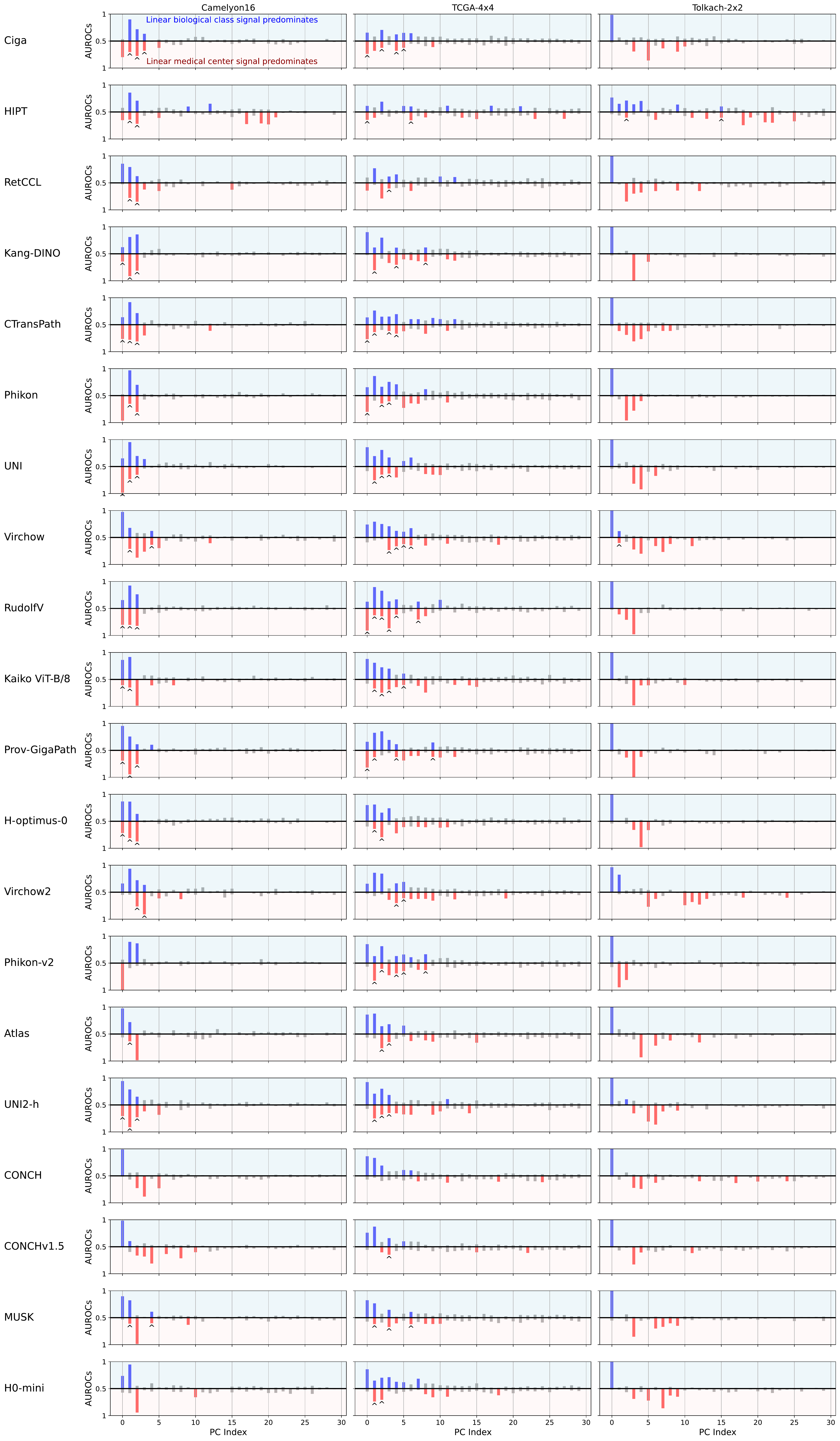}
    \caption{\textbf{Biological and medical center separability per PC}. Separability is measured by pairwise AUROCs, with $1$ indicating strong separability. For many FMs,  both signals were highly separable and polysemantic (corresponding PCs with both AUROCs $>0.6$ are marked (\^{})).}
    \label{fig:pca_aurocs}
\end{figure}

The results are shown in Figure~\ref{fig:pca_aurocs}. All models encoded both biological and medical center information in their early PCs, with the highest separability often occurring within the first 5 to 10 PCs --- sometimes even approaching perfect separation. Thus, the largest variations in the embeddings align with variations in the biological or medical center label. However, the degree of prominence and polysemy of these signals varied across models and datasets. For Tolkach ESCA, there is a strong encoding of the biological signal in the very first PC and almost no polysemy. This indicates that distances between patches in the embedding space mostly reflect biologically relevant similarities. This matches the good robustness indices and high clustering scores (cf.~Figure~\ref{fig:robidx_clusidx_overview}b). In contrast, for TCGA, both signals are encoded deeply, and principal directions are polysemantic. This provides an explanation for the low robustness indices and clustering scores, as the data has complex embedding representations that reflect medically relevant and confounding information equally. For Camelyon, we observe a strong encoding in the first PCs and, depending on the FM, some polysemy. In general, the representation spaces of older models and \mrudolfv{} include polysemantic directions, in contrast to \mconch{}, \mconchONEFIVE{}, and \mhmini{} whose principal directions mostly encode either biological or center information.
Due to the polysemy, selectively isolating biological signals without retaining confounding information might become difficult.
Accordingly, robustification of FMs with polysemantic directions is precarious; removing these features to eliminate confounding signals will also discard relevant biological information, which reduces the utility of the representation.

Similarly, as the directions of greatest variance hold both relevant and irrelevant information, pairwise distances are affected by both and reflect differences in biology and artifacts. This explains the clustering performance and robustness indices observed in Figures~\ref{fig:robustness_index}~and~\ref{fig:robidx_clusidx_overview}, which verified that pairwise distances in the representation space reflect not only morphological differences of the patches, but also variations in their medical center origin.

While the previously described approaches already illuminated the structure of representation spaces, advanced Explainable AI (XAI) techniques might deliver deeper insights. Concept-based XAI (e.g.~\cite{DBLP:journals/natmi/AchtibatDEBWSL23,chormai2024disentangled}), for instance, may reveal human-understandable concepts encoded by the latent features within these spaces. When applied to data shift analysis (e.g.~\cite{naumann2025wasserstein}), such methods can uncover directions along which distributional differences are most pronounced, enabling assessment of whether they reflect meaningful variation or spurious signals---such as class separation driven by confounders like medical center bias. These techniques thus support robustness analysis by revealing whether observed class differences and learned representations are shaped by genuine biological signals or underlying biases.

\FloatBarrier
\section{Further downstream experiment results}

\subsection{Patch retrieval from heterogeneous databases} \label{sec:apx_retrieval_results}


We further assessed the efficacy of unsupervised patch retrieval in unrobust feature spaces: given a patch of interest, retrieve a biologically similar patch from a database using the foundation model representations. This clinically relevant setting has recently gained attention \cite{Dippel2024RudolfV, tizhoosh2024retrieval, alfasly2025retrievalval}. Although retrieval is more commonly performed on the slide level, algorithms are regularly based on patch-level distances.


Identical to the supervised downstream model experiments (see Section~\ref{sec:downstream_experiments}, Figure~\ref{fig:sketch-downstream}, and Section~\ref{sec:methods_downstream}), we sampled patch databases with increasing correlation between the medical center and biological class, and evaluated whether patches retrieved from the database had the same biological class as the query patches. For three PathoROB datasets (Camelyon, TCGA~4x4, Tolkach ESCA), we considered splits with increasing levels of correlation between medical centers and biological targets (see Figures~\ref{fig:camelyon_dataset_splits},~\ref{fig:tcga_dataset_splits},~\ref{fig:tolkach_dataset_splits}). As a retrieval database, we used the FM embeddings of all patches of the respective training split. For each query patch from the test set (both in-domain and out-of-domain), we retrieved the most similar patch from the database based on the cosine distance between the two embedding vectors. We then assessed whether the retrieved patch belonged to the same biological class as the query patch, resulting in an accuracy score for the test set (= the set of query patches). From these accuracies, we derived the average performance drop measure and correlated it with the robustness index (see Section~\ref{sec:methods_downstream}). Notice that in practice, the patch retrieval experiments correspond to exchanging the linear probing downstream model for a $k$-nearest neighbors model with $k=1$.

\begin{figure}[h!]
    \centering
    \includegraphics[width=0.95\linewidth]{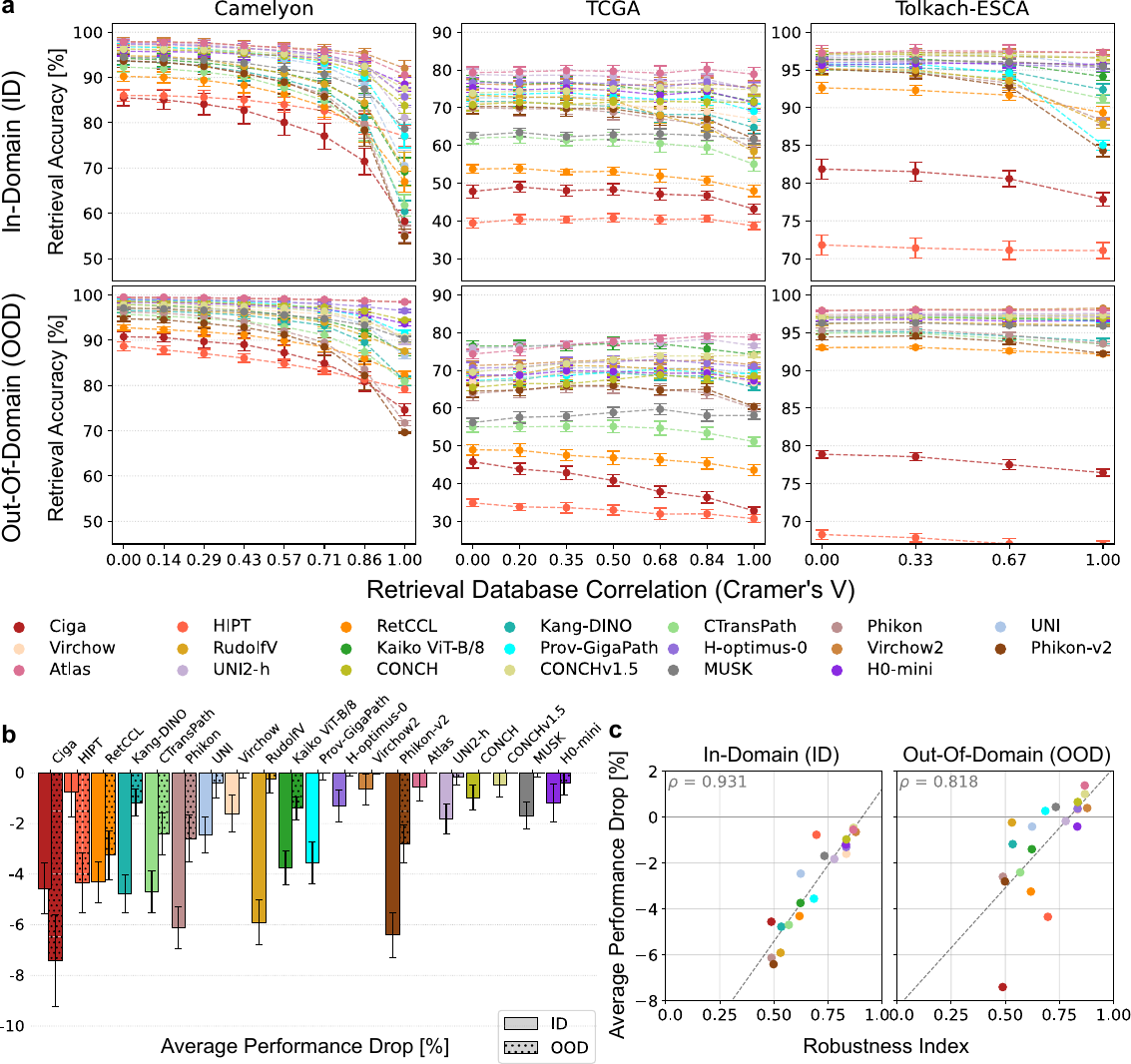}
    \caption{\textbf{Retrieval performances from retrieval databases with heterogeneous data contributions across medical centers}. \textbf{a}~In-/out-of-domain retrieval performance on data from seen/unseen hospitals. Retrieval accuracies are reported with 95\% confidence intervals over 20 resampling repetitions for each model–dataset combination. \textbf{b}~In-/out-of-domain average retrieval performance drops per model over all datasets and repetitions with 95\% confidence intervals. \textbf{c}~Correlation between the robustness index averages and the in-/out-of-domain average retrieval performance drops of the models (p-values: $0.00004$, $0.00008$.}
    \label{fig:sketch-downstream-retrieval}
\end{figure}

%
We observed retrieval performance drops when retrieval databases were heterogeneous across medical centers (Figure~\ref{fig:sketch-downstream-retrieval}), meaning that retrieval is prone to errors if the retrieval database cannot be guaranteed to have balanced data contributions from all medical centers.
For Camelyon, for example, the in-domain retrieval performance dropped from 85\%-98\% to 55\%-92\%. For TCGA and Tolkach ESCA, the decreases were less steep, but still present for all foundation models. On out-of-domain data, the best foundation models (\mvirchow{}, \mhoptimus{}, \mvirchowTWO{}, \mconch{}, \mconchONEFIVE{}, \mmusk{}) enabled stable performances on imbalanced retrieval databases, whereas other models also suffered from diminishing retrieval accuracy.
At the same time, the magnitude of the average performance drop was strongly correlated with the robustness of the foundation model ($\rho=0.931$ for ID; $\rho=0.818$ for OOD), and more robust foundation models led to more stable retrieval results.

\subsection{Downstream model experiments on robustified FM representations} \label{sec:apx_robustification_downstream_results}

\begin{figure}[h!]
    \centering    \includegraphics[width=0.9\linewidth]{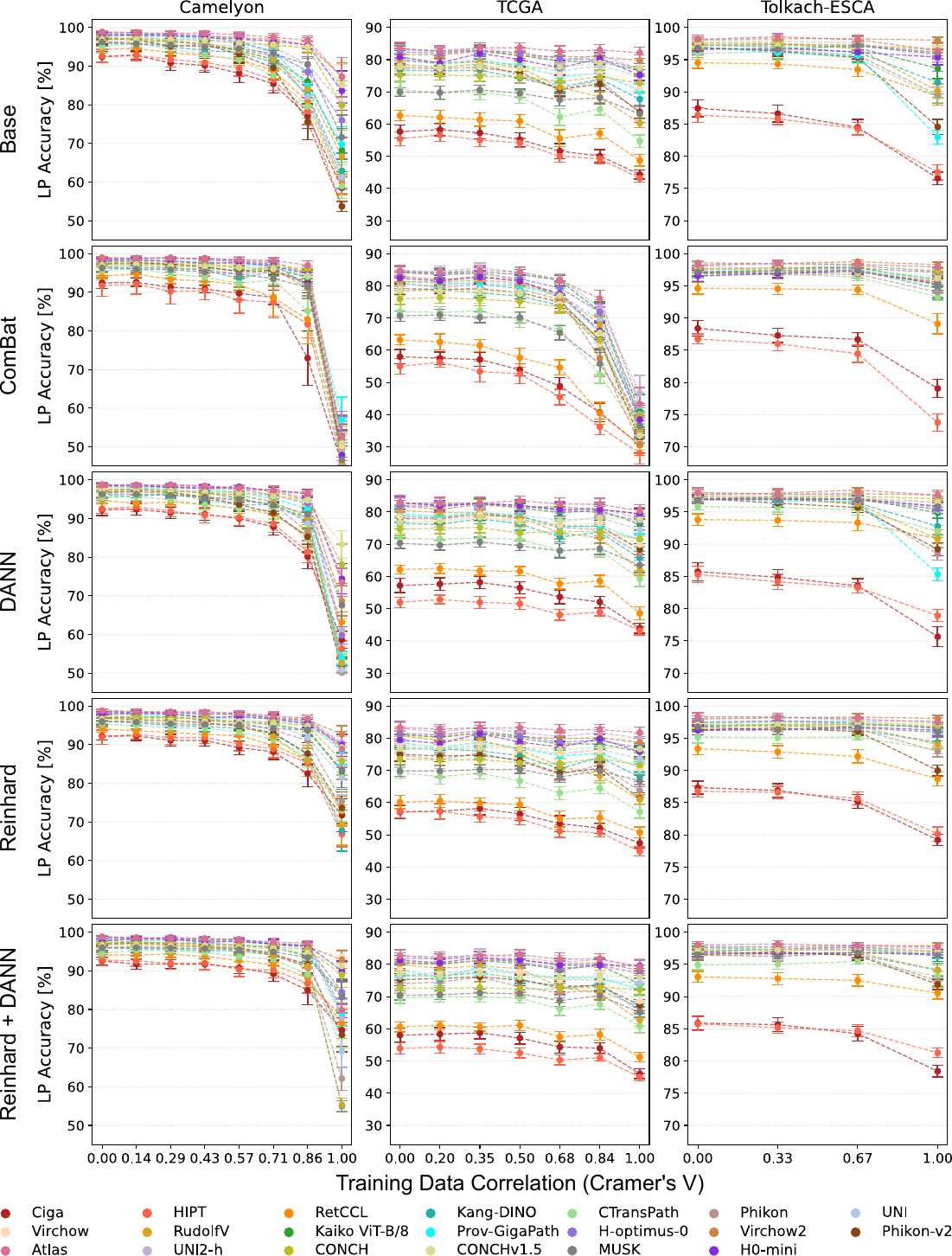}
    \caption{\textbf{Downstream model generalization performance on robustified feature spaces}. In-domain generalization performance on data from seen hospitals for different post-hoc robustification methods. Linear probing accuracies are reported with 95\% confidence intervals over 20 resampling repetitions for each model–dataset combination.}
    \label{fig:mitigation_downstream_per_model}
\end{figure}

\begin{figure}[h!]
    \centering    \includegraphics[width=0.95\linewidth]{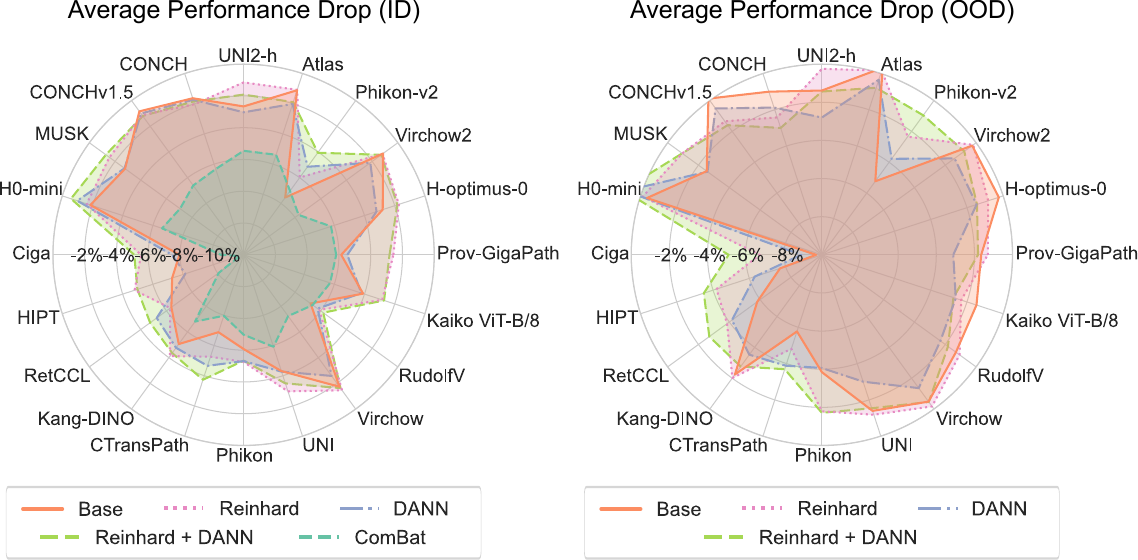}
    \caption{\textbf{Downstream model generalization performance on robustified feature spaces}. The effect of post-hoc robustness improvement methods on downstream performance drops (in-domain and out-of-domain) across foundation models. For in-domain performance, ComBat is included. For out-of-domain performance, however, ComBat led to non-competitive results far below the Base setting.}
    \label{fig:mitigation_downstream_all}
\end{figure}

We reran the experiments to assess Clever Hans learning of downstream models under heterogeneous training data (Section~\ref{sec:robustification}, Figure~\ref{fig:sketch-mitigation}c, Methods Section~\ref{sec:methods_downstream}) on the robustified embedding spaces. Figure~\ref{fig:mitigation_downstream_per_model} provides further insights on how robustification methods impact downstream models and Clever Hans learning~\cite{clever-hans, geirhos2020shortcut, hermannfoundations, unsupervised-clever-hans, Howard2021, mahmood-demographic}.

ComBat~\cite{johnson2007combat, murchan2024combat} showed in-domain generalization performance improvements on training data with low or moderate spurious correlations. It even mostly outperformed Reinhard and DANN up until Split 6 ($V=0.71$) out of 8 (Camelyon) and Split 4 ($V=0.5$) out of 7 (TCGA). This is in line with previously reported results \cite{murchan2024combat}. Nevertheless, the method exacerbated generalization performance drops for fully correlated data in two datasets, specifically when the medical centers did not contribute any data for at least one biological class. In Camelyon, for example, all FMs drop below 60\% performance, which is considerably worse than without any robustification (Base setting). This suggests that, in this case, ComBat also removes relevant biological information. Consequently, ComBat may be a promising robustification option if training data imbalances are guaranteed to be limited, and test data are known to come from the same medical centers as the training data.

The DANN method~\cite{DBLP:journals/jmlr/GaninUAGLLML16} also led to better generalization performance for training data with low or moderate spurious correlations. On fully correlated training data ($V=1$), though, it did not yield improvements over the baseline setting without embedding space robustification. This is not surprising, as the medical center-based domain loss does not provide complementary information to the biological class prediction loss in this case. Therefore, using DANN alone is arguably not advisable when some medical centers did not contribute any data for some biological classes.

In contrast, Reinhard stain normalization~\cite{reinhard} strongly improved the accuracy under fully correlated training data ($V=1$) in many cases (although not all cases). For Camelyon, in particular, the linear probing accuracies in the last split were between 67\%-93\% --- substantially higher than for the baseline setting with 53\%-87\% average accuracy. At the same time, the generalization performances on splits with low to moderate correlations did not always improve compared to Base, indicating that stain normalization may also remove some biologically relevant information. Therefore, it is arguably most effective in extremely unbalanced scenarios.

Using Reinhard and DANN sequentially combined the advantages of both methods, i.e., it led to improved performances across foundation models, datasets, and splits. Therefore, it is arguably the best overall choice when the composition of the training data is unknown or variable. However, the experiments also show that only using one of the two may yield better results in specific situations. For example, only applying Reinhard stain normalization performed better on Split 8 ($V=1$) in Camelyon.

Figure~\ref{fig:mitigation_downstream_all} gives an overview of the overall generalization performances of all robustification methods in the downstream model evaluation. Notably, ComBat exacerbated performance drops on both in-domain and out-of-domain data due to its poor performance on strongly correlated training data. On unseen out-of-distribution test data -- where the base performance drops were lower than on in-distribution data -- we find that DANN only led to better performance for a few FMs. Reinhard stain normalization, in contrast, could increase performance for 14 out of 20 FMs. Stain normalization was most effective for lower-performing models, as the prediction ability of the higher-performing models was already saturated. Notably, \mhmini{} in combination with domain-adversarial training enabled stable downstream models on out-of-distribution data.

\subsection{Clustering performance independent of the silhouette score-based $K$ selection}
\label{sec:apx_upperbound_cs}

\begin{figure}[h!]
    \centering
    \includegraphics[width=0.82\linewidth]{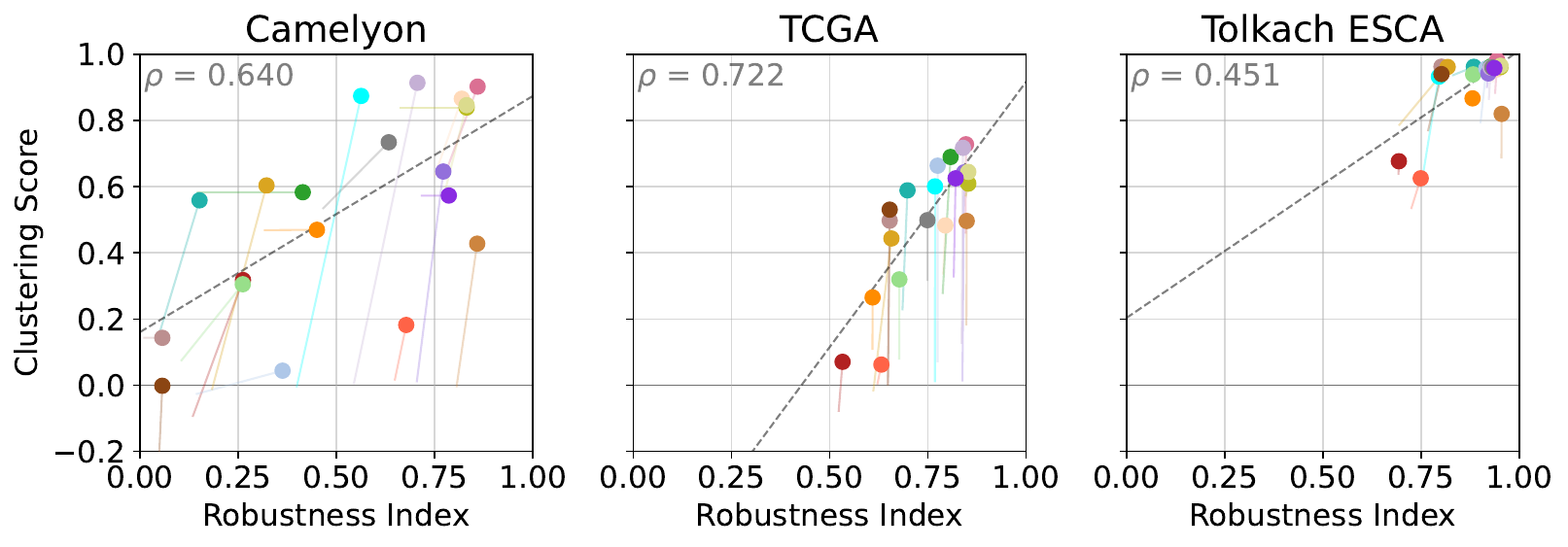}
    \includegraphics[width=0.82\linewidth]{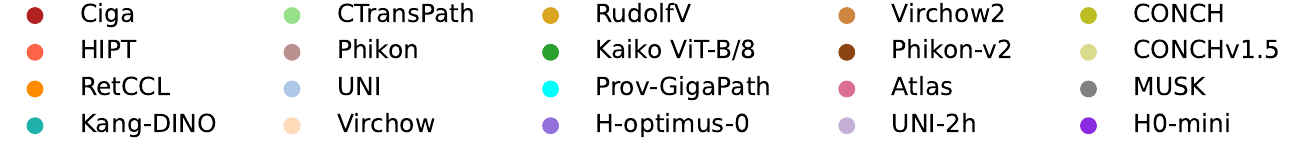}
    \caption{\textbf{Maximum clustering scores obtainable with $K$-means and maximum robustness index per dataset.} The number of clusters $K$ was selected by maximizing the clustering score, rather than by automatic selection. Similarly, the number of neighbors $k$ for the robustness index was chosen to maximize the robustness index. Selecting $K$ using the silhouette score seemed to have a strong impact on the clustering performances for Camelyon and TCGA, while the robustness index showed only minor differences to its default setting. Yet, both datasets remained challenging to cluster. Strong correlations between the two maximal scores were observed with p-values $0.0029$, $0.0007$, and $0.0463$ resp.}
    \label{fig:apx_upperbound_cs}
\end{figure} 

The initial experimental setup was designed to emulate a realistic scenario where the number of clusters $K$ is determined automatically by maximizing the silhouette score. Although the silhouette score is not the final optimization objective, it serves as a useful heuristic for selecting $K$, particularly in scenarios such as subtyping, where the ground-truth number of clusters is typically unavailable. However, like any heuristic, the silhouette score may fail in certain cases.

To investigate inconsistencies observed between the robustness index and the clustering score --- specifically on the Camelyon dataset, where some FMs exhibited high robustness indices but low clustering scores --- we conducted an ablation study to remove the effect of a potentially suboptimal $K$ selection. In this analysis, we performed $K$-means clustering and evaluated the clustering score across a range of $K \in [2, 100]$. The highest observed clustering score within this range serves as an upper-bound on the clustering score attainable with $K$-means clustering that is independent of $K$.
Similarly, the number of neighbors $k$ for evaluating the robustness index was not set to the median of the optimal values across FMs (for details see Sup.~Note~\ref{sec:apx_ri_resultsperset}), but was instead selected individually for each FM by maximizing the robustness index. This yields an upper-bound on the robustness index that is independent of $k$.

As shown in Figure~\ref{fig:apx_upperbound_cs}, this upper-bound analysis revealed further insights. On the Tolkach ESCA dataset, already high clustering scores improved further, with many FMs achieving near-perfect clustering scores close to 1. The TCGA dataset, despite some improvement, remained the most challenging, reflecting an inherent difficulty in distinguishing certain tumor types. In particular, in most FM representation spaces, the LUAD-LUSC combination, especially from Asterand and Christiana Healthcare, remained difficult to cluster sensibly, while BRCA and COAD consistently formed well-separated clusters.

In contrast, stark discrepancies between the clustering scores, obtained with the silhouette score selection of $K$, and their upper-bounds were observed in the Camelyon dataset. For several FMs, such as \mrudolfv{}, \mprovgigapath{}, \muniTWO{}, \mhoptimus{}, and \mvirchowTWO{}, clustering performance improved dramatically, with scores increasing from near 0 (reflecting a clustering based on biological information and the medical center origin jointly) to over 0.8 (indicating a compliance of the clustering with the tumor status). For these FMs, the silhouette score determined highly suboptimal values (e.g., $K > 26$) and exhibited instability, with low and fluctuating scores across the initially tested range of $K$.

Exceptions to this trend were noted for \mphikonTWO{} and \muni{}, where clustering performance remained poor regardless of $K$. For these FMs, even an optimal $K$ choice yielded clustering scores close to 0. Their learned representation spaces appear to be dominated by medical center information rather than biological signals.

One potential reason for the silhouette score's failure in selecting appropriate $K$ values is its sensitivity to its assumptions on the cluster structure. The silhouette score favors clusterings with high intra-cluster cohesion and well-separated inter-cluster distances. In cases where clusters have heterogeneous densities or varying degrees of compactness, the silhouette score tends to penalize even semantically meaningful partitions, leading to poor selections of $K$. The inability of the silhouette score to identify a suitable $K$ for some FMs (\mrudolfv{}, \mprovgigapath{}, \muniTWO{}, \mhoptimus{}, and \mvirchowTWO{}), despite the datasets of the $2\times2$ pairings being completely balanced in terms of sample size, suggests the presence of heterogeneous subgroups of embeddings within the learned representation spaces. These subgroups appear not to correspond to either tumor status or the medical center origin, indicating that these FMs’ representation spaces may be structured by an additional source of variation.

\subsection{Clustering experiments on robustified FM representations}
\label{sec:apx_robustified_clustering}

\begin{figure}[h!]
    \centering
    \includegraphics[width=0.82\linewidth]{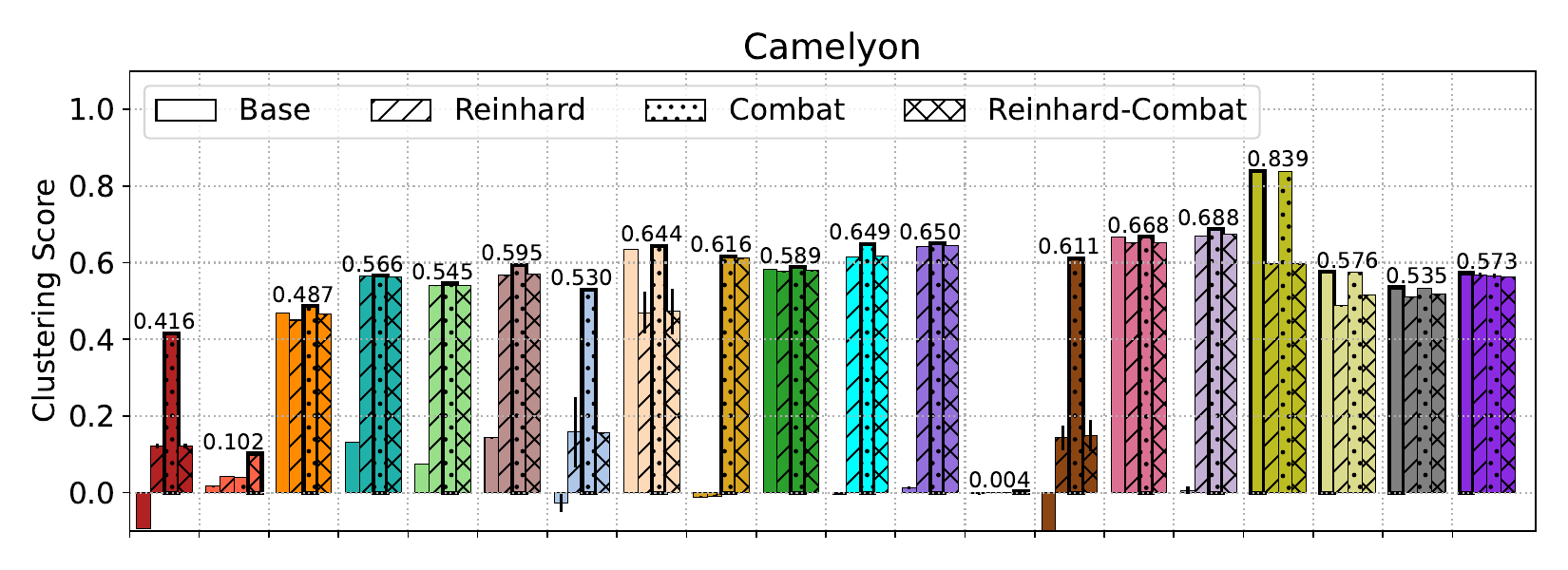}
    \includegraphics[width=0.82\linewidth]{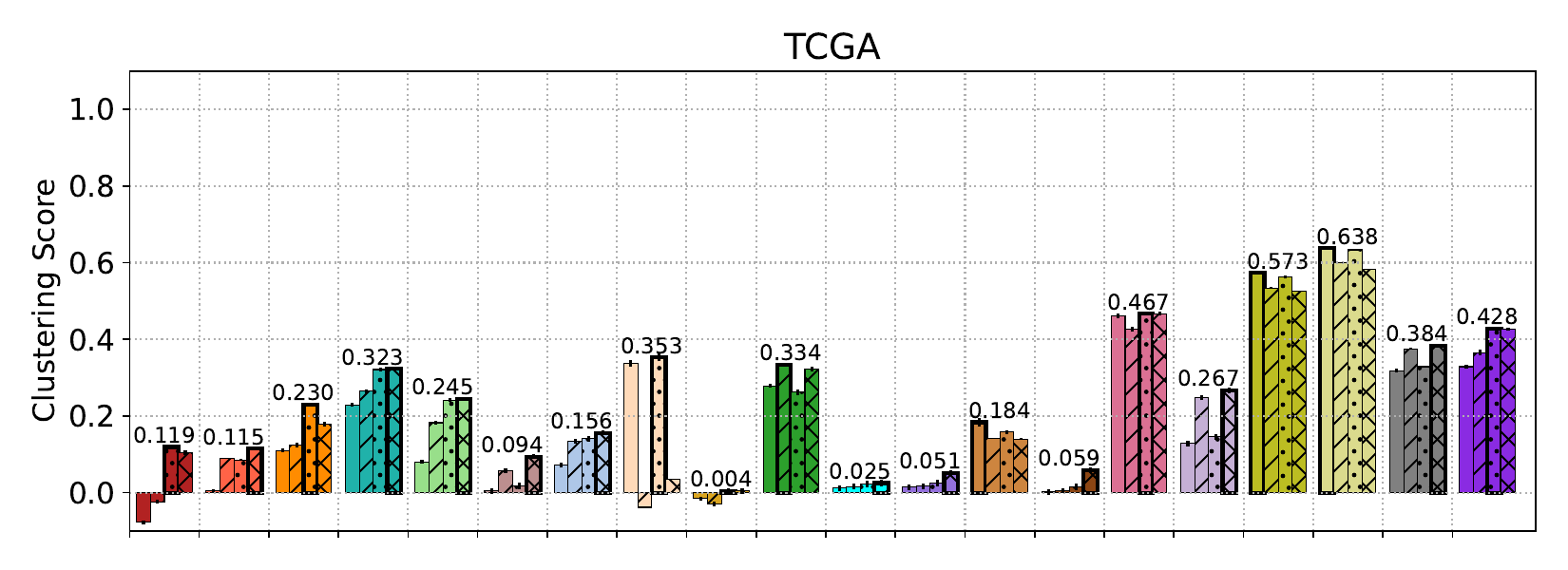}
    \includegraphics[width=0.82\linewidth]{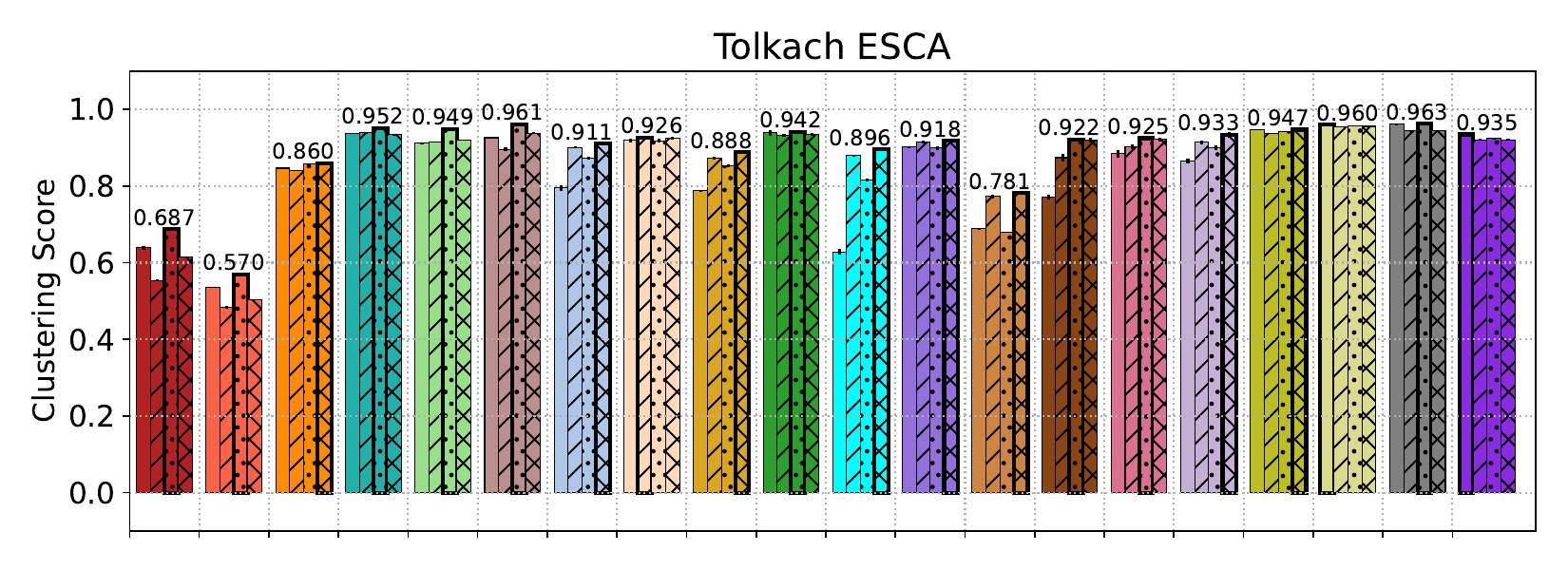}
    \includegraphics[width=0.82\linewidth]{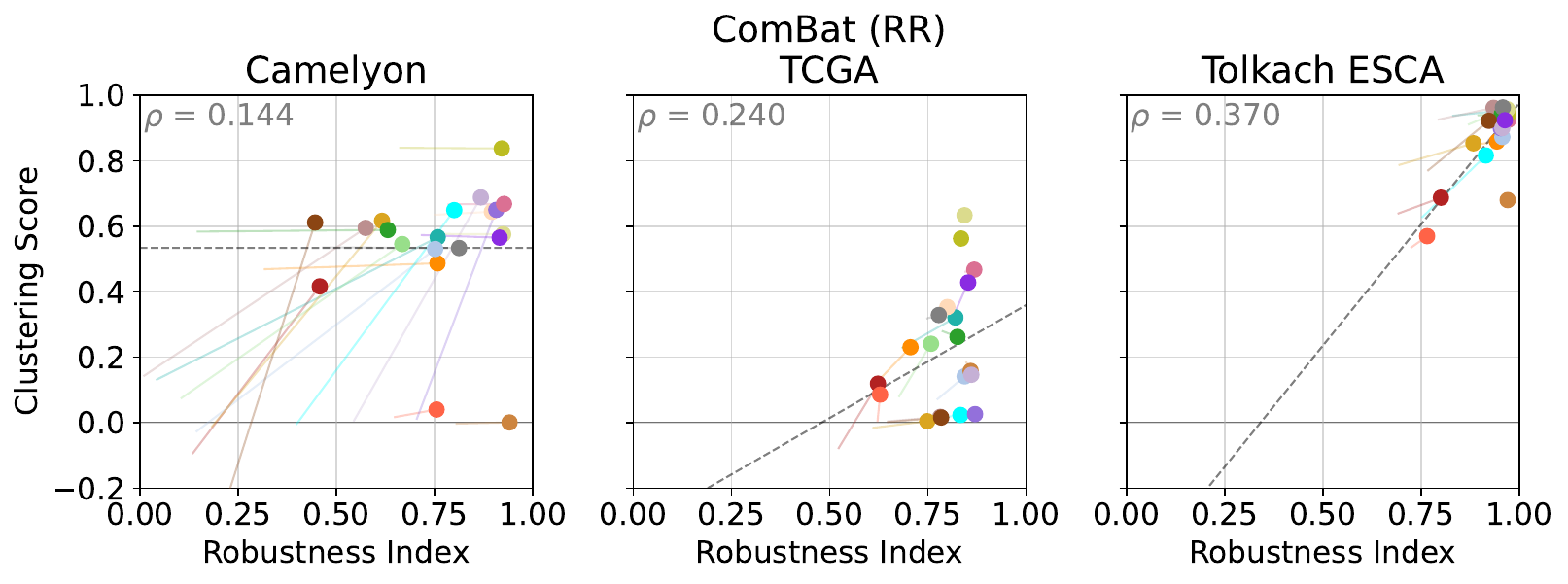}
    \includegraphics[width=0.82\linewidth]{figures/clustering/overview_combat_lgd.pdf}
    \caption{\textbf{Effect of robustication on the clustering performance per dataset.} Robustifying the representation spaces led to strong improvements in clustering scores across FMs. (bottom) The correlation between the robustness index and the clustering score was no longer significant on the ComBat (RR) robustified representation spaces (p-values $0.5555$, $0.3227$, and $0.1189$ resp.), since ComBat is most effective on the less robust models.}
    \label{fig:apx_clustering_mitigation}
\end{figure}

The clustering experiments described in Sup.~Note~\ref{sec:apx_clustering_score_details} were reevaluated on the robustified embedding spaces. To ensure comparability with the baseline results (i.e., without embedding space robustification), the experimental setup was kept identical. The outcomes are summarized in Figure~\ref{fig:apx_clustering_mitigation}. Overall, the impact of robustification on clustering performance varied considerably depending on both the dataset and the foundation model under evaluation.

\paragraph{Dataset-wise results}
For the Tolkach ESCA dataset, most FMs already exhibited relatively high clustering scores prior to robustification. Consequently, only minor improvements were observed, with the best-performing strategy --- most frequently ComBat --- yielding an average relative increase of $7.11\%$. In contrast, models with initially low clustering scores (e.g., \mciga{}, \mhipt{}, \mprovgigapath{}, and \mvirchowTWO{} with $<0.7$) benefited more substantially, achieving an average improvement of $17.51\%$. The largest relative gain ($42.61\%$) was observed for \mprovgigapath{} with the combined application of ComBat and Reinhard. These findings suggest that biologically relevant patterns are inherently well-represented in the Tolkach ESCA dataset and are effectively captured by most FMs.

The TCGA dataset presented a greater challenge. It had the lowest baseline clustering scores across all datasets, and even after robustification, the average clustering score remained low ($0.25$). This suggests that medical center-specific information is strongly embedded and not easily removed by current robustification strategies. The feature space analysis (Sup.~Note~\ref{sec:apx_feature-space-analysis}) supports this, revealing that TCGA embeddings span a more entangled subspace, where the medical center origin is tightly coupled with biologically relevant features. Nevertheless, the combined ComBat and Reinhard strategy yielded a fourfold increase in clustering scores, particularly benefiting FMs with negative or near-zero baseline scores. However, for models such as \mvirchowTWO{}, \mconch{}, and \mconchONEFIVE{}, robustification sometimes disrupted biologically meaningful groupings, reducing clustering scores by an average of $-5.57\%$.

In the Camelyon dataset, robustification led to a marked unification and improvement in clustering outcomes across FMs, with an average $22$-fold increase. It had a strong impact on both the clustering score and the robustness index, making the scores more aligned. This suggests that while the medical center origin was encoded, it is relatively straightforward to remove. For recent FMs (e.g., \matlas{}, \mconch{}, \mconchONEFIVE{}, \mmusk{}, and \mhmini{}) that already exhibited biologically structured embeddings, robustification had a negligible impact. In contrast, models whose embeddings were initially influenced by confounding variables (e.g., \mhipt{}, \muni{}, \mrudolfv{}, \mphikonTWO{}) saw substantial improvements. ComBat was generally the most effective method on this dataset.

\paragraph{Robustification method performance}
ComBat, and in some cases its combination with Reinhard, proved most effective, particularly for embedding spaces heavily influenced by medical center information. When performance diverged between ComBat and Reinhard, their combination typically yielded results aligned with those from Reinhard alone, with a few notable exceptions (e.g., \mrudolfv{} on Camelyon, \mciga{}, \mkangdino{}, and \mctranspath{} on TCGA). Robustification was particularly beneficial for FMs with low or negative baseline clustering scores, enabling a shift toward biologically meaningful groupings. In contrast, FMs with embeddings already structured according to biological differences showed limited or even adverse responses to robustification.

\paragraph{Foundation model-wise results}
Among all models, \mconch{}, \mconchONEFIVE{}, and \matlas{} consistently achieved the best clustering performance, reflecting a strong alignment of their embedding spaces with biological similarity; both with and without robustification. \mhmini{}, \mmusk{}, and \mkaiko{} also ranked highly. Conversely, older models such as \mciga{} and \mhipt{} performed poorly, with embeddings strongly reflecting medical center origin. While most FMs benefited from robustification, \mrudolfv{}, \mphikon{} variants, \mvirchowTWO{}, \mprovgigapath{}, and \mhoptimus{} exhibited inconsistent behavior across datasets and robustification strategies.



\end{document}